\documentclass[a4paper,11pt]{article}
\usepackage{jheppub} 
\usepackage{subfigure, subcaption, graphicx}
\usepackage{multicol}
\usepackage{lineno}
\usepackage{placeins}
\usepackage{booktabs}  
\usepackage{parskip}

\title{\boldmath Deconfinement from Thermal Tensor Networks:\\
Universal CFT signature in (2+1)-dimensional $\mathbb{Z}_N$ lattice gauge theory}

\usepackage{tikz,xcolor}
\definecolor{lime}{HTML}{A6CE39}
\DeclareRobustCommand{\orcidicon}{%
	\begin{tikzpicture}
	\draw[lime, fill=lime] (0,0) 
	circle [radius=0.16] 
	node[white] {{\fontfamily{qag}\selectfont \tiny ID}};	\draw[white, fill=white] (-0.0625,0.095) 
	circle [radius=0.007];	\end{tikzpicture}
	\hspace{-2mm}}
\foreach \x in {A,...,Z}{%
	\expandafter\xdef\csname orcid\x\endcsname{\noexpand\href{https://orcid.org/\csname orcidauthor\x\endcsname}{\noexpand\orcidicon}}
}

\author[1]{Adwait Naravane\orcidN{},}
    \affiliation[1]{Department of Physics and Astronomy, Ghent University, Krijgslaan 281, 9000 Gent, Belgium}
    \emailAdd{adwait.naravane@ugent.be}
\author[2]{Yuto Sugimoto\orcidY{},}
\emailAdd{sugimoto@nucl.phys.tohoku.ac.jp}
\affiliation[2]{Department of Physics, Tohoku University, Sendai 980-8578, Japan}
\author[3,4]{Shinichiro Akiyama\orcidS{},}
\affiliation[3]{Center for Computational Sciences, University of Tsukuba, Tsukuba, Ibaraki
    305-8577, Japan}
\affiliation[4]{Graduate School of Science, The University of Tokyo, Bunkyo-ku, Tokyo, 113-0033, Japan}
\emailAdd{akiyama@ccs.tsukuba.ac.jp}
\author[1]{Jutho Haegeman,}
\emailAdd{jutho.haegeman@ugent.be}
\author[1]{Atsushi Ueda\orcidU\,}
\emailAdd{Atsushi.Ueda@ugent.be}

\preprint{UTHEP-818, UTCCS-P-175}
    
\abstract{
    Tensor networks offer a sign-problem-free approach to study lattice gauge theories, but extracting precise universal information associated with the deconfinement transition remains challenging. 
    In this work, we study the deconfinement transition of (2+1)-dimensional $\mathbb{Z}_N$ lattice gauge theories at finite temperature using a thermal tensor network approach, where the partition functions at finite temperature are formulated as three-dimensional tensor networks. 
    These tensor networks are first contracted in the temporal direction, and the subsequent coarse-graining in the spatial directions yields a renormalized transfer matrix, the spectrum of which directly encodes the universal conformal field theory data. 
    In particular, by numerically extracting the central charge and scaling dimensions, we verify that the universality class of the thermal deconfinement transition matches the prediction of the Svetitsky--Yaffe conjecture for $N=2,3,5$. 
    Moreover, we show that the $\mathbb{Z}_5$ theory at finite temperature exhibits an intermediate phase with an emergent U(1) symmetry. 
    Critical couplings are determined via Gu--Wen ratios and agree with existing Monte Carlo simulations. 
    Finally, extrapolating these critical couplings at finite temperature enables us to determine the deconfinement transition points for $N=2,3$ at zero temperature.
}

\begin{document}
\maketitle
\flushbottom

\section{Introduction}
\label{sec:intro}

Gauge theories and gauge invariance are concepts of universal importance across modern quantum physics, ranging from condensed matter physics to the standard model of particle physics.
An almost defining feature of gauge theories is the phenomenon of confinement~\cite{Wilson:1974sk}, whereby charged constituents cannot exist in isolation. 
As local gauge symmetry cannot be spontaneously broken~\cite{Elitzur:1975im}, the transition between confined and deconfined phases must be identified through non-local topological objects such as the Wilson loop. 
In particular, the thermal Wilson loop, which is known as the Polyakov loop, serves as an order parameter for the center symmetry, which is a prototypical example of a 1-form symmetry~\cite{tHooft:1977nqb,Polyakov:1978vu,Gaiotto:2014kfa}.

While gauge theories were initially conceived as quantum field theories, the lattice formulation of gauge theories have proven instrumental in their study.
The lattice provides a UV regulator of the theory that preserves gauge invariance exactly, thereby yielding a non-perturbative definition of the theory. Furthermore, Monte Carlo (MC) simulations have played a central role in investigating lattice gauge theories, particularly in the context of quantum chromodynamics (QCD).
Creutz achieved a pioneering milestone in 1980, by demonstrating the coexistence of confinement and asymptotic freedom in the SU(2) Yang--Mills theory via the MC simulation~\cite{Creutz:1980zw}.
Creutz also reported a series of MC studies on the Abelian lattice gauge theories~\cite{Creutz:1979zg,Creutz:1979he,Bhanot:1980pc,PhysRevB.22.3370}.
Despite their simplicity, the $\mathbb{Z}_{N}$ lattice gauge theories provide a minimal yet non-trivial setting in which the confinement and deconfinement phases are realized.
They can also provide relevant insights into the SU($N$) gauge theories, because $\mathbb{Z}_{N}$ is the center of SU($N$).
In addition, duality transformations between the $\mathbb{Z}_{N}$ lattice gauge theories and $\mathbb{Z}_{N}$ spin models constitute another striking feature: they enable us understand the phase structure via well-known statistical systems~\cite{Balian:1974ir,Ukawa:1979yv,Savit:1979ny}.
Svetitsky and Yaffe further conjectured that the thermal phase transition in a $(d+1)$-dimensional lattice gauge theory with gauge group $G$ belongs to the same universality class as that of a $d$-dimensional spin model whose global symmetry is given by the center of $G$~\cite{Svetitsky:1982gs}.

Recently, tensor networks, originally developed and extensively used in condensed matter physics, have been attracting growing interest as an alternative computational approach in high-energy physics, offering a complementary perspective to traditional MC simulations.
Tensor-network methods are free of the sign problem. This makes tensor networks especially attractive for QCD at finite density, although methods for treating dynamical gauge degrees of freedom are still under development.
Early studies focused on the Schwinger model, which is regarded as one of the most QCD-like theories in two dimensions. To date, many studies have used the Hamiltonian formalism~\cite{Byrnes:2002nv,Banuls:2015sta,Buyens:2015tea,Buyens:2017crb,Dempsey:2023gib,Itou:2024psm,ArguelloCruz:2024xzi,Fujii:2024reh} and the Lagrangian formalism~\cite{Shimizu:2014uva,Shimizu:2014fsa,Shimizu:2017onf,Butt:2019uul,Kanno:2024elz}.
More recently, it has been demonstrated that MC simulations of the Schwinger model with a topological term, formulated via bosonization, provide results that are consistent with tensor-network calculations~\cite{Ohata:2023sqc, Ohata:2023gru}.

For the further development of tensor-network methods, particularly with an eye toward their application to QCD at finite density, it is essential to address their efficiency when applied to higher-dimensional lattice gauge theories.
In this context, tensor networks have been applied to investigate abelian gauge theories not only in (2+1) dimensions~\cite{Unmuth-Yockey:2018xak,Felser:2019xyv,Bender:2023gwr,Wu:2025aly, TagliacozzoVidalGauge, canals2024tensornetworkformulationlattice, Ba_uls_2020} but also in (3+1) dimensions~\cite{Magnifico:2020bqt,Akiyama:2022eip,Akiyama:2023hvt}, and QCD at finite density in the strong-coupling limit~\cite{Sugimoto:2025vui,Sugimoto:2026wnw}.

Despite this progress, applying tensor networks to deconfinement transitions in gauge theories remains challenging. 
The deconfined phase encompasses the high temperature regime, which is where tensor networks traditionally perform well in the case of spin systems, because of the reduced amount of correlations.
In gauge theories, however, the difficulty originates from the increased number of dynamical gauge degrees of freedom in the deconfined phase. 
In non-Abelian gauge theories such as the SU($N$) Yang-Mills theory, the effective number of states scales as $N^2$, while even in the U(1) theory, the presence of massless photons makes tensor-network computations more demanding than in the confined phase.
Consequently, the applications of tensor networks to study deconfinement transitions are limited to a few studies on the $\mathbb{Z}_{N}$ gauge theories with $N=2$~\cite{Kuramashi:2018mmi}, and $N=3$~\cite{Robaina:2020aqh,Emonts:2020drm}. 
For $N>4$, however, substantially richer physics emerges. According to the Svetitsky--Yaffe conjecture, the universality class of the finite-temperature transition is governed by the $N$-state clock model. For $N>4$, this entails an intermediate critical phase with an emergent U(1) symmetry, separating the confined and deconfined phases by the Berezinskii--Kosterlitz--Thouless (BKT) transitions. Historically, characterizing these transitions has proven notoriously difficult for MC simulations. 
In this sense, this constitutes an ideal testing ground for the capabilities of tensor networks. 

In this work, we study the phase structure of the (2+1)-dimensional pure $\mathbb{Z}_N$ lattice gauge theories, for $N=2,3,5$, using the thermal tensor network renormalization approach proposed in Ref.~\cite{Ueda:2025mhu}.
Our central objective is to numerically identify the nature of the transition from the universal conformal field theory (CFT) data, thereby providing a quantitative test of the Svetitsky--Yaffe conjecture. A key advantage of tensor networks is their ability to directly access such universal information, which lies beyond standard local observables. Furthermore, we demonstrate that the critical gauge couplings at finite temperature are sufficiently accurate to admit a controlled extrapolation to a critical coupling for the transition point at zero temperature.

This paper is organized as follows. 
In Section~\ref{sec:model}, we formulate the $\mathbb{Z}_{N}$ lattice gauge theory in terms of tensor networks and describe its duality with the $N$-state clock model.
Sec.~\ref{sec:methods} summarizes the tensor network algorithms that were used, , in which we introduce an improved thermal tensor network renormalization algorithm based on optimal projectors along the temporal direction. 
Section~\ref{sec:results} presents th numerical results for CFT data, including the central charge and scaling dimensions, as well as a ratio of partition functions~\cite{Gu:2009dr}, and determine the critical gauge coupling in the zero-temperature limit.
Section~\ref{sec:summary} is devoted to the summary and outlook.

\section{Tensor network formulation}
\label{sec:model}

We consider the $\mathbb{Z}_N$ lattice gauge theory defined by the standard Wilson gauge action,
\begin{align}
\label{eq:wilson_S}
    S[U_{n,\mu\nu}] 
    = 
    \beta
    \sum_{n\in\Lambda_{2+1}}\sum_{\mu<\nu}
    \left[
        1 - \Re U_{n,\mu\nu}
    \right],
    \qquad \text{with} \quad
    U_{n,\mu\nu}=
    U_{n,\mu}U_{n+\hat{\mu},\nu}U_{n+\hat{\nu},\mu}^{*}U_{n,\nu}^{*},
\end{align}
where the link variable $U_{n,\mu}\in\mathbb{Z}_{N}$ is defined on each link, with $n$ denoting the lattice site and $\hat{\mu}$ the unit vector in the $\mu$ direction.
The inverse gauge coupling is represented by $\beta$.
Since we consider the theory on a (2+1)-dimensional lattice $\Lambda_{2+1}$, there are three directions labeled by $\mu=x,y,z$.
We also label these directions by $1$, $2$, and $3$, with the identification $x\equiv1$, $y\equiv2$, and $z\equiv3$.
The lattice extents in each direction are denoted by $L_{x}$, $L_{y}$, and $L_{z}$, and periodic boundary conditions are imposed in all directions.
The $z$-direction is identified as the temporal direction.
We parametrize the link variables as
\begin{align}
\label{eq:parametrization}
    U_{n,\mu}=
    \exp\left(
        \frac{2\pi{\rm i}}{N}a_{n,\mu}
    \right),
\end{align}
with a modulo-$N$ integer $a_{n,\mu}$.
With this parametrization, the Wilson gauge action in Eq.~\eqref{eq:wilson_S} is given by
\begin{align}
\label{eq:wilson_S_param}
    S[f_{n,\mu\nu}] 
    = 
    \beta
    \sum_{n\in\Lambda_{2+1}}\sum_{\mu<\nu}
    \left[
        1 - \cos \left(
            \frac{2\pi}{N}f_{n,\mu\nu}
        \right)
    \right],
\end{align}
with the modulo-$N$ plaquette variable $f_{n,\mu\nu}=a_{n,\mu}+a_{n+\hat{\mu},\nu}-a_{n+\hat{\nu},\mu}-a_{n,\nu}$.
The partition function is therefore defined as the sum of the Boltzmann weights over all configurations of the link variables:
\begin{align} 
\label{eq:part_Z}
    Z(\beta) = \sum_{\{a_{n,\mu}\}} \exp(-S[f_{n,\mu\nu}]).
\end{align}

\subsection{Original gauge theory}
The partition function in Eq.~\eqref{eq:part_Z} can be written as a contraction of a (2+1)-dimensional tensor network. While there are many possible choices for the local building-block tensor, contraction algorithms typically perform best when the formulations preserve the global symmetry explicitly. 
A standard way to do this is via the character expansion~\cite{Liu:2013nsa}, in which each Boltzmann weight on a plaquette is expanded in terms of the characters of $\mathbb{Z}_N$, which reads
\begin{align}
\label{eq:char_ex}
    {\rm e}^{\beta\Re U_{n,\mu\nu}}
    =
    \sum_{r}F_{r}(\beta)\chi_{r}(U_{n,\mu\nu})
    ,
\end{align}
where the summation runs over all irreducible representations $r$.
In our parametrization in Eq.~\eqref{eq:wilson_S_param}, the above expansion is explicitly given by the following Fourier expansion:
\begin{align}
\label{eq:fourier_ex}
    \exp\left(
        \beta\cos \left(
            \frac{2\pi}{N}f_{n,\mu\nu}
        \right)
    \right)
    =
    \sum_{m=0}^{N-1}
    I^{(n,\mu,\nu)}_{m}(\beta){\rm e}^{
        2 \pi {\rm i} 
        f_{n,\mu\nu}m/N
    },
\end{align}
with
\begin{align}
\label{eq:def_Im}
    I^{(n,\mu,\nu)}_{m}(\beta) 
    = 
    \frac{1}{N}\sum^{N-1}_{f_{n,\mu\nu}=0} {\rm e}^{
        \beta \cos(2 \pi f_{n,\mu\nu}/N) 
    }
    {\rm e}^{
        -2 \pi{\rm i}  f_{n,\mu\nu} m/N
    }.
\end{align}
\footnote{
Although the function $I^{(n,\mu,\nu)}_{m}(\beta)$ is defined on a plaquette specified by $(n,\mu,\nu)$, it is more natural to regard the dual integer variable $m$ as being associated with the plaquette, i.e., to denote it as $m^{(n,\mu,\nu)}$.
For later convenience, however, we deliberately adopt the former convention, and use the notation $m^{(n,\mu,\nu)}$ only when deriving the tensor-network representation of the dual spin model.
}
Eq.~\eqref{eq:fourier_ex} now allows us to carry out the sum over $a_{n,\mu}$ in the partition function~\eqref{eq:part_Z} individually.
The partition function is then represented as
\begin{equation} 
\label{eq:TN_AC}
    Z(\beta)
    = 
    {\rm tTr}
    \left[
        \prod_{n, \mu} A^{(n, \mu)}
        \prod_{n, \mu < \nu} C^{(n, \mu, \nu)} 
    \right],
\end{equation}
where $A^{(n, \mu)}$ is a four-leg tensor defined on each link given by
\begin{equation} 
\label{eq:A_link}
    A^{(n,\mu)}_{m_1 m_2 m_3 m_4} 
    = 
    \sum_{a_{n,\mu}=0}^{N-1} {\rm e}^{2 \pi {\rm i} a_{n,\mu} (m_1 + m_2 - m_3 - m_4)/N}
    =
    \delta_{0,m_1 + m_2 - m_3 - m_4~{\rm mod}~N}
    .
\end{equation}
The tensor $C^{(n, \mu, \nu)}$ sits on each plaquette, playing a role in restoring the original plaquette interaction via
\begin{equation} 
\label{eq:C_plaq}
    C^{(n, \mu, \nu)}_{m_1 m_2 m_3 m_4}
    = 
    I^{(n, \mu, \nu)}_{m_{1}}(\beta)
    \delta_{m_{1},m_{2},m_{3},m_{4}},
\end{equation}
where $\delta_{m_{1},m_{2},m_{3},m_{4}}$ takes the value 1 when all $m_{i}$ are identical, and 0 otherwise.

In previous numerical studies~\cite{Bazavov:2015kka,Kuramashi:2018mmi,Unmuth-Yockey:2018xak,Akiyama:2022eip,Kuwahara:2022ubg,Akiyama:2023hvt}, the tensor networks ~\eqref{eq:TN_AC} are first converted into a uniform tensor networks on a hypercubic lattice by contracting the tensors $A^{(n, \mu)}$ and $C^{(n, \mu, \nu)}$ into a local tensor $T^{(n)}$ at each lattice site $n$, and the resulting network is then approximately contracted.

\begin{figure}
    \centering
    \subfigure[Magnetic PEPO]{
        \includegraphics[width=0.4\linewidth]{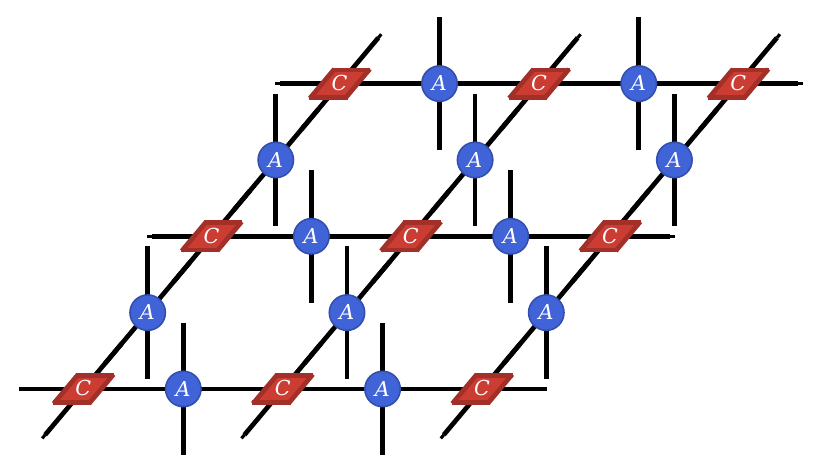}
    }
    \subfigure[Electric PEPO]{
        \includegraphics[width=0.4\linewidth]{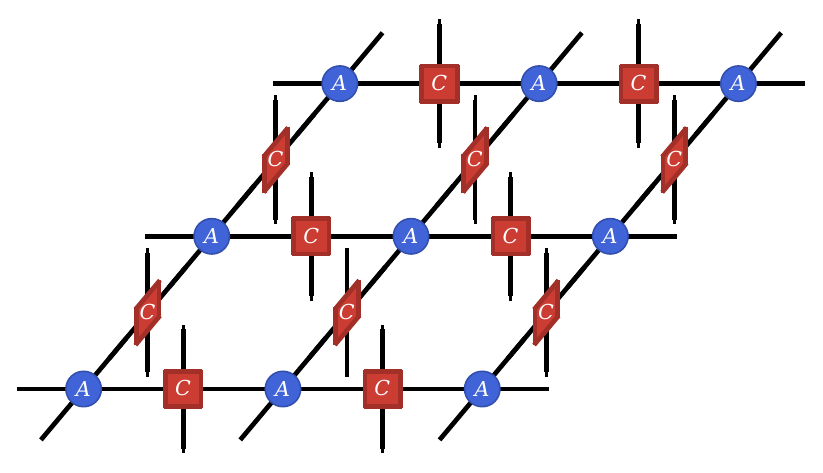}
    }
    \caption{
        Graphical representation of two types of PEPOs constructing Eq.~\eqref{eq:TN_AC}.
        Blue tensors refer to the link constraint tensors $A^{(n, \mu)}$, and red square tensors refer to plaquette tensors $C^{(n, \mu, \nu)}$.
    }
    \label{fig:InitialPEPOs}
\end{figure}

\begin{figure}[tb]
    \centering
    \includegraphics[width=0.6\linewidth]{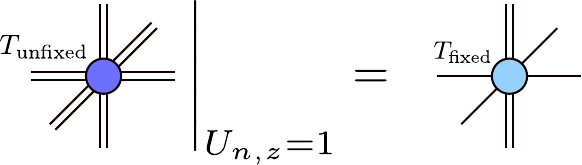}
    \caption{Electric PEPO is removed by imposing the temporal gauge condition.
    }
    \label{fig:T_temporal}
\end{figure}

In contrast, in the present study, we do not introduce such a local tensor $T^{(n)}$.
Instead, we regard the tensor networks~\eqref{eq:TN_AC} as a stack of two types of projected entangled-pair operators (PEPOs) as shown in Fig.~\ref{fig:InitialPEPOs}.  
This viewpoint has already appeared in Ref.~\cite{Meurice:2020gcd} in the discussion of conservation laws based on tensor-network formulations of the U(1) and $\mathbb{Z}_N$ gauge theories within the transfer-matrix formalism.
However, to the best of our knowledge, there has been no previous example in which this perspective has been directly applied to actual numerical calculations.\footnote{In particular, in the numerical simulations, the resulting tensors can be constructed and manipulated as $\mathbb{Z}_N$-symmetric tensors, which reduces the computational cost and memory overhead.}
We refer to the PEPO in Fig.~\ref{fig:InitialPEPOs}(a) as ``magnetic," since it consists of spatial plaquette interactions.
On the other hand, the PEPO in Fig.~\ref{fig:InitialPEPOs}(b) is referred to as ``electric," as it contains only the temporal link variables.
When the local tensor $T^{(n)}$ is constructed explicitly, one has to contract tensor networks with spatial bond dimension $N^{2}$. 
In contrast, within our PEPO representation, the system can be treated as tensor networks with spatial bond dimension $N$, and therefore, a higher efficiency of the approximate contraction is achieved, as we will see.
Moreover, as the plaquette tensors $C^{(n, \mu, \nu)}$ are non-zero only when all the indices match, the physical information from the electric PEPO can be easily pushed into the magnetic PEPO in the practical calculations. 
The electric PEPO can be interpreted as a Gauss-law projector.  
In practice, we introduce two types of local tensors,
\begin{align}
\label{eq:T_ele}
    (T^{(n)}_{\text{ele}})_{x y z_1 z_2 x' y' z'_1 z'_2}
    =
    \sum_{m_{1}, m_{2}}
    C^{(n, x, z)}_{xz_{1}m_{1}z'_{1}}
    C^{(n, y, z)}_{yz_{2}m_{2}z'_{2}}
    A^{(n,z)}_{m_{1}m_{2}x'y'}
    ,
\end{align}
and
\begin{align}
\label{eq:T_mag}
    (T^{(n)}_{\text{mag}})_{x y z_1 z_2 x' y' z'_1 z'_2}
    =
    \sum_{m_{1}, m_{2}}
    A^{(n,x)}_{xz_{1}m_{1}z'_{1}}
    A^{(n,y)}_{yz_{2}m_{2}z'_{2}}
    C^{(n, x, y)}_{m_{1}m_{2}x'y'},
\end{align}
with spatial and temporal bond dimensions $N$ and $N^2$, respectively.
These two tensors rewrite Eq.~\eqref{eq:TN_AC} as 
\begin{equation} 
\label{eq:TN_ele_mag}
    Z(\beta)
    = 
    {\rm tTr}
    \left[
        \prod_{n} 
        T^{(n)}_{\text{ele}}
        T^{(n)}_{\text{mag}}
    \right].
\end{equation}
If we perform the tensor contraction between $T^{(n)}_{\text{ele}}$ and $T^{(n)}_{\text{mag}}$ along the $z$-direction, we will encounter a local tensor $T^{(n)}$ shown in the left-hand side in Fig.~\ref{fig:T_temporal}, which is the same with the tensor introduced in the so-called asymmetric formulation~\cite{Liu:2013nsa}.
We further discuss another way to derive Eq.~\eqref{eq:TN_ele_mag} in Appendix~\ref{app:TN_alt}.

\begin{figure}[tb]
    \centering
    \includegraphics[scale=0.5]{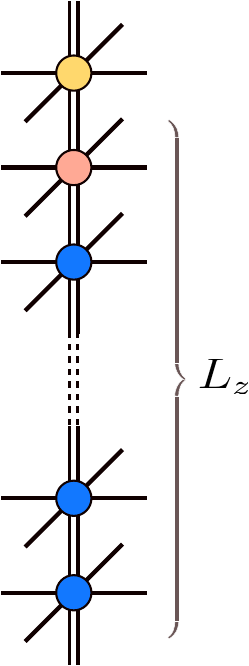}
    \caption{
        Schematic illustration of the tensor-network structure along the temporal direction at each spatial site when the temporal gauge is imposed.
        Yellow and red tensors denote Eqs.~\eqref{eq:T_ele} and \eqref{eq:T_mag}, respectively.
        Note that the top and bottom indices are contracted due to the periodic boundary condition.
    }
    \label{fig:unitcell_gauge}
\end{figure}

In our numerical computations, we adopt the temporal gauge to set $U_{n,z}=1$, as depicted in Fig.~\ref{fig:T_temporal}.
Under this gauge condition, the $A$-tensors turn into rank one tensors, i.e. the outer product $A = \bigotimes_{i=1,2,3,4} (\sum_{m_i=0}^{N-1} |m_i\rangle)$.
Contracting these different factors into the corresponding $C$ tensors reduced the electric PEPO into a trivial product of bond tensors on the vertical legs, which can then be absorbed into the magnetic PEPO layer, giving rise to the tensors $T_{\text{fixed}}$ with a single $N$-dimensional index in the spatial directions.
We note, however, that not all temporal link variables can be fixed at each spatial site in our setup.
Since the periodic boundary condition is imposed in the temporal direction with finite $L_{z}$, one temporal link variable must remain unfixed and be integrated over in the partition function.\footnote{An alternative interpretation can be given to this procedure. Because of the $\delta_{m_1,m_2,m_3,m_4}$ structure of the $C$ tensor, the associated weight $I_m(\beta)$ can be shifted to the vertical bond tensor and absorbed in the magnetic PEPO, which now contains the Boltzmann weight of both the magnetic and the electric plaquettes. The remaining ``electric'' PEPO, containing no Boltzmann weights, is then reduced to a pure Gauss-law projector, which commutes with the ``magnetic'' layers and squares to itself, so that these layers can be squashed into a single projector to enforce the Gauss law at one timeslice in the network.
}
The situation is illustrated in Fig.~\ref{fig:unitcell_gauge}.

We also remark that temporal lattice spacing plays the same role as a Trotter error introduced by the Suzuki--Trotter decomposition in the Hamiltonian formalism.
In principle, this error can be systematically removed by taking the temporal continuum limit.
This requires working with an anisotropic lattice action that introduces two types of gauge couplings, one for the spatial and the other for the temporal directions.
To take the temporal continuum limit, both couplings must be tuned appropriately while keeping the temperature fixed.
In the present study, however, we restrict ourselves to the isotropic lattice action and do not take the temporal continuum limit, in order to facilitate a direct comparison with Monte Carlo simulation results.
Since taking the temporal continuum limit makes the present tensor network approach directly comparable to the Hamiltonian formalism, we leave this investigation for future work.

\subsection{Duality with clock model}

In addition to the tensor-network representation in Eq.~\eqref{eq:TN_ele_mag} for the $\mathbb{Z}_N$ gauge theories, we also derive a tensor-network representation for the dual $N$-state clock models in three dimensions.
As we explain, these dual models solve the gauge constraint exactly, so that the resulting representation is expected to be more efficient from a practical viewpoint.

We can regard the tensors $A^{(n,\mu)}$ in Eq.~\eqref{eq:A_link} as constraints on the tensor network in Eq.~\eqref{eq:TN_AC}.
Recalling that the dual integer variable $m$ introduced in Eq.~\eqref{eq:fourier_ex} is labeled by plaquettes and can be denoted as $m^{(n,\mu,\nu)}$, the tensors $A^{(n,x)}$, $A^{(n,y)}$, and $A^{(n,z)}$ have non-vanishing entries only when the equations
\begin{align}
    m^{(n,x,y)}
    +
    m^{(n,x,z)}
    -
    m^{(n-\hat{y},x,y)}
    -
    m^{(n-\hat{z},x,z)}
    &=
    0 \mod N,\\
    m^{(n,y,z)}
    +
    m^{(n-\hat{x},x,y)}
    -
    m^{(n-\hat{z},y,z)}
    -
    m^{(n,x,y)}
    &=
    0 \mod N,\\
    m^{(n-\hat{x},x,z)}
    +
    m^{(n-\hat{y},y,z)}
    -
    m^{(n,x,z)}
    -
    m^{(n,y,z)}
    &=
    0 \mod N,
\end{align}
are satisfied, respectively.
These constraints can be exactly resolved by introducing $\mathbb{Z}_{N}$ dual spin variables~\cite{Bhanot:1980pc}.
We define the dual lattice $\tilde{\Lambda}_{2+1}$ with sites $n'$ that are located at the centers of the unit cells of the original lattice $\Lambda_{2+1}$.
Letting $s(n')$ be a $\mathbb{Z}_{N}$ spin variable on a dual lattice site $n'$, we introduce
\begin{align}
\label{eq:reexpress_m}
    m^{(n,\mu,\nu)}
    =
    \frac{N}{2\pi{\rm i}}
    \varepsilon_{\mu\nu\lambda}
    \ln\left(
        s(n+\hat{\mu}/2+\hat{\nu}/2+\hat{\lambda}/2)
        s(n'+\hat{\mu}/2+\hat{\nu}/2-\hat{\lambda}/2)^{-1}
    \right),
\end{align}
where $\varepsilon_{\mu\nu\lambda}$ is the antisymmetric Levi-Civita tensor.
It is easy to see that Eq.~\eqref{eq:reexpress_m} does solve the above constraints.
Thus, Eq.~\eqref{eq:part_Z} is now simplified as
\begin{align}
\label{eq:Z_new}
    Z(\beta) = \sum_{\{s\}} \prod_{n'\in\tilde{\Lambda}_{2+1}}\prod_{\mu<\nu} I_{\ell^{(\mu,\nu)}}(\beta),
\end{align}
where we now denote the right-hand side of Eq.~\eqref{eq:reexpress_m} as the value $\ell^{(\mu,\nu)}$ associated to the $\hat{\lambda}$-oriented link of the dual lattice that intersects the $(n,\mu,\nu)$ plaquette of the original lattice, and $I_{\ell^{(\mu,\nu)}}(\beta)$ is defined in the same way with Eq.~\eqref{eq:def_Im}.
The new partition function in Eq.~\eqref{eq:Z_new} can be identified as that of the $N$-state clock model on a three-dimensional cubic lattice $\tilde{\Lambda}_{2+1}$. 
The tensor-network representation is straightforwardly obtained as
\begin{align}
\label{eq:TN_dualS}
    Z(\beta)={\rm tTr}
    \left[\prod_{n'\in\tilde{\Lambda}_{2+1}} T^{(n')}\right],
\end{align}
where
\begin{equation}
    T^{(n')}_{xyz x'y'z'} 
    = 
    \sum_{k = 0}^{N-1} 
    M_{k x} M_{k y} M_{k z} 
    M^\dagger_{x' k} M^\dagger_{y' k} M^\dagger_{z' k},
\end{equation}
with the following $N\times N$ matrix,
\begin{align}
    M_{kf} 
    = \frac{1}{\sqrt{N}} {\rm e}^{(\beta/2)\cos(2 \pi f/N)} {\rm e}^{-2 \pi {\rm i}fk/N},
\end{align}
parametrizing the $\mathbb{Z}_{N}$ spin variable $s$ in the same manner with Eq.~\eqref{eq:parametrization}.\footnote{
One can recognize that the indices in $T^{(n')}_{xyz x'y'z'}$ correspond to the original plaquette variables $f_{\mu\nu}$, which ultimately stems from the fact that the representation of the clock model in Eq.~\eqref{eq:TN_dualS} is formulated in terms of variables that are once again dual to the natural spin variables of the clock model.
}

\section{Methods}
\label{sec:methods}
    The tensor-network representations in Eqs.~\eqref{eq:TN_ele_mag} and \eqref{eq:TN_dualS} can now be approximately contracted along the temporal and the spatial directions to obtain partition functions as well as the CFT data. 
    This task can be accomplished using the spatial coarse-graining schemes such as the tensor renormalization group (TRG)~\cite{Levin:2006jai} and tensor network renormalization (TNR)~\cite{Evenbly:2015ucs, Evenbly2017, Gu:2009dr, Hauru2018, looptnr, Bal2017, Homma2024, Ueda:2025mhu}.
    To approach the low-temperature region, the temporal lattice extent should be enlarged.
    One way to carry it out is to construct local projectors that achieve approximate contractions along the temporal direction.
    These local projectors can be efficiently computed using the standard higher-order TRG (HOTRG) method~\cite{HOTRG} when the fundamental six-leg tensor is hermitian in the spatial directions. 
    When there is no such hermiticity, as in the case of Eq.~\eqref{eq:TN_ele_mag}, the projectors must be constructed by taking into account the absence of reflection symmetries~\cite{Fishman:2018lnr,boundarytrg}.

\begin{figure}[tb]
        \centering
        \includegraphics[width=0.99\linewidth]{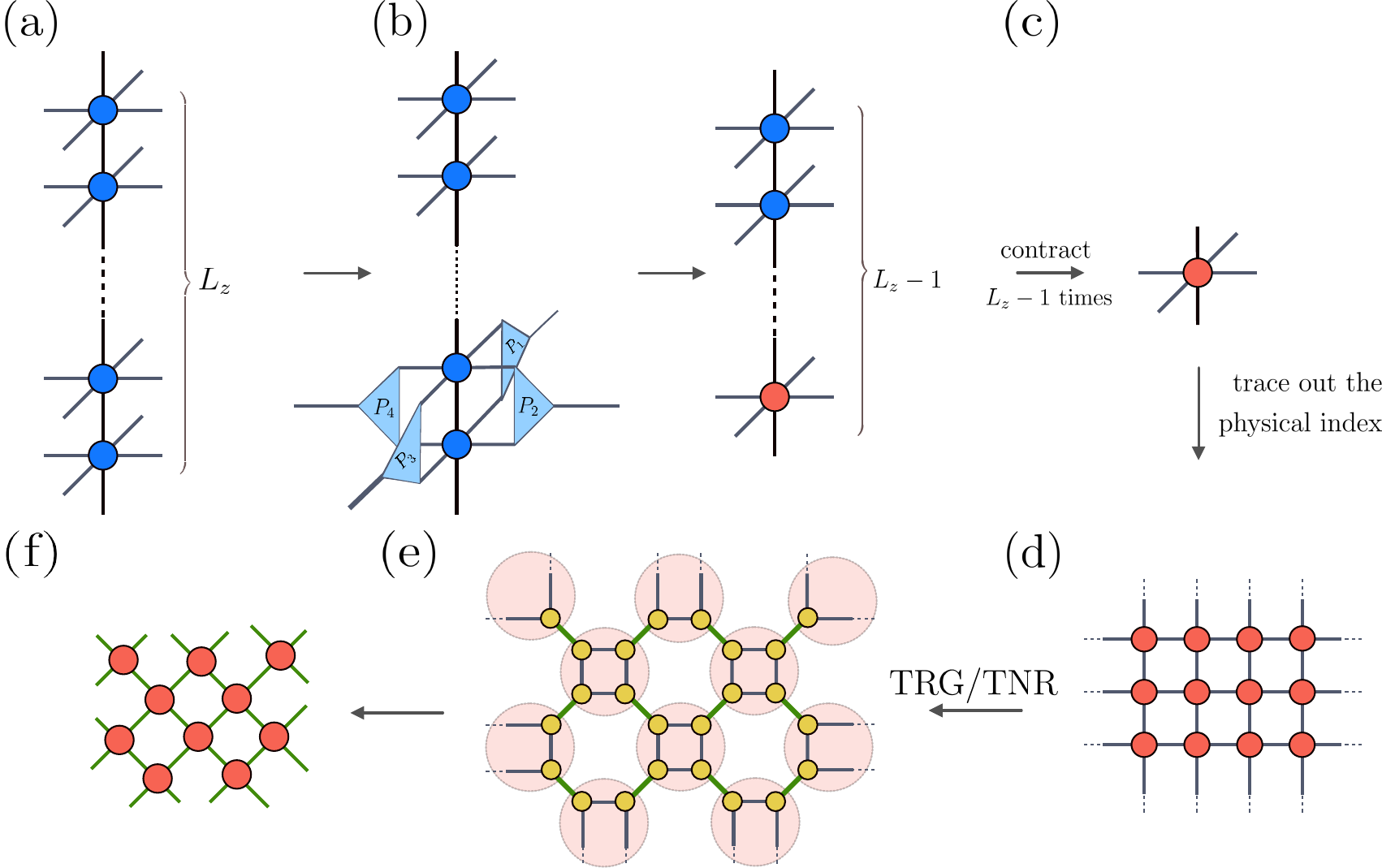}
        \caption{
            (a) Partition function in Eq.~\eqref{eq:TN_dualS} at height $L_z$, consisting of columns of six-legged tensors. 
            (b) Sequential coarse-graining of the column along the vertical direction using projectors in the TTNR method, whose details are given in Appendix~\ref{appendix:projector}, resulting in a single effective tensor. 
            (c) Tracing out the vertical indices reduces the three-dimensional tensor network of height $L_z$ to an effective two-dimensional infinite network, as shown in (d). 
            (e) Coarse-graining of the two-dimensional tensor network using TRG/TNR methods.
            Since the geometry of the tensor network in (f) is identical to that in (d), the coarse-graining procedure can be iterated to evaluate the partition function. As the number of tensors reduces by half, the network of $2^s$ tensors becomes a single tensor after $s$ steps.
            The procedure is identical for the partition function in Eq.~\eqref{eq:TN_ele_mag}.
        }
        \label{fig:flowchartTTNR}
    \end{figure}

    Here, we perform the contraction along the $z$ direction, identified as the imaginary-time direction, using the thermal TNR (TTNR) method~\cite{Ueda:2025mhu}.
    In the conventional strategy based on HOTRG~\cite{Kuramashi:2018mmi,Akiyama:2021xxr,Akiyama:2021glo}, the lattice size in the temporal direction grows exponentially with the number of coarse graining steps, whereas it increases only linearly in the TTNR approach.
    Since our primary target is phase transitions at finite temperature, the TTNR method is expected to be particularly efficient.
    Using TTNR, we first carry out the contraction along the temporal direction and reduce the original three-dimensional tensor-network representation to an effective two-dimensional one, which is represented by a four-leg tensor.
    The resulting two-dimensional tensor network is then further contracted using the established TRG/TNR methods for two-dimensional partition functions.
    In the present work, we use Loop-TNR~\cite{looptnr} and bond-weighted TRG~(BW-TRG)~\cite{BWTRG}. 
    Repeating their coarse-graining procedures ultimately yields a single four-leg tensor that encodes the original full tensor network. 
    Our computational strategy is illustrated in Fig.~\ref{fig:flowchartTTNR}.
    Furthermore, we construct the projectors shown in Fig.~\ref{fig:flowchartTTNR}(b), which provide an optimal truncation for the loop formed by tensors extending along the temporal direction at each spatial site.\footnote{
    At each spatial site, this loop consists of $L_{z}+1$ tensors in the representation of Eq.~\eqref{eq:TN_ele_mag} imposing the temporal gauge, as illustrated in Fig.~\ref{fig:unitcell_gauge}, and $L_{z}$ tensors in that of Eq.~\eqref{eq:TN_dualS}.
    }
    The algorithmic details are presented in Appendix~\ref{appendix:projector}. 
    We introduce two types of bond dimensions in our algorithm. 
    During the first stage of the algorithm, the accuracy of the temporal contraction is controlled by bond dimension $\chi_{{\rm TTNR}}$ in the spatial direction, while the bond dimension during the spatial coarse-graining using Loop-TNR or BW-TRG in the second stage is denoted as $\chi_{\text{Loop-TNR}}$ or $\chi_{\text{BW-TRG}}$, respectively.

    The resulting four-leg tensor captures the essential physics of the original model. 
    First, the partition function is obtained by tracing over the legs in both the $x$ and $y$ directions. More importantly, universal data can be extracted from a renormalized transfer matrix, which we construct by tracing only the legs in the $x$ direction. Geometrically, the spectrum of this transfer matrix corresponds to that of a partition function on a cylinder, so it gives us direct access to the low-energy physics. In particular, we can read out the central charge and scaling dimension of the underlying CFT~\cite{Gu:2009dr} at criticality and the renormalization group (RG) flow in its vicinity~\cite{Ueda_2023}. 

\section{Results}
\label{sec:results}

\subsection{Universal data from finite-temperature simulations}
\label{subsec:universal_data}

\begin{figure}
    \centering
    \includegraphics[width=0.48\linewidth]{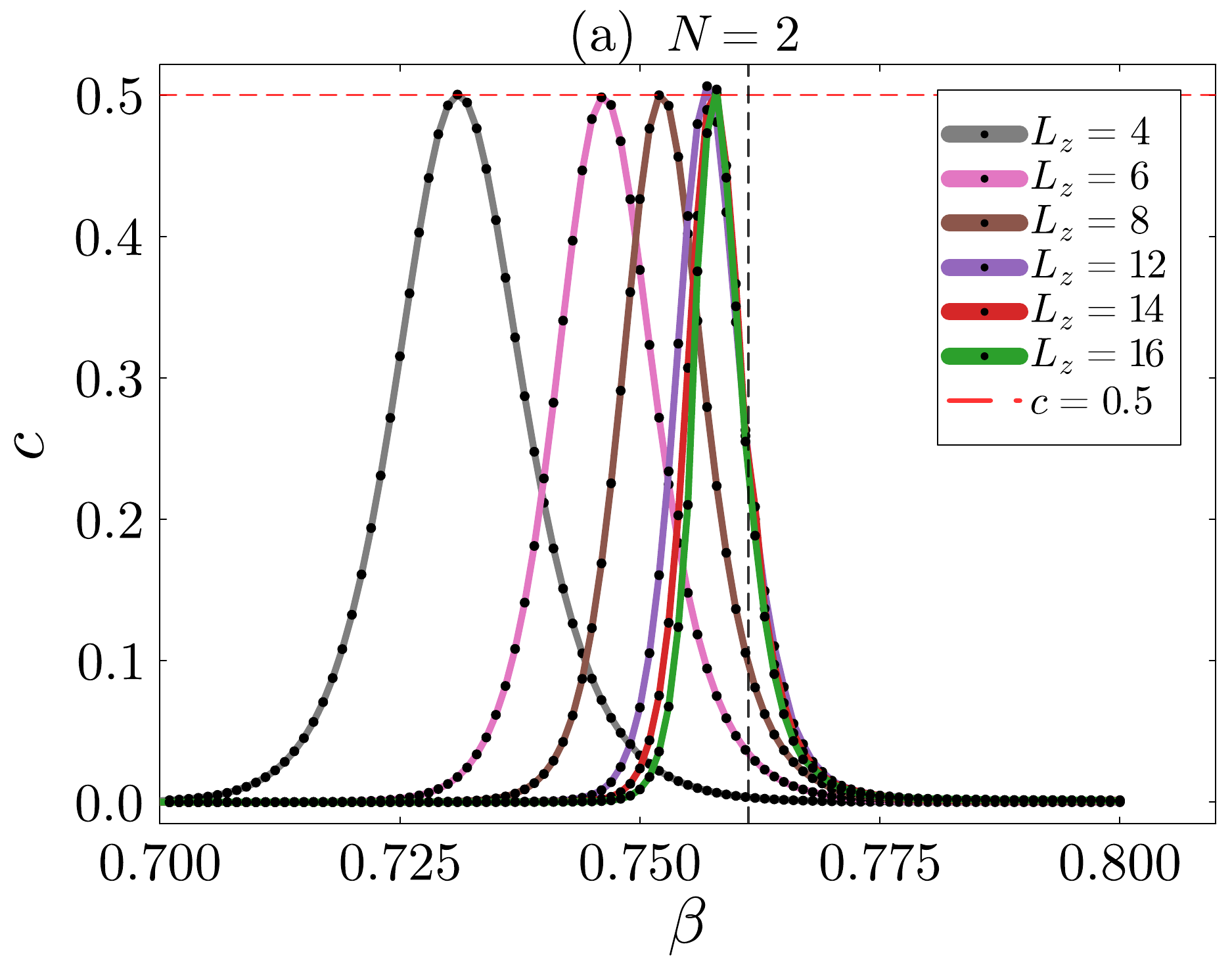}
    \includegraphics[width=0.48\linewidth]{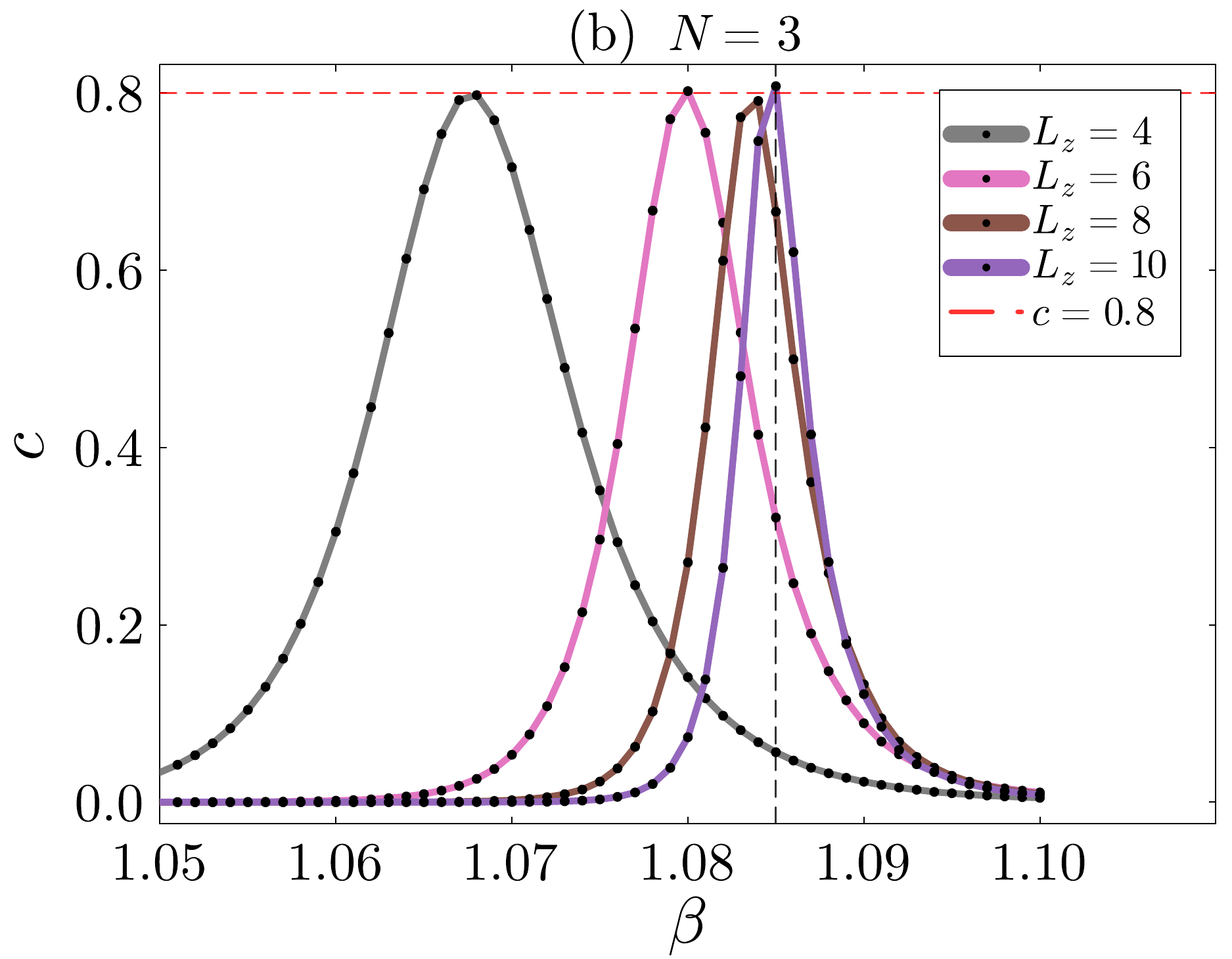}
    \caption{
        Central charge of (a) $N=2$ and (b) $N=3$ dual spin models with $(\chi_{{\rm TTNR}},\chi_{\text{Loop-TNR}})=(48, 36)$. The central charges are obtained from the fixed-point tensor using the prescription in Ref.~\cite{Gu:2009dr}.
        The values $c=0.5$ and $c=0.8$ are consistent with the two-dimensional Ising and three-state Potts CFTs, respectively.
        Dashed vertical lines indicate the critical inverse gauge couplings $\beta_c$ at vanishing temperature ($L_z \to \infty$) obtained using MC simulations~\cite{Borisenko:2013xna}.
    }
    \label{fig:centralchargevsbeta}
\end{figure}

We use the central charge extracted using Loop-TNR to locate the critical inverse gauge coupling $\beta_{c}(L_{z})$ in the $\mathbb{Z}_N$ dual spin models.
The central charge takes a nonzero value only at criticality and vanishes in both confinement and deconfinement phases.
We obtain $\beta_{c}(L_{z})$ using the peak of the central charge as shown in Fig.~\ref{fig:centralchargevsbeta} at finite spatial lattice extent $L$, where we set $L_{x}=L_{y}=L$.

In practice, a Gaussian fit is considered and $\beta_c(L_z)$ is extrapolated from the critical points at different coarse-graining steps $s$, i.e., at different spatial sizes $L = \sqrt{2}^s$~\cite{Ueda_2023}. 
Alternatively, $\beta_c(L_z)$ can also be extrapolated from scaling dimensions. 
Using the known values of scaling dimensions for two-dimensional CFTs, we extract critical inverse gauge couplings at finite spatial volume $\beta_c(L_z,L)$ as the points when the scaling dimensions cross these known values.
The critical point $\beta_c(L_z, L\to\infty)$ is then extrapolated by fitting $\beta_c(L_z, L) = a + b (L)^{x_\epsilon}$ where $x_\epsilon$ is the scaling dimension of the thermal operator. 

\begin{figure*}[ht]
\begin{multicols}{2}
    \includegraphics[width=\linewidth]{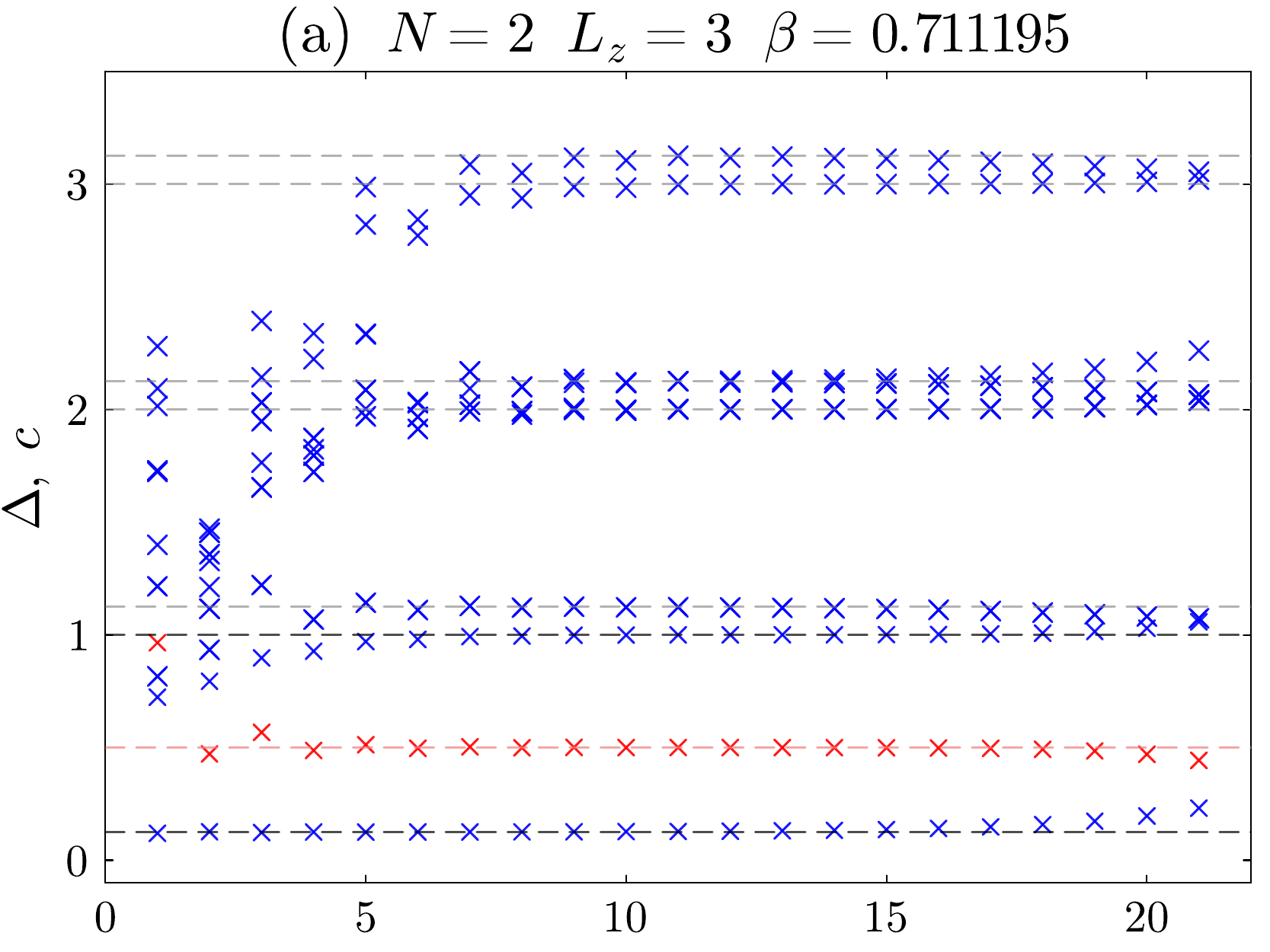}\par 
    \includegraphics[width=\linewidth]{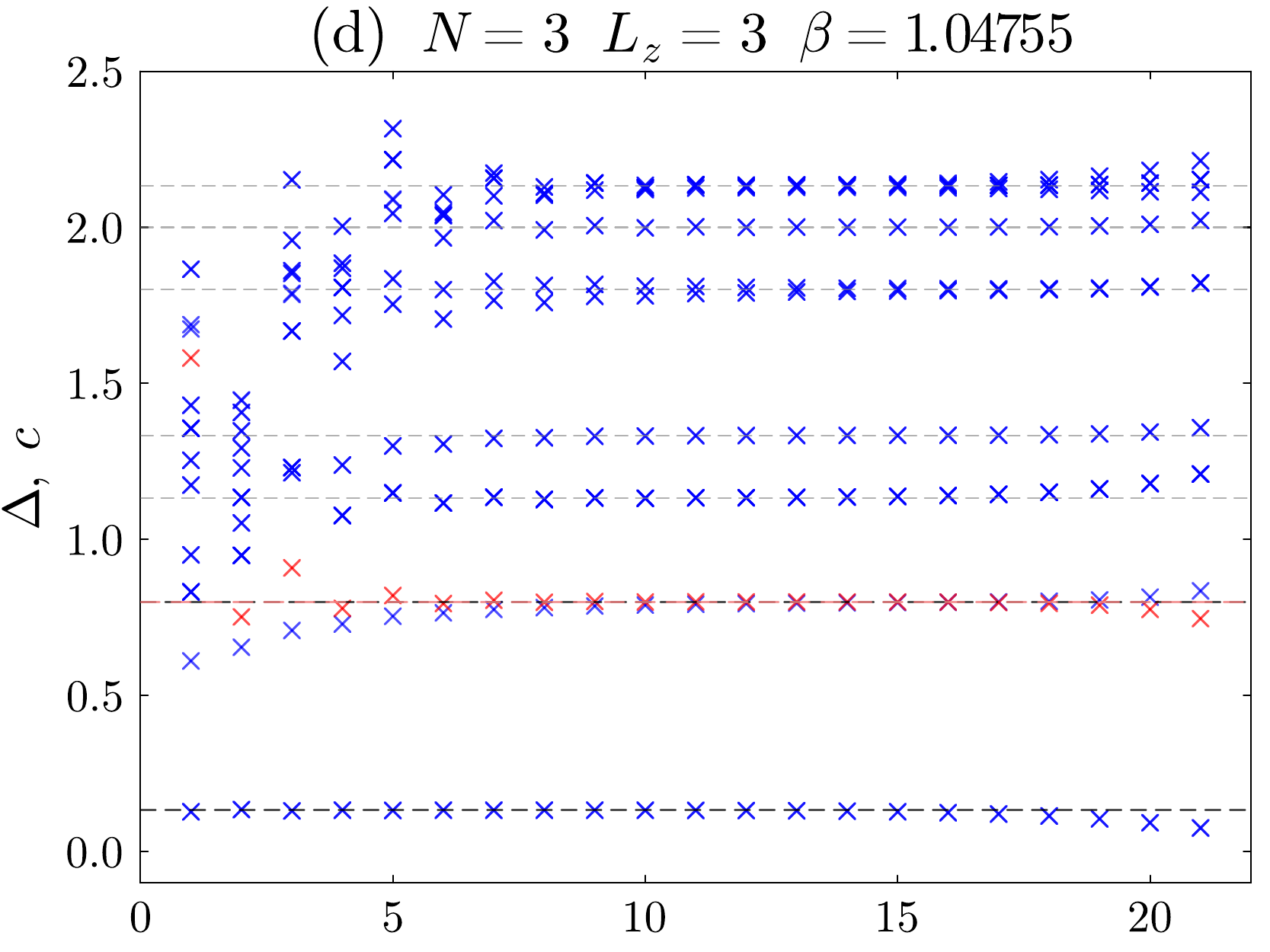}\par
    
    \end{multicols}
\begin{multicols}{2}
    
    \includegraphics[width=\linewidth]{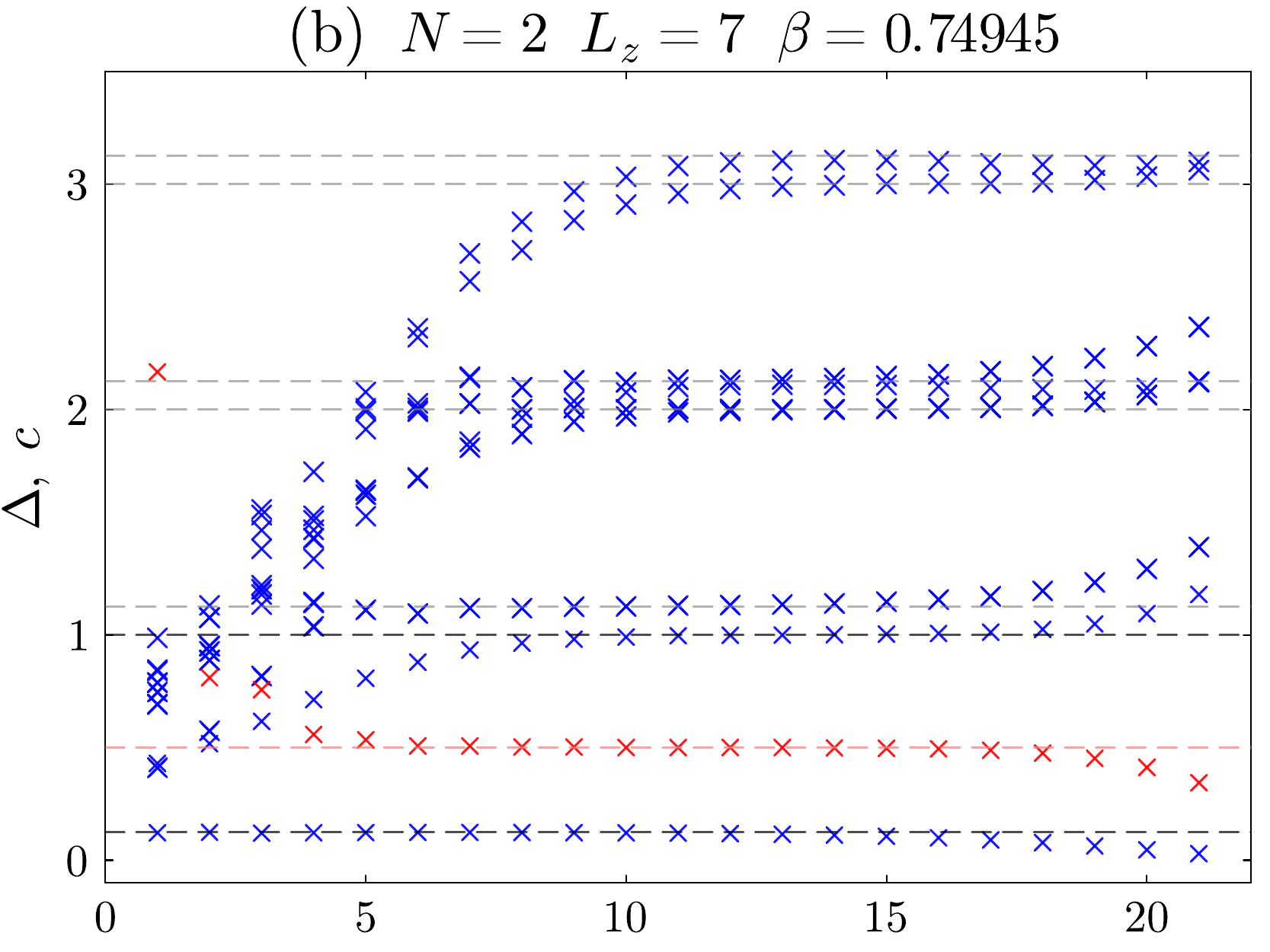}\par
    \includegraphics[width=\linewidth]{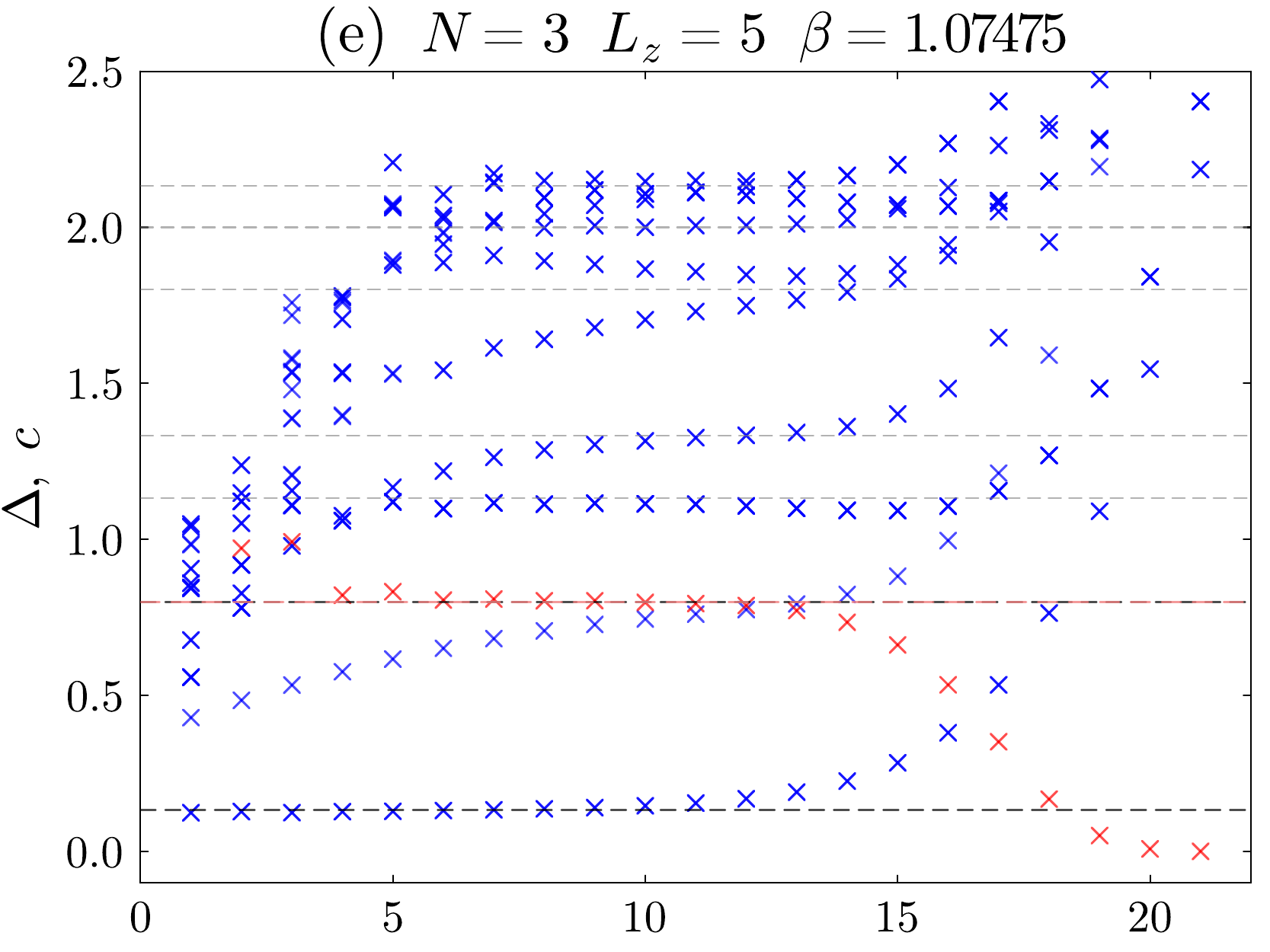}\par
    
\end{multicols}
\begin{multicols}{2}
    \includegraphics[width=\linewidth]{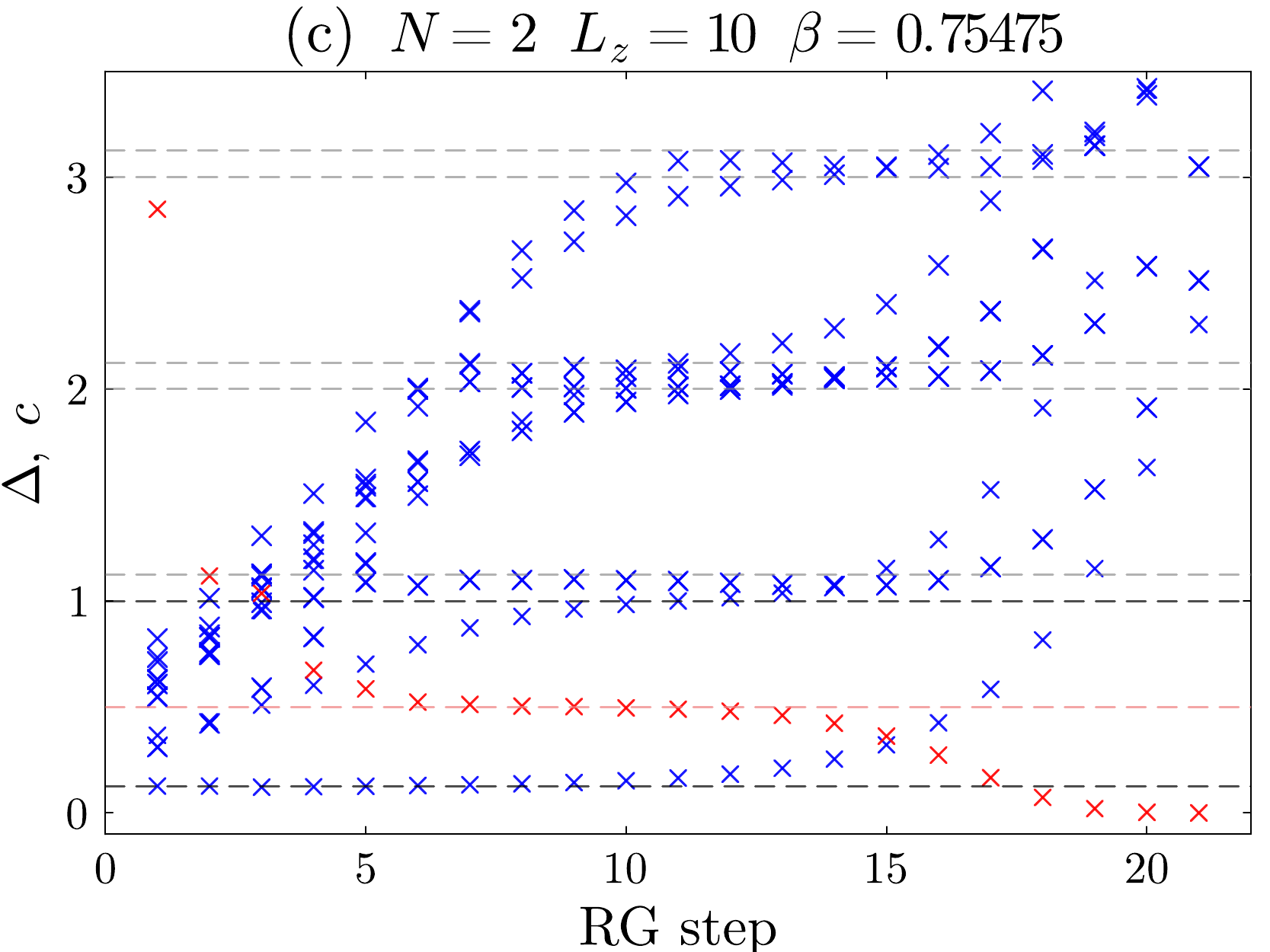}
    \includegraphics[width=\linewidth]{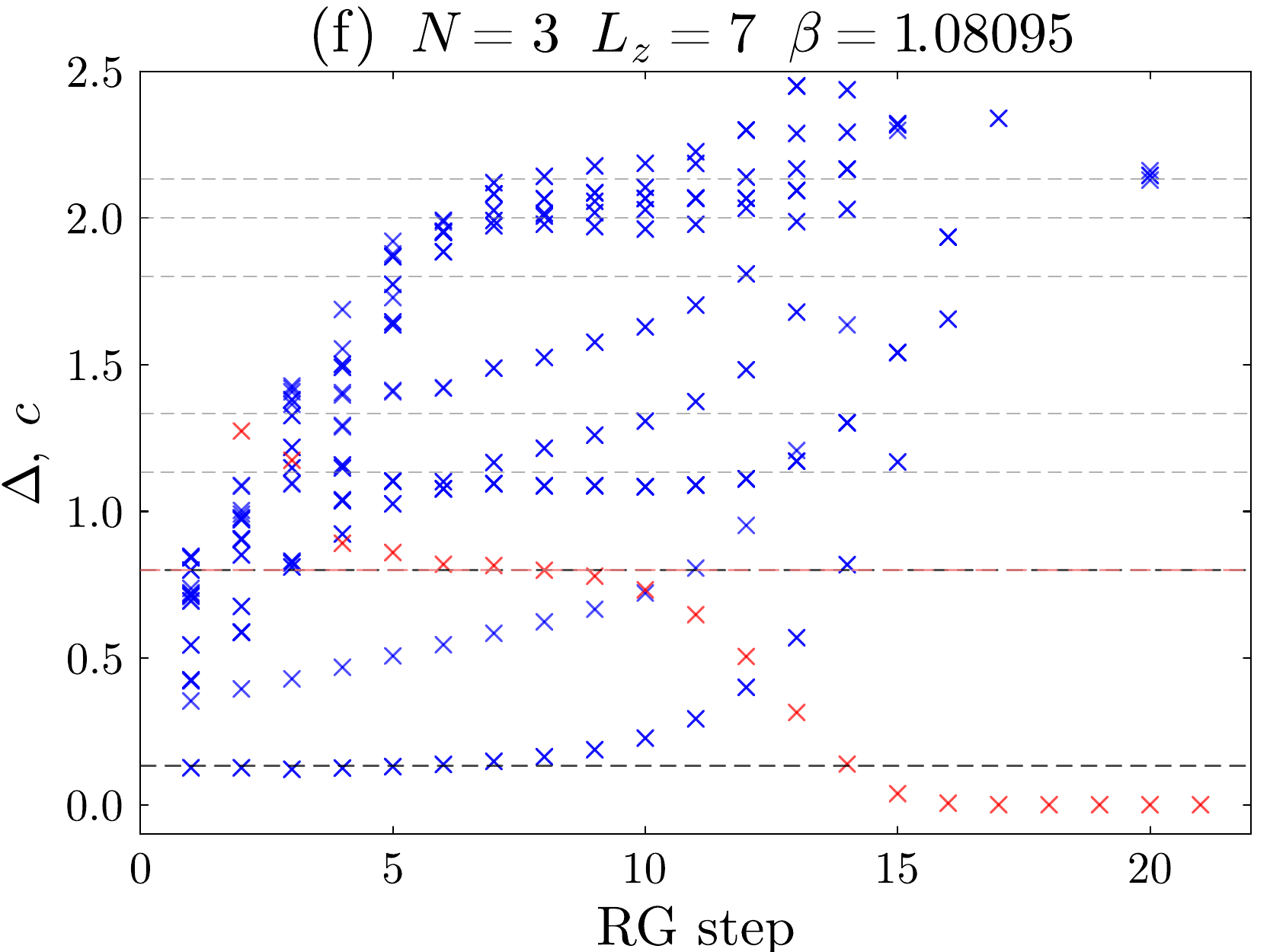}
\end{multicols}
\caption{
    Central charge $c$ (red) and scaling dimensions $\Delta$ (blue) along the RG steps for $N=2$ and $N=3$  with $(\chi_{{\rm TTNR}},\chi_{\text{Loop-TNR}})=(81,48)$. Dashed lines denote the known values of the CFT scaling dimensions and central charge.
}
\label{fig:UniversaldataZ2Z3}
\end{figure*}

In the cases of the $\mathbb{Z}_2$ and $\mathbb{Z}_3$ dual spin models, the universal data at the extrapolated points show remarkable stability and correspond to the two-dimensional Ising and three-state Potts conformal field theories, respectively, as shown in Fig.~\ref{fig:UniversaldataZ2Z3}.
These observations provide direct evidence for the Svetitsky--Yaffe conjecture for (2+1)-dimensional $\mathbb{Z}_2$ and $\mathbb{Z}_3$ gauge theories, as their thermal transitions belong to the same universality classes as those of two-dimensional spin models with the same $\mathbb{Z}_2$ and $\mathbb{Z}_3$ symmetries.
We observe that the universal data becomes less stable along the RG steps for higher values of $L_z$. 
We expect that the stability against the Loop-TNR step can be improved by filtering out the local entanglement~\cite{lyu2025latticereflectionsymmetryTNR} in the contraction along the temporal direction, which we leave for future work.

\begin{figure}[ht]
    \centering
    \includegraphics[width=0.49\linewidth]{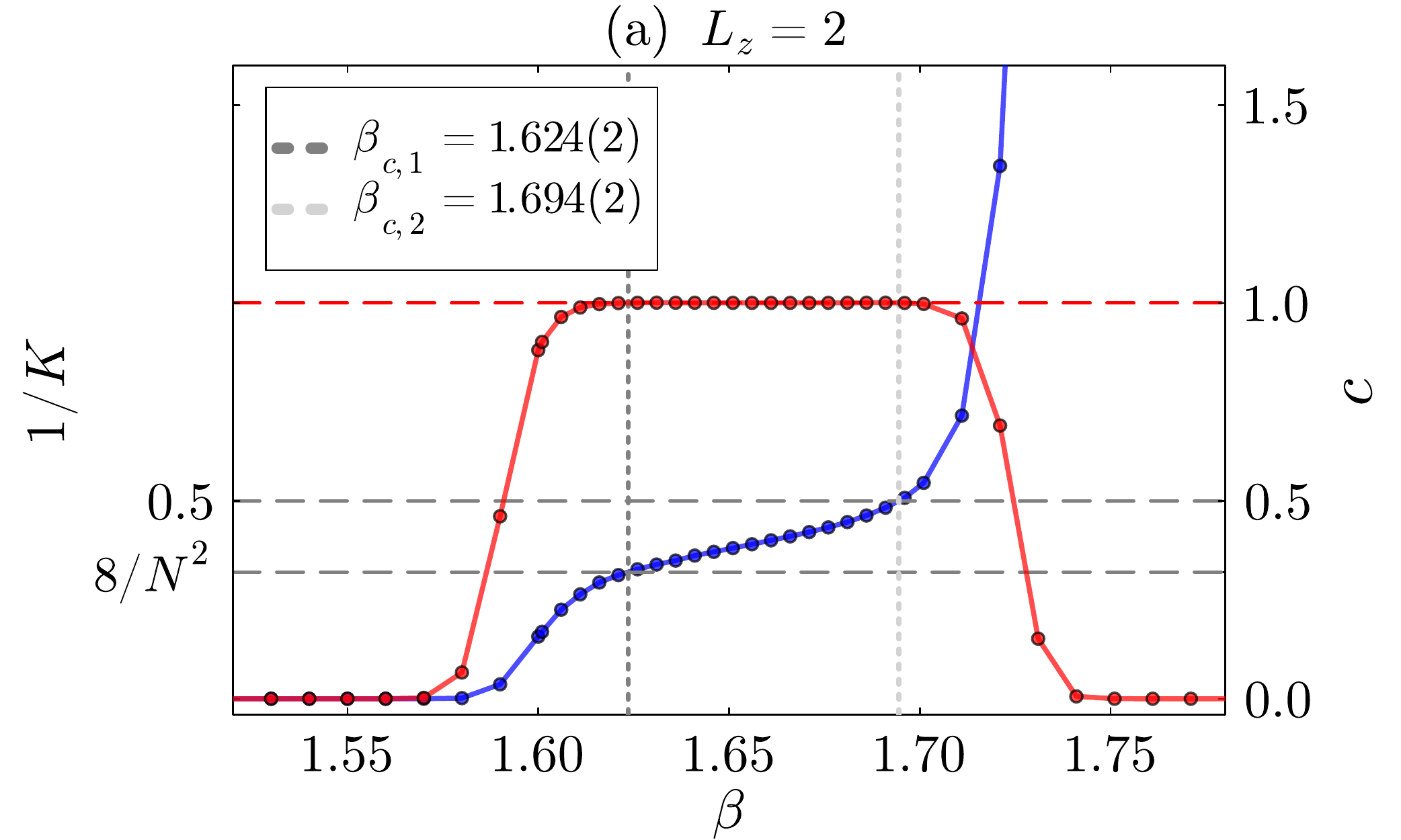}
    \includegraphics[width=0.5\linewidth]{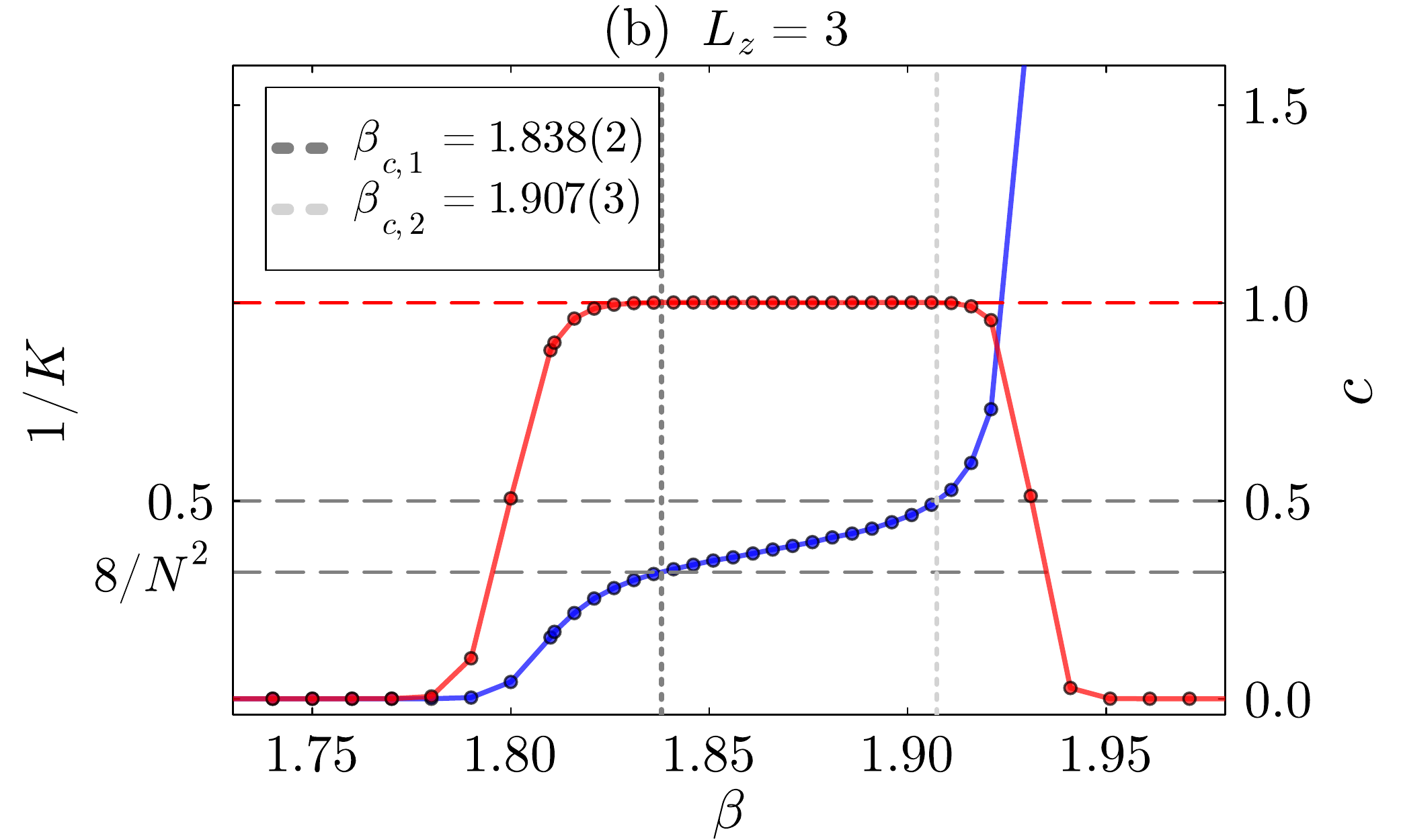}
    \includegraphics[width=0.49 \linewidth]{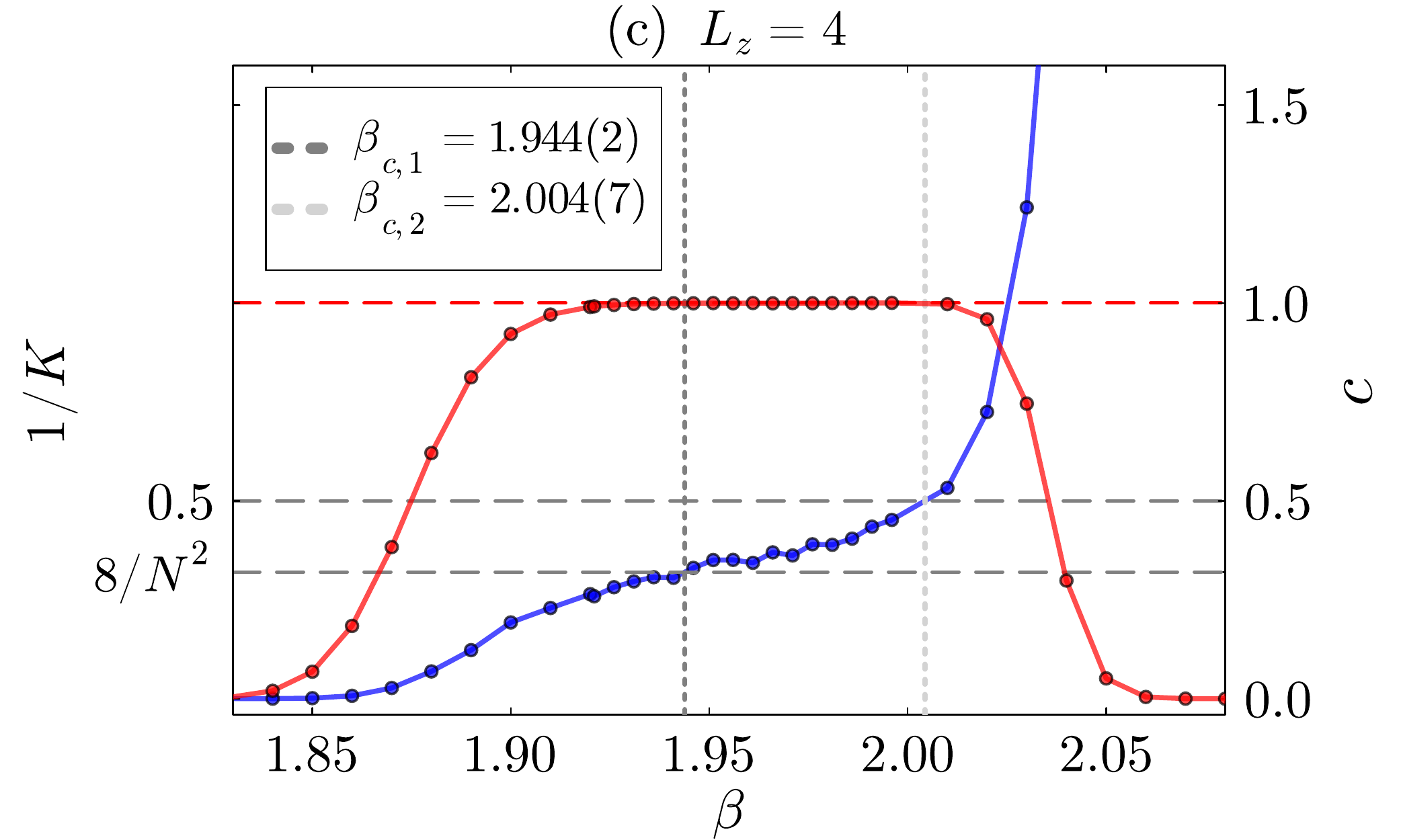}
    \includegraphics[width=0.5\linewidth]{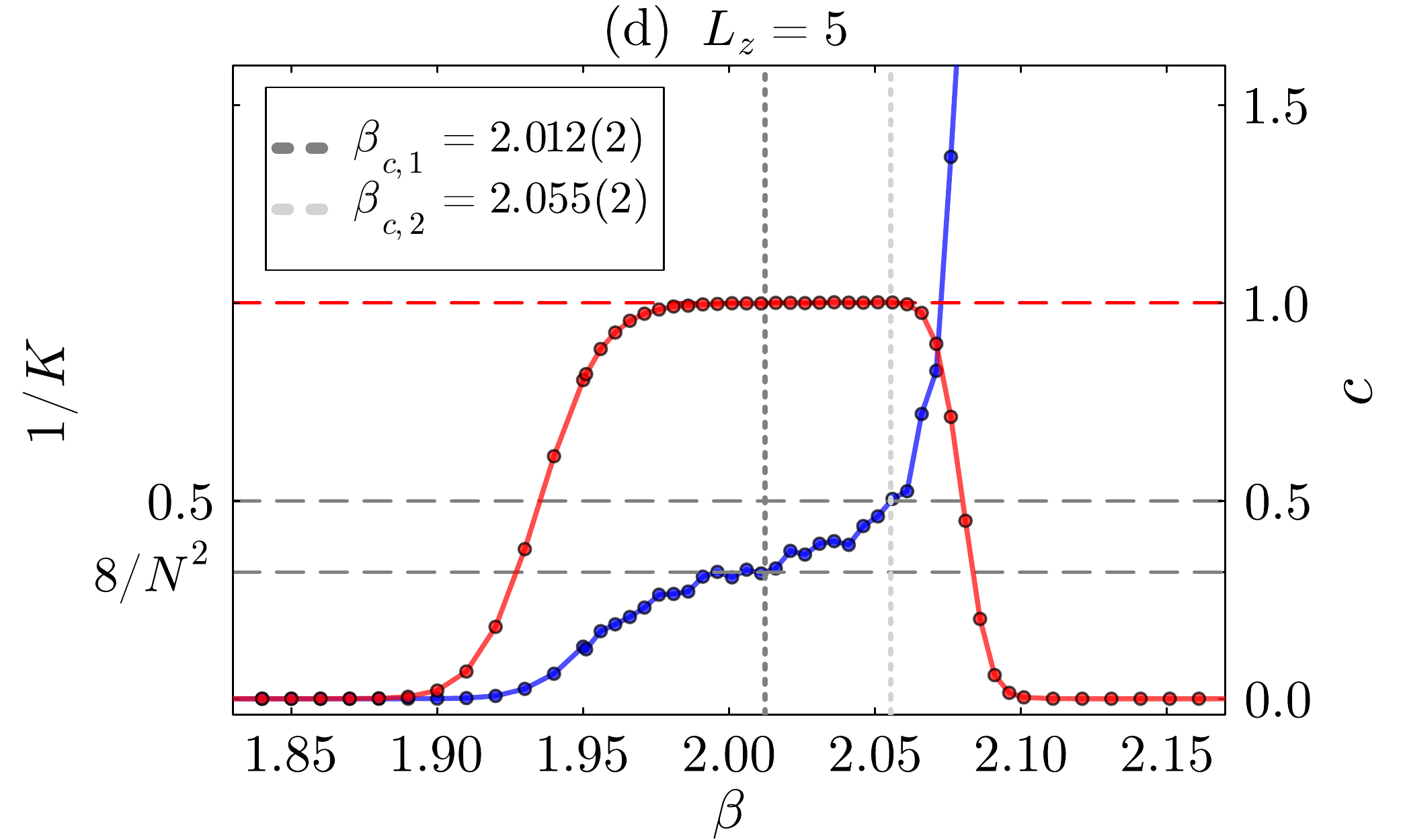}
    \caption{Central charge $c$ (red) and Luttinger parameter $K$ (blue) for $N=5$ with $(\chi_{{\rm TTNR}},\chi_{\text{Loop-TNR}})=(80, 80)$. 
    The spatial lattice size is $L=\sqrt{2}^{20}$ for (a), (b),\ and $L=\sqrt{2}^{15}$ for (c), (d). 
    The Luttinger parameter passes through $1/K = 8/25$ for the first BKT transition and $1/K = 0.5$ for the second BKT transition, indicating that the middle phase is described by the Tomonaga--Luttinger liquid field theory.
    Grey vertical dashed lines denote the estimates of $\beta_{c,1}$ and $\beta_{c,2}$ from LoopTNR.
    }
    \label{fig:Universaldataz5}
\end{figure}

For $N=5$, we observe in Fig.~\ref{fig:Universaldataz5} that the central charge plateaus at $c = 1$ for $L_{z}=2,3,4,5$, suggesting that the system belongs to the universality class of the five-state clock model, which exhibits two BKT transitions and an intermediate phase with an emergent U(1) symmetry described by the Tomonaga--Luttinger liquid field theory~\cite{Luttinger:1963zz,Tomonaga:1950zz}.
Fig.~\ref{fig:Universaldataz5} also depicts the (inverse) Luttinger parameter extracted from the first scaling dimension through the relation $1/K=(4\Delta_{1})$.
We find that the first scaling dimension $\Delta_{1}$ reaches the value $2/25$ at a specific values of the gauge coupling $\beta_{c,1}$, which agrees with the prediction $K=N^{2}/8$ for the Luttinger parameter and identifies the transition as that between the $\mathbb{Z}_{N}$ symmetry-broken phase and the massless phase for $N=5$.
As elaborated upon in Sec.~\ref{subsec:duality}, the former phase corresponds to the confinement phase in the original $\mathbb{Z}_{N}$ gauge theory.
We also observe that $1/K$ takes the value $0.5$ at a specific value $\beta_{c,2}$, which corresponds to the second BKT transition point.
These observations provide direct evidence for the Svetitsky--Yaffe conjecture for the (2+1)-dimensional $\mathbb{Z}_5$ gauge theory.

Having confirmed the Svetitsky--Yaffe conjecture based on the universal data, we now proceed to compare the resulting critical couplings with those obtained in previous studies.
Since detailed comparisons for $N=2$ and $N=3$ are given in Sec.~\ref{subsec:zero_temp}, we restrict ourselves here to the case of $N=5$.
In Table~\ref{tab:Z5}, we show our estimates of $\beta_{c,1}$ and $\beta_{c,2}$, which are determined as the points at which $1/K$ takes the values $8/25$ and $1/2$, respectively.
In Table~\ref{tab:Z5}, we also compare our estimates to those from previous MC simulations for $L_{z}=2$ and $4$~\cite{Borisenko:2014vva}, which are likewise based on the dual spin model.

\begin{table}[htb]
  \caption{
    Estimates of critical inverse gauge couplings for $N=5$.
  }
  \label{tab:Z5}
  \centering
  \begin{tabular}{c||cc|cc}\hline
    & \multicolumn{2}{c}{$\beta_{c,1}$} & \multicolumn{2}{c}{$\beta_{c,2}$} \\ \cline{2-5}
    $L_z$ & Loop-TNR & MC~\cite{Borisenko:2014vva} & Loop-TNR & MC~\cite{Borisenko:2014vva} \\ \hline
    2  & 1.624(2) & 1.617(2) & 1.694(2) & 1.6972(14) \\
    3  & 1.838(2) &          & 1.907(3) &           \\
    4  & 1.944(2) & 1.943(2) & 2.004(7) & 1.9885(15) \\ 
    5 & 2.012(2) &        &   2.055(2) &    \\ \hline
  \end{tabular}
\end{table}

\subsection{Duality between the gauge theories and spin models}
\label{subsec:duality}

Here, we utilize the ratios of partition functions to study the phases in the $\mathbb{Z}_{N}$ gauge theory at finite temperature.
Denoting the partition function defined on an $L_{x}\times L_{y}$ lattice with periodic boundary  conditions by $Z_{L_{x}\times L_{y}}$, we introduce the following ratio of partition functions:
\begin{equation}
\label{eq:GW_ratio}
    X=\frac{Z_{L_{x}\times L_{y}}^{2}}{Z_{L_{x}\times 2L_{y}}}.
\end{equation}
We call $X$ the Gu--Wen ratio, which computes the ground-state degeneracy, simply from partition functions, when the system has a global $\mathbb{Z}_{N}$ symmetry~\cite{Gu:2009dr}.\footnote{
Recently, several studies have further highlighted the usefulness of various ratios of partition functions within tensor network calculations~\cite{Morita:2024lwg,Morita:2025hsv}, as well as partition functions defined on non-orientable geometries, such as the Klein bottle and $\mathrm{RP}^{2}$~\cite{Shimizu:2024ipw}, and twisted partition functions~\cite{Maeda:2025ycr,Akiyama:2026dzg}.
}
Eq.~\eqref{eq:GW_ratio} can be efficiently calculated using the renormalized tensor whose tensor trace estimates $Z_{L_{x}\times L_{y}}$.
The strong-coupling regime tends to preserve the $\mathbb{Z}_N$ symmetry. 
In contrast, in the large-$\beta$ region, the link variables become effectively frozen, leading to a nonzero expectation value of the Polyakov loop, which results in the spontaneous breaking of the $\mathbb{Z}_N$ center symmetry.
Furthermore, the structure of the partition function in this regime can be understood as follows. 
For sufficiently weak coupling and small $L_{z}$, it effectively reduces to the trace over the Gauss-law projector onto the gauge-invariant subspace. 
On an $L_{x}\times L_{y}$ spatial lattice, there are $2L_{x}\times L_{y}$ link variables, each taking $N$ possible values. 
These variables are subject to $L_{x}\times L_{y}$ Gauss-law constraints, of which $L_{x}\times L_{y}-1$ are linearly independent due to periodic boundary conditions. 
The dimension of the gauge-invariant Hilbert space is therefore $N^{L_{x}\times L_{y}+1}$, which equals the trace of the Gauss-law projector.
Consequently, the Gu--Wen ratio in Eq.~\eqref{eq:GW_ratio} evaluates to $X=(N^{L_{x}\times L_{y}+1})^{2}/N^{L_{x}\times 2L_{y}+1}=N$.
This has indeed been confirmed in the finite-temperature $\mathbb{Z}_{2}$ gauge theory at $L_{z}=3$~\cite{Kuramashi:2018mmi}.
Alternatively, invoking the duality between the $\mathbb{Z}_{N}$ gauge theories and $N$-state clock models as shown in Sec.~\ref{sec:model}, this result can be understood from the dual spin model perspective. 
In that description, the deconfined phase of the gauge theory corresponds to the spontaneously broken $\mathbb{Z}_N$ phase of the spin system. The Gu--Wen ratio then counts the number of symmetry-broken sectors, yielding $X=N$.
In the following, we compute the Gu--Wen ratio using BW-TRG as a diagnostic both for Eqs.~\eqref{eq:TN_ele_mag} and \eqref{eq:TN_dualS}, to confirm the duality between the original $\mathbb{Z}_{N}$ gauge theory and the dual $N$-state clock model.
As in Sec.~\ref{subsec:universal_data}, we always set $L_{x}=L_{y}=L$.

\begin{figure}[htbp]
    \centering
    \includegraphics[width=0.48\linewidth]{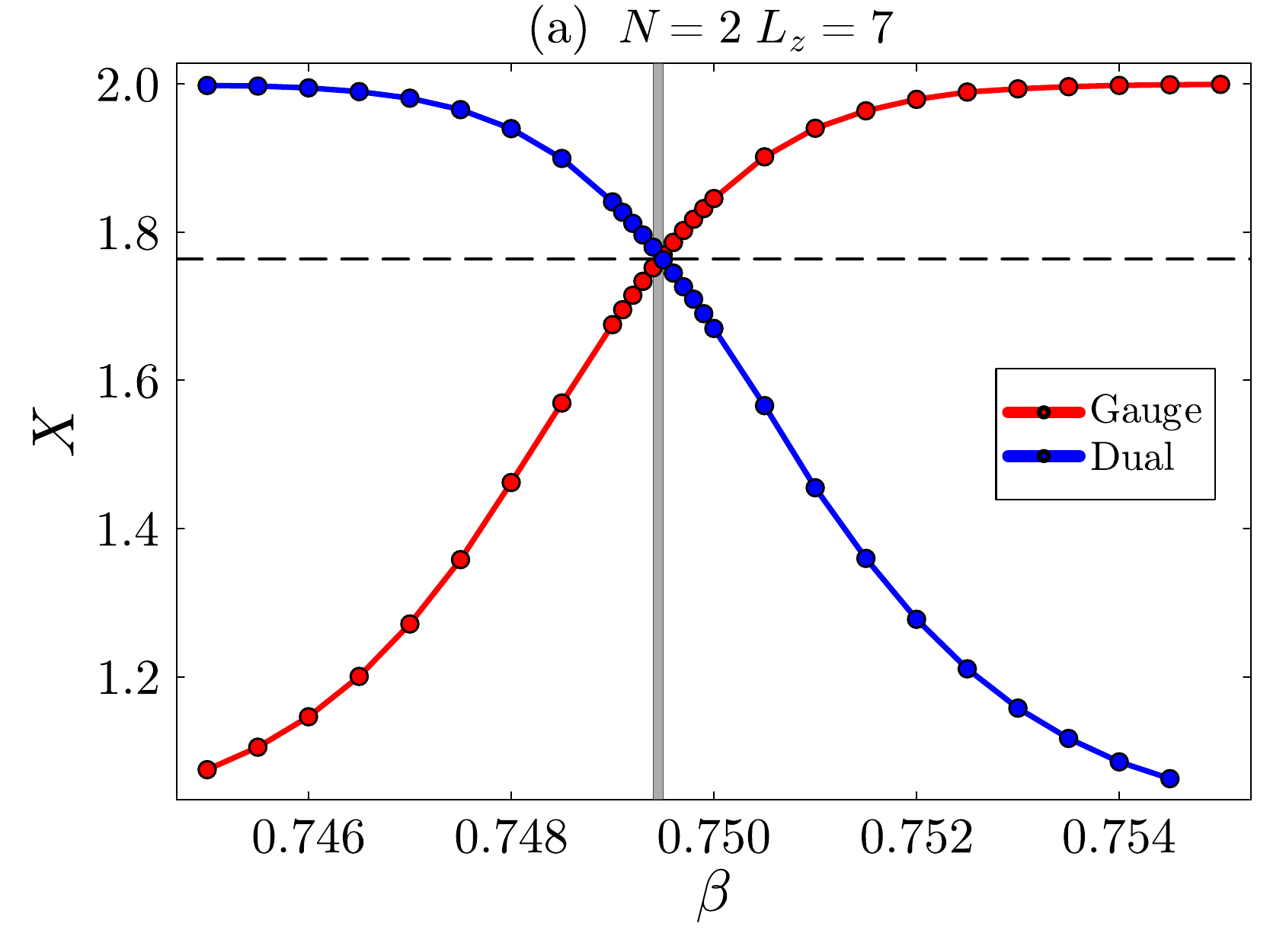}\\
    \includegraphics[width=0.48\linewidth]{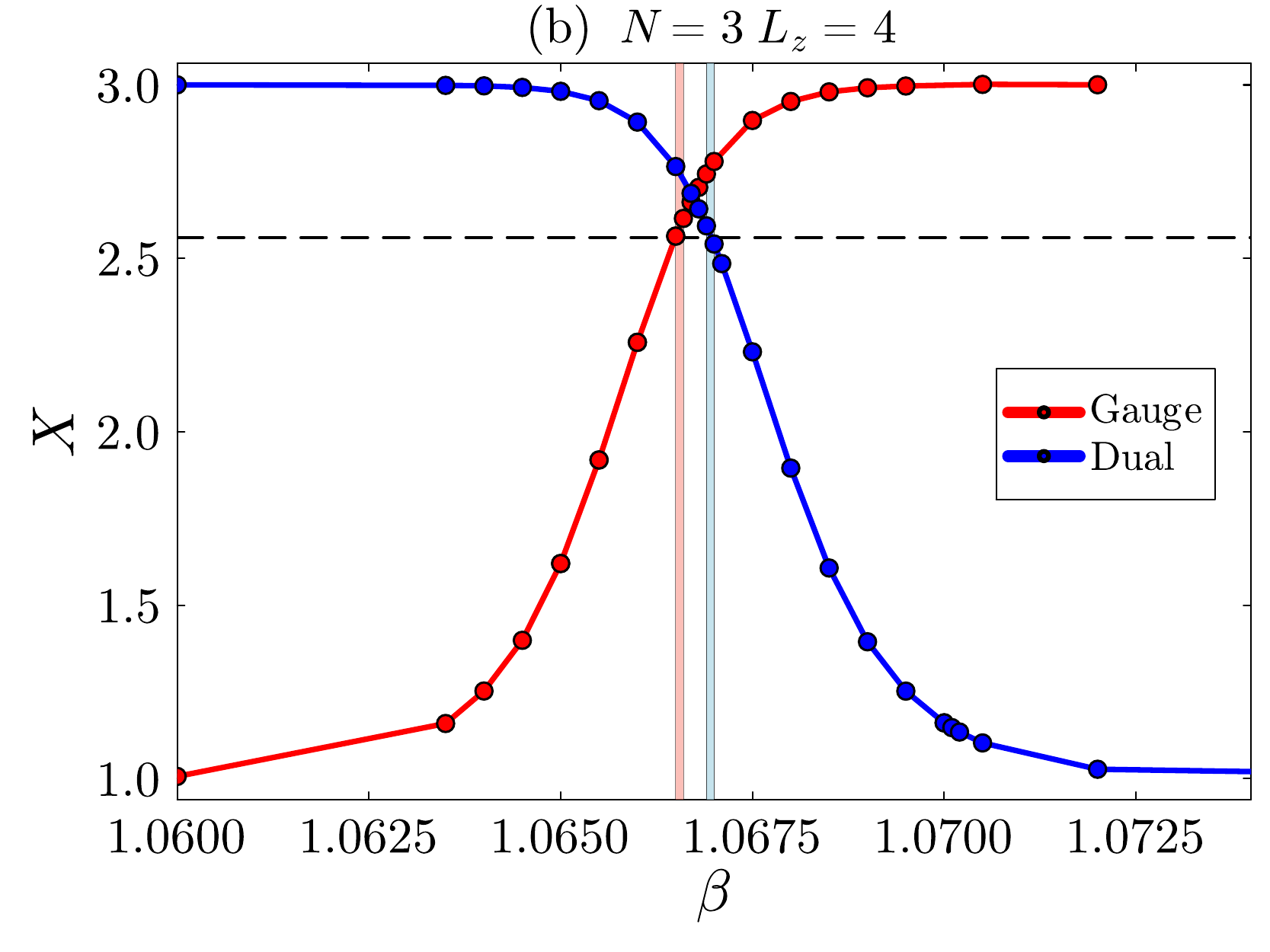}\\
    \includegraphics[width=0.48\linewidth]{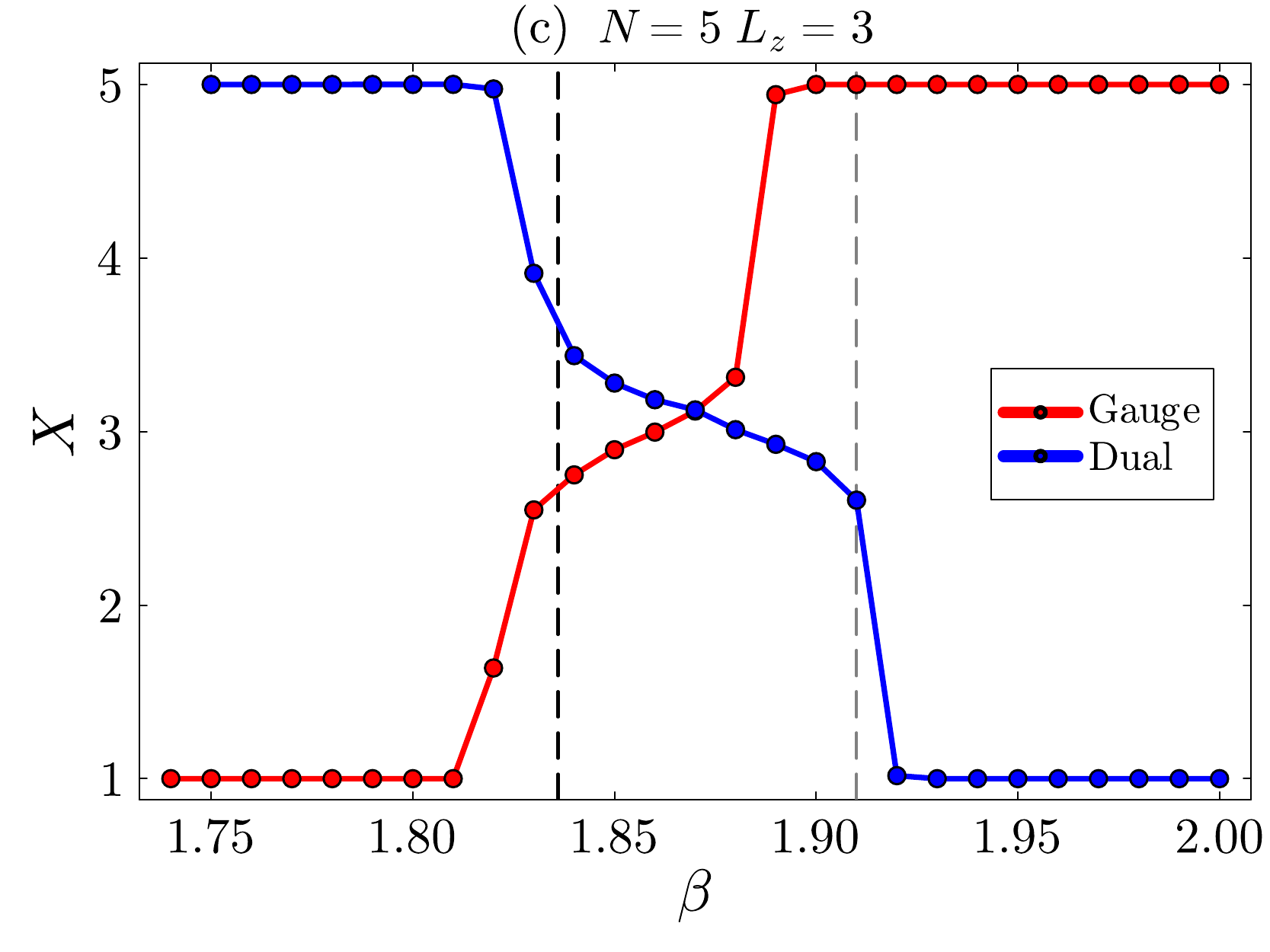}
    \caption{
        Gu--Wen ratios $X$ for
        (a) $N=2$ at $(L_{z},L)=(7,2^{6})$ with $(\chi_{{\rm TTNR}},\chi_{\text{BW-TRG}})=(120,64)$.
        (b) $N=3$ at $(L_{z},L)=(4,2^{6})$ with $(\chi_{{\rm TTNR}},\chi_{\text{BW-TRG}})=(120,81)$ for gauge theory and $(\chi_{{\rm TTNR}},\chi_{\text{BW-TRG}})=(120,100)$ for dual theory. 
        (c) $N=5$ at $(L_{z},L)=(3,2^{15})$ computed with $(\chi_{{\rm TTNR}},\chi_{\text{BW-TRG}})=(40,81)$ for gauge theory and $(\chi_{{\rm TTNR}},\chi_{\text{BW-TRG}})=(90,120)$.
        The horizontal dashed lines denote the universal values for $X$~\cite{Morita:2025hsv}. 
        The vertical bands indicate the estimates of the critical couplings obtained from the thermodynamic $X$. 
        For (b), the gray band denotes the critical coupling estimated using the gauge theory representation, while the light blue one is from the dual spin model.
        For (c), the vertical dashed lines denote the estimates of two transition points $\beta_{c,1}=1.838(2)$ and $\beta_{c,2}=1.907(3)$ from Loop-TNR.
    }
    \label{fig:gu-wen}
\end{figure}

Fig.~\ref{fig:gu-wen}(a) shows the resulting $X$ for $N=2$ at $L=2^{6}$ and $L_{z}=7$ as a function of the inverse gauge coupling $\beta$.
As expected, the Gu--Wen ratio takes the value 1 in the confinement phase ($\beta<\beta_{c}$), while it becomes 2 in the deconfinement phase ($\beta>\beta_{c}$) for the original gauge theory.
In contrast, for the dual spin model, the situation is reversed: a unique ground state appears for $\beta>\beta_{c}$, while a twofold degeneracy is realized for $\beta<\beta_{c}$.
The crossing point of the Gu--Wen ratios corresponds to the critical point.
At criticality, the resulting $X$ agrees with the universal value obtained from the modular-invariant partition function on a torus~\cite{Morita:2025hsv}.
This confirms the duality between two representations in Eqs.~\eqref{eq:TN_ele_mag} and \eqref{eq:TN_dualS} when $N=2$.

We perform a similar comparison in the case of $N=3$ as shown in Fig.~\ref{fig:gu-wen}(b), where the Gu--Wen ratio is computed at $L=2^6$ and $L_{z}=3$.
In the symmetry-broken phase, the Gu--Wen ratio clearly results in $X=3$.
In the dual spin model, strong and weak coupling are interchanged, as we observed for $N=2$. 
We also see the Gu--Wen ratios approach a universal value at criticality, despite small deviations in their crossing points.

We now move on to the case with $N=5$. 
For the original gauge theory, we observe a unique ground state in the strong coupling region and fivefold degeneracy in the weak coupling region.
In contrast to the $\mathbb{Z}_{2}$ and $\mathbb{Z}_{3}$ theories, there is an intermediate-coupling region, where the Gu--Wen ratio is found to vary continuously and take nontrivial values, as shown in Fig.~\ref{fig:gu-wen}(c).
Deep inside the intermediate-coupling region, the value of the Gu--Wen ratio is largely insensitive to the system volume, which is another typical signal of the BKT phase~\cite{Li:2020sbg}, in addition to the plateau of $c=1$ and characteristic values for the Luttinger parameter $K$ we observed in Fig~\ref{fig:Universaldataz5}.
We further compare the transition points $\beta_{c,1}$ and $\beta_{c,2}$ estimated by Loop-TNR in Sec.~\ref{subsec:universal_data} with $X$ obtained from BW-TRG, as shown in Fig.~\ref{fig:gu-wen}(c).
Since these $\beta_{c,1}$ and $\beta_{c,2}$ from Loop-TNR are estimated based on the dual spin representation, the behavior of $X$ in the dual spin model is more consistent with these transition points than the corresponding result obtained from the original gauge representation.
Since we do not compare our results with a CFT prediction for the $N=5$ case, we briefly discuss the finite-bond-dimension effects.
We have confirmed that our choice of $(\chi_{\rm TTNR},\chi_{\rm BW\text{-}TRG})=(90,120)$ is sufficiently large to suppress finite-bond-dimension effects associated with both the temporal and spatial bond dimensions in the dual spin model.
In particular, we find that the value of the Gu--Wen ratio is well converged even inside the BKT phase as shown in Figure~\ref{fig:chittnrchitrgdependence}.
For the case of original gauge theory, see Appendix~\ref{appendix:projector}.

\begin{figure}[htbp]
  \centering
    \includegraphics[width=0.4\linewidth]{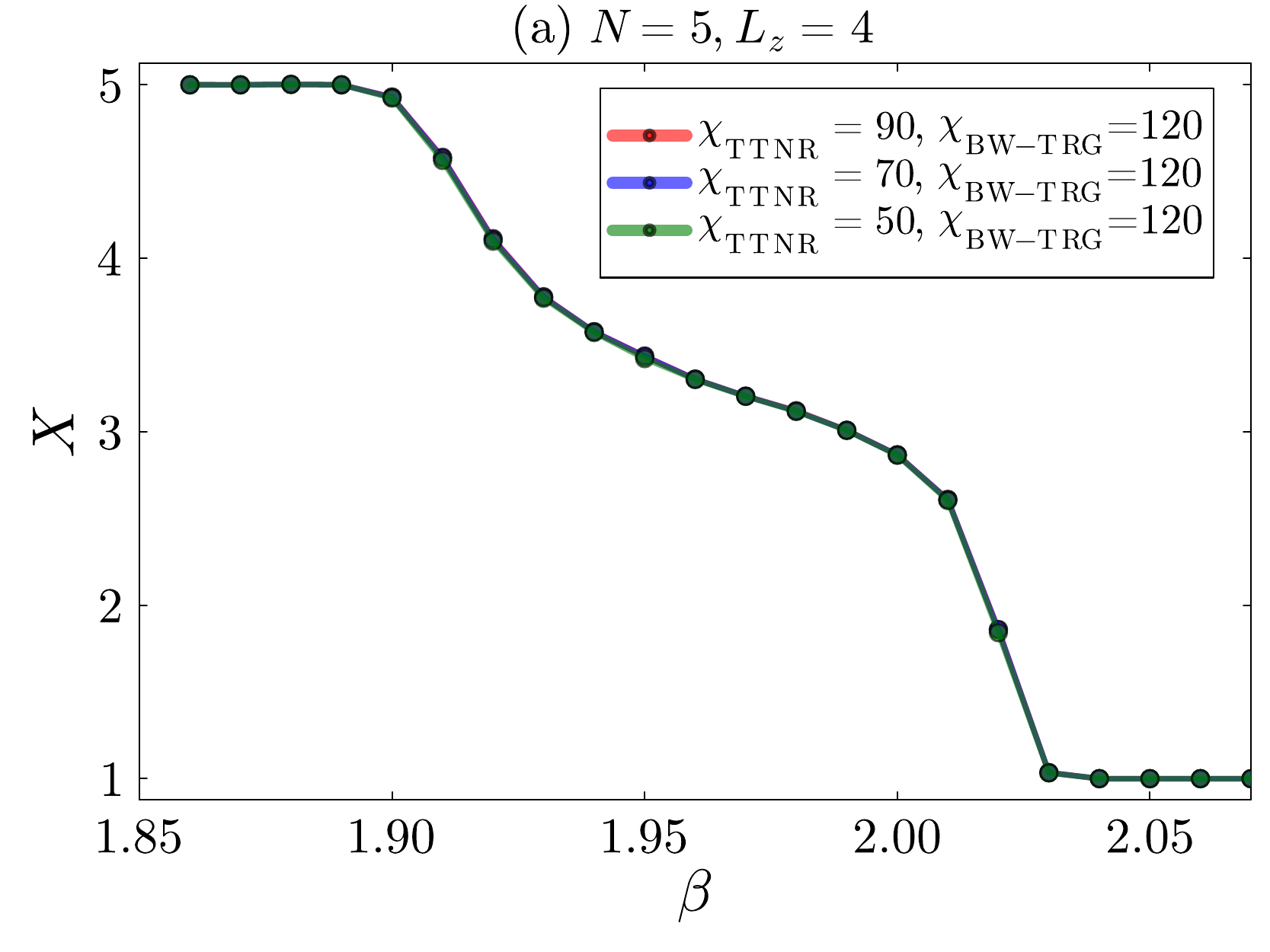}
    \includegraphics[width=0.4\linewidth]{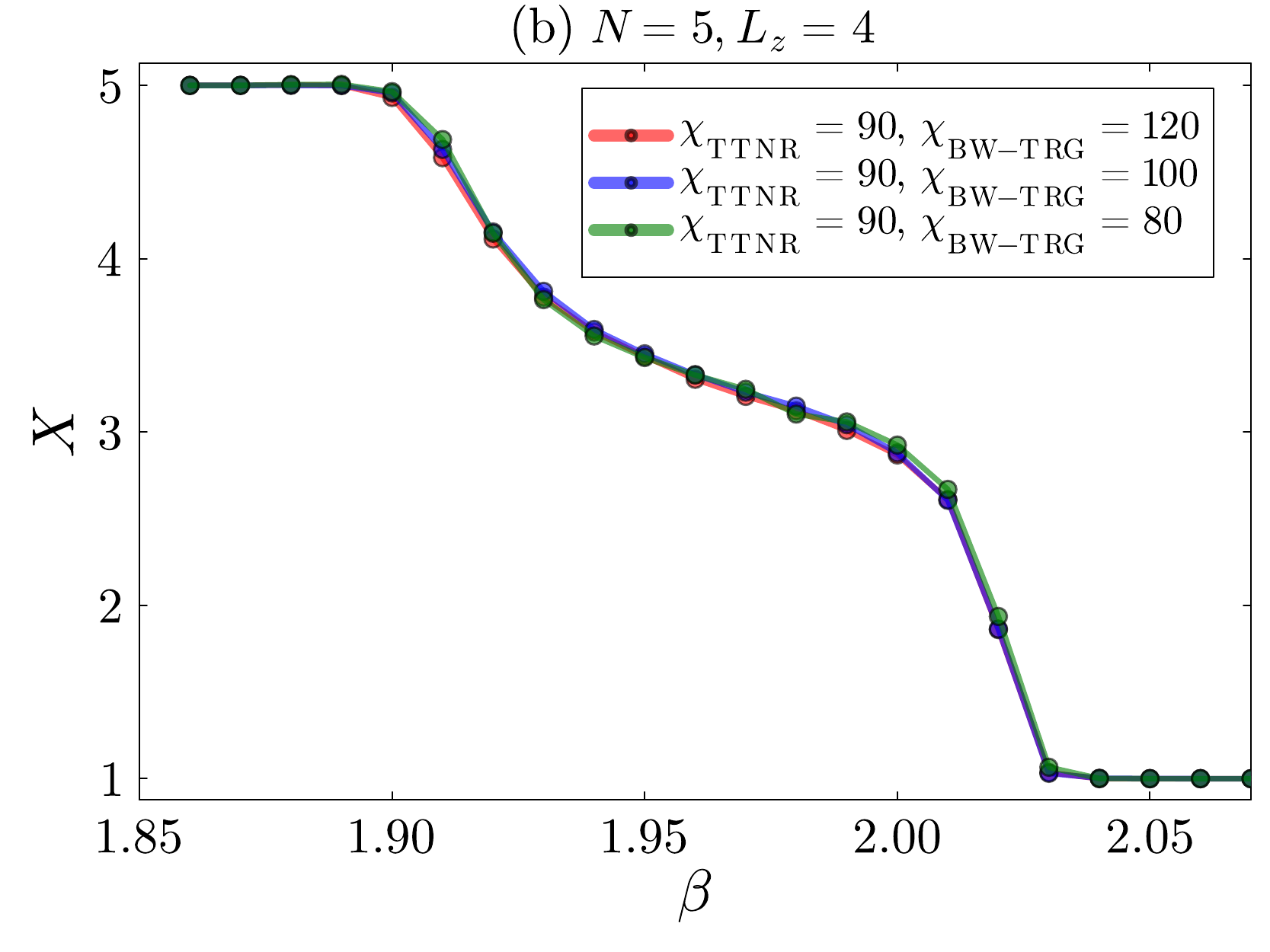}
  \caption{(a) $\chi_{\text{TTNR}}$ and (b) $\chi_{\text{BW-TRG}}$ dependences of Gu--Wen ratio for the $N=5$ dual theory.}
  \label{fig:chittnrchitrgdependence}
\end{figure}

\subsection{Critical points in the zero-temperature limit}
\label{subsec:zero_temp}

When $N=2,3$, the discontinuity in the Gu--Wen ratio in the thermodynamic limit locates the critical point.
Table~\ref{tab:Z2} summarizes the resulting critical couplings $\beta_{c}(L_{z})$ for $N=2$. 
The temporal lattice extent $L_{z}$ is in the first column, and the estimates of the critical coupling based on the original gauge theory and the dual spin model are in the second and third columns, respectively.
Values of the critical coupling $\beta_c$ are obtained from the Gu--Wen ratio $X$ in the limit $L \to \infty$, where it shows an almost exact step-like behavior between the asymptotic values $1$ and $N$, thus enabling us to extract the value of $\beta_c$ with a very small error bar.

We confirm that the critical points estimated in this way are in agreement with the peak positions of the specific heat, as presented in Appendix~\ref{app:specific_heat}.
The two critical points determined from the original gauge theory and dual spin model are consistent with each other at the same $L_{z}$ within the estimated errors.
Slight deviations are observed only for $L_{z}=9,11$, which may be attributed to the finite bond dimension used in our computations.
Our estimates are also consistent with those obtained from previous MC simulations~\cite{Caselle:1995wn} and TRG calculations~\cite{Kuramashi:2018mmi}.
Table~\ref{tab:Z3} shows the critical couplings for $N=3$. 
At $L_{z}=2,4$, we confirm that our estimates are consistent with those from previous MC simulations~\cite{Brower:1991ip}.
We observe that $\beta_{c}(L_{z})$ obtained from the different representations agree typically up to three digits for $L_{z}>4$.
Since the dual tensor-network representation in Eq.~\eqref{eq:TN_dualS} has a smaller temporal bond dimension than the original representation in Eq.~\eqref{eq:TN_AC}, we expect that the former can achieve better accuracy at the same bond dimensions $\chi_{{\rm TTNR}}$ and $\chi_{\text{BW-TRG}}$.

\begin{table}[htb]
  \caption{
    Critical inverse gauge couplings $\beta_{c}(L_{z},L\to\infty)$ for $N=2$.
    For finite $L_{z}$, the thermodynamic Gu--Wen ratios are employed to obtain these estimates with the bond dimensions $(\chi_{{\rm TTNR}},\chi_{\text{BW-TRG}})=(64,120)$.
}
  \label{tab:Z2}
  \centering
  \begin{tabular}{c||cccccc}\hline
    & \multicolumn{6}{c}{$\beta_c$} \\ \cline{2-7}
    $L_z$ & Gauge & Dual spin & MC~\cite{Caselle:1995wn} & TRG~\cite{Kuramashi:2018mmi} & MC~\cite{Borisenko:2013xna}  & MC~\cite{Ferrenberg:2018zst} \\ \hline
    2  & 0.65605(5) & 0.65605(5) & 0.65608(5) & 0.656097(1) & & \\
    3  & 0.71115(5) & 0.71115(5) & 0.71102(8) & 0.711150(4) & & \\
    4  & 0.73105(5) & 0.731065(5) & 0.73107(2) & & & \\
    5  & 0.74055(5) & 0.74065(5) & 0.74057(3) & 0.740730(3) & & \\
    6  & 0.74605(5) & 0.74605(5) & 0.746035(8) & & & \\
    7  & 0.74945(5) & 0.74945(5) & 0.74947(2) & & & \\
    8  & 0.75175(5) & 0.75185(5) & 0.75180(1) & & & \\
    9  & 0.75335(5) & 0.75355(5) & & & & \\
    10 & 0.75465(5) & 0.75475(5) & 0.75472(1) & & & \\
    11 & 0.75555(5) & 0.75585(5) & & & & \\
    12 & 0.75655(5) & 0.75655(5) & 0.756427(6) & & & \\
    13 & 0.75705(5) & 0.75715(5) & & & & \\
    14 & 0.75755(5) & 0.75765(5) & 0.757527(8) & & & \\
    $\infty$ & 0.76139(13) & 0.76157(13) & & & 0.761395(4) & 0.761413292(11) \\ \hline
  \end{tabular}
\end{table}

\begin{table}[htb]
  \caption{
    Critical inverse gauge couplings $\beta_{c}(L_{z},L\to\infty)$ for $N=3$.
    For finite $L_{z}$, the thermodynamic Gu--Wen ratios are employed to obtain these estimates with the bond dimensions $(\chi_{{\rm TTNR}},\chi_{\text{BW-TRG}})=(81,120)$ for gauge theory and $(\chi_{{\rm TTNR}},\chi_{\text{BW-TRG}})=(100,120)$ for dual theory.
  }
  \label{tab:Z3}
  \centering
  \begin{tabular}{c||ccccc}\hline
    & \multicolumn{5}{c}{$\beta_c$} \\ \cline{2-6}
    $L_z$ & Gauge & Dual spin & MC~\cite{Brower:1991ip} & MC~\cite{Borisenko:2013xna} & MC~\cite{Bazavov:2007tw}\\ \hline
    2  & 0.98205(5) & 0.98205(5) & 0.9822(5) & & \\
    3  & 1.04755(5) & 1.04755(5) &  & & \\
    4  & 1.06655(5) & 1.06695(5) & 1.0668(5) & & \\
    5  & 1.07295(5) & 1.07475(5) & & & \\
    6  & 1.07685(5) & 1.07875(5) & & & \\
    7  & 1.07925(5) & 1.08095(5) & & & \\
    8  & 1.08055(5) & 1.08215(5) & & & \\
    9  & 1.08145(5) & 1.08305(5) & & & \\
    10 & 1.08245(5)  & 1.08335(5) & & & \\
    11 & 1.08295(5) & 1.08345(5) & & & \\
    $\infty$ & 1.08538(16)  & 1.08455(8) & & 1.0844(2) & 1.084314(8)\\ \hline
  \end{tabular}
\end{table}

\begin{figure}[tb]
    \centering
    \includegraphics[width=0.48\linewidth]{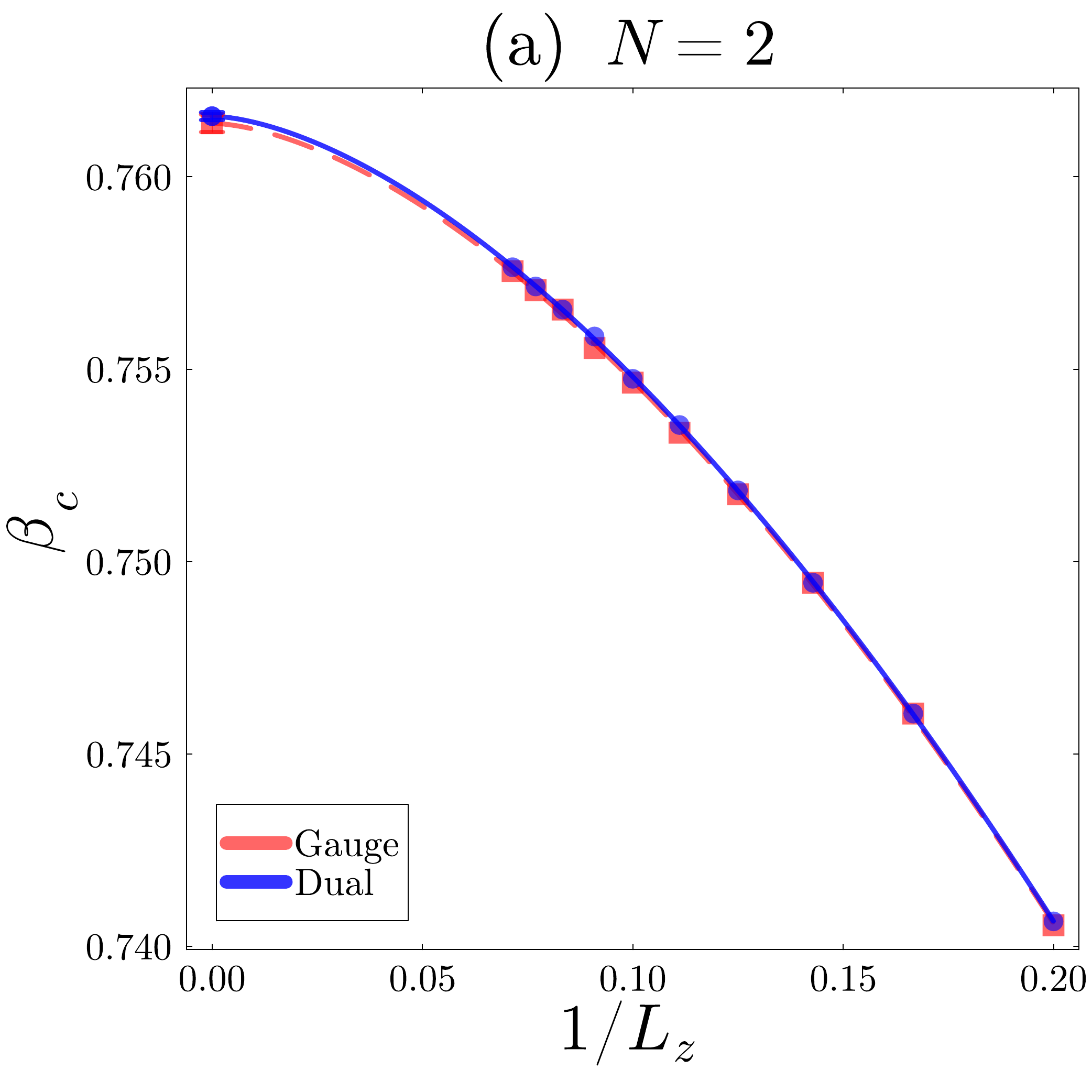}
    \includegraphics[width=0.48\linewidth]{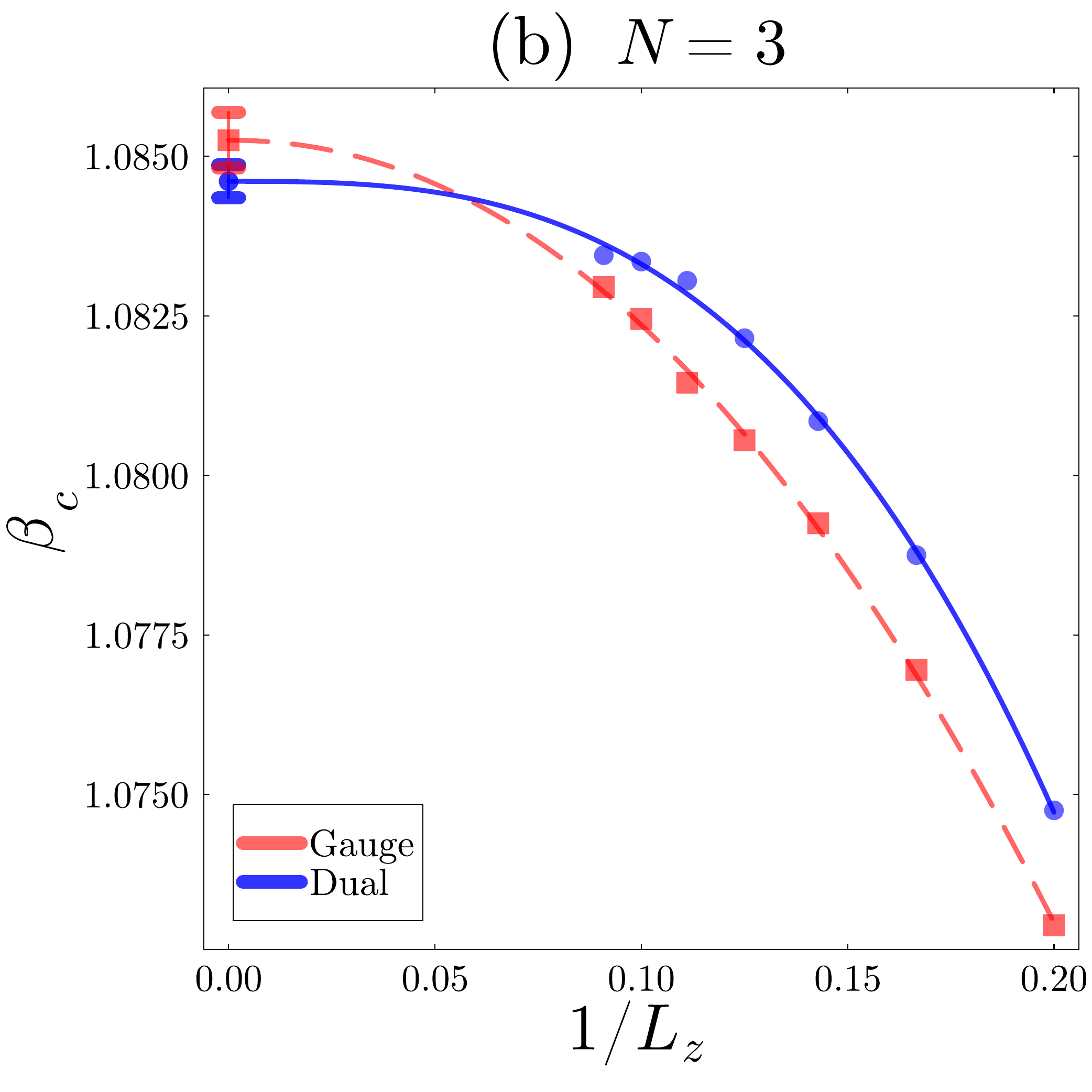}
    \caption{
        Zero-temperature extrapolation of the critical inverse gauge couplings for (a) $N=2$ and (b) $N=3$.
        Gray and blue points denote the critical couplings obtained from the gauge theory representations and dual spin representations, respectively.
    }
    \label{fig:zero_temp_lim}
\end{figure}

Finally, we extrapolate these resulting critical couplings to the zero-temperature limit.
This will serve as a non-trivial benchmark to test our TTNR approach, because the values of $\beta_{c}(L_{z}\to\infty)$ are available thanks to the duality argument~\cite{Balian:1974ir,KorthalsAltes:1978tp,Savit:1979ny}, which transforms the critical temperature of the three-dimensional $\mathbb{Z}_{N}$ spin systems into $\beta_{c}$ of the (2+1)-dimensional $\mathbb{Z}_{N}$ lattice gauge theory at vanishing temperature when $N<5$.
These critical temperatures are accurately determined by MC simulations~\cite{Bazavov:2007tw,Ferrenberg:2018zst,Wada:2025ycz}.
We assume that the finite-$L_{z}$ effects take the form,
\begin{align}
\label{eq:fit}
    \beta_{c}(L_{z})
    =
    \beta_{c}^{*}
    +
    cL_{z}^{-1/\nu},
\end{align}
where $\beta_{c}^{*}$ denotes the critical inverse gauge coupling in the zero-temperature limit, and $\nu$ is the critical exponent of the corresponding three-dimensional CFT.
As shown in Fig.~\ref{fig:zero_temp_lim}(a), for $N=2$, the data for $\beta_{c}(L_{z})$ are well described by Eq.~\eqref{eq:fit} for $L_{z}\ge 5$.
We obtain
\begin{align}
\label{eq:beta_c_gauge}
    \beta_{c}^{*} = 0.76139(13),~~~\nu=0.611(8),
\end{align}
from the original gauge theory and
\begin{align}
\label{eq:beta_c_spin}
    \beta_{c}^{*} = 0.76157(13),~~~\nu=0.614(8),
\end{align}
from the dual spin model.
These $\beta_{c}^{*}$ are consistent with previous MC simulations based on the $\mathbb{Z}_{2}$ gauge theory~\cite{Borisenko:2013xna} and critical coupling obtained from the critical temperature of the three-dimensional Ising model~\cite{Ferrenberg:2018zst} via the duality relation.\footnote{
One can relate the critical temperature $T_{c}$ of the three-dimensional ${N}$-state clock model to $\beta_{c}^{*}$ via the following equations:
\begin{align}
    \beta_{c}^{*}
    =
    -\frac{1}{2}\ln\tanh T_{c}^{-1},
\end{align}
for $N=2$, and
\begin{align}
    \beta_{c}^{*}
    =
    -\frac{2}{3}\ln\frac{1-{\rm e}^{-3 T_{c}^{-1}/2}}{1+2{\rm e}^{-3T_{c}^{-1}/2}},
\end{align}
for $N=3$.
}
As shown in Appendix~\ref{appendix:projector}, our TTNR algorithm that introduces the optimal temporal projectors is crucial to observe this consistency.
To the best of our knowledge, this is the first successful determination of the deconfinement transition point in lattice gauge theory at the vanishing temperature using tensor-network methods based on the Lagrangian formalism.
This suggests that the TTNR approach enables the identification of criticality in the zero-temperature limit, as anticipated in the original work~\cite{Ueda:2025mhu}. 
Although we observe a deviation of the resulting critical exponent $\nu$ from the MC estimate $\nu=0.629912(86)$~\cite{Ferrenberg:2018zst} and the conformal bootstrap $\nu=0.629971(4)$~\cite{Kos:2016ysd}, this discrepancy can be attributed to the fact that our algorithm does not directly optimize the contraction of the full three-dimensional tensor network.

Fig.~\ref{fig:zero_temp_lim}(b) shows the fit to the $\beta_{c}(L_{z})$ data for $N=3$ with $L_{z}\ge 5$.
From this fit, we obtain
\begin{align}
    \beta_{c}^{*} = 1.08538(16),~~~\nu=0.491(13),
\end{align}
for the original gauge theory, and
\begin{align}
    \beta_{c}^{*} = 1.08455(8),~~~\nu=0.334(7),
\end{align}
from the dual spin model.
The latter estimate of $\beta_{c}^{*}$ is closer to the values obtained by MC simulations for the $\mathbb{Z}_{3}$ gauge theory~\cite{Borisenko:2013xna} and the three-state Potts model~\cite{Bazavov:2007tw} via the duality relation.
Since the three-state Potts model exhibits a first-order phase transition in three dimensions~\cite{PhysRevLett.43.799}, $\nu=1/3$ is expected. 
Although the value of $\nu$ obtained from the original gauge theory deviates from $1/3$, that extracted from the dual spin model is consistent with this expectation.
This provides further evidence that TTNR can effectively locate the critical coupling in the zero-temperature limit beyond $N=2$, and that increasing the bond dimension will reduce the discrepancy between the two representations.

\section{Summary and Outlook}
\label{sec:summary}

We have investigated deconfinement transitions for $\mathbb{Z}_2, \mathbb{Z}_3$ and $\mathbb{Z}_5$ gauge theories and their dual spin models using the TTNR method.
In this work, we improved the TTNR method by introducing an optimal projection along the temporal direction.
The resulting tensors were then coarse-grained spatially using TRG/TNR algorithms such as Loop-TNR and BW-TRG.
Using Loop-TNR, we identified the universality classes of the finite-temperature phase transitions and confirmed the Svetitsky--Yaffe conjecture.
In particular, we found strong evidence that the finite-temperature $\mathbb{Z}_{5}$ gauge theory exhibits an intermediate phase with an emergent U(1) symmetry described by the Tomonaga--Luttinger liquid theory.
We also studied the duality between the $\mathbb{Z}_N$ gauge theories and $N$-state clock models using the Gu--Wen ratio, which explicitly demonstrates that the role of the original gauge coupling is interchanged in the dual spin models.
By extrapolating the critical couplings obtained at finite temperature, we determined the zero-temperature critical gauge couplings for $N=2$ and $3$, finding good agreement with MC simulations.
Our results provide the first successful determination of the deconfinement transition points in lattice gauge theory at the vanishing temperature using the TRG/TNR approach.
An interesting direction for future work is to extend our TTNR algorithm to (2+1)-dimensional SU($N$) lattice gauge theories at finite temperature, as existing tensor network studies are currently restricted to the high-temperature regime with $L_{z}=1$~\cite{Yosprakob:2024sfd}.

Although the TTNR has been improved in the present work, further refinements are possible by incorporating an ``entanglement-filtering" scheme for three-dimensional tensor networks.
Such a scheme is particularly important in three dimensions, where entanglement associated with local structures, such as corner triple lines (CTLs) and edge double lines (EDLs), grows linearly with the system volume, in contrast to the constant scaling observed in two dimensions~\cite{lyu2025latticereflectionsymmetryTNR}.
Implementing entanglement filtering would therefore enable simulations to reach larger extents in the temporal direction while requiring a smaller bond dimension.

An important future direction is to extend the present study to gauge theories at finite temperature with continuous symmetry groups such as $\mathrm{U}(1)$ and $\mathrm{SU}(N)$. 
Although the path integrals for these theories are formally formulated as tensor networks with infinite bond dimension, the character expansion allows us to truncate them to a finite bond dimension~\cite{Liu:2013nsa}.
While the convergence of the character expansion becomes slower toward the weak-coupling regime, it may still be applicable in the context of studying finite-temperature deconfinement transitions, which occur at finite gauge coupling. 
In particular, motivated by the Svetitsky--Yaffe conjecture, this provides a promising setting because the character expansion preserves the same global symmetry as the original theory and is therefore suited to our approach based on CFT data.
The main challenge lies in the difficulty of performing the dual transformation used in this work for $\mathrm{SU}(N)$ gauge theories, which makes the numerical cost significantly higher.
Nevertheless, our blocking procedure for the $\mathbb{Z}_{N}$ theories is expected to remain useful.
This is because our $\mathbb{Z}_N$ formulation can be viewed as working in the representation category $\mathrm{Rep}(\mathbb{Z}_N)$ where the bond space corresponds to the Fourier basis of the gauge degrees of freedom. 
The same structural principles carry over to $\mathrm{SU}(N)$, which suggests that the extension to non-Abelian gauge groups does not require fundamentally new algorithmic ingredients. 
This opens the door to applying the present framework to broader class of lattice gauge theories with non-Abelian symmetries, which we leave for future work.

\acknowledgments
We acknowledge helpful discussions with Nick Bultinck. A.U. thanks Etsuko Itou for the inspiring discussion during the conference at YITP.
A part of the numerical calculation was carried out using the computational resources of SQUID provided by Osaka University through the HPCI System Research Project (Project ID: hp250055, G16353). We also used the Barcelona Supercomputing cluster under the EuroHPC program (Project ID: EHPC-DEV-2025D09-030
) to run parts of the simulations.
We also used the supercomputer Pegasus under the Multidisciplinary Cooperative Research Program of Center for Computational Sciences, University of Tsukuba and Yukawa-21 at YITP in Kyoto University. 
A.~N.\ is supported by FWO doctoral fellowship (grant No. 11A8E26N). 
Y.~S.\ is supported by Graduate Program on Physics for the Universe (GP-PU), Tohoku University and JSPS KAKENHI (Grant-in-Aid for JSPS Fellows) Grant Number 25KJ0537.
S.~A.\ acknowledges the support from JSPS KAKENHI Grant Numbers JP23K13096 and JP25H01510, the Center of Innovations for Sustainable Quantum AI (JST Grant Number JPMJPF2221), the Endowed Project for Quantum Software Research and Education, the University of Tokyo~\cite{qsw}, and the Top Runners in Strategy of Transborder Advanced Researches (TRiSTAR) program conducted as the Strategic Professional Development Program for Young Researchers by the MEXT.
J.~H.\ and A.~N.\ acknowledge support from the European Research Council (ERC) under the European Union’s Horizon 2020 program (grant agreement No. 101125822).
A.~U.\ is supported by FWO Junior Postdoctoral Fellowship (grant No. 3E0.2025.0049.01) and the Watanabe Foundation. Numerical computations were partially carried out on EuroHPC (EHPC-DEV-2025D12-166). Tensor network simulations were carried out with the help of the Ghent Quantum group softwares TensorKit \& TNRKit. \cite{vanthilt2026practicalintroductiontensornetwork, devos2025tensorkitjljuliapackagelargescale}
 
We benefited from discussions at the YITP workshop (YITP-I-25-02) on ``Recent Developments and Challenges in Tensor Networks: Algorithms, Applications to Science, and Rigorous Theories."

\appendix

\section{Another tensor network formulation}
\label{app:TN_alt}

\begin{figure}[tb]
    \centering
    \includegraphics[width=0.4\linewidth]{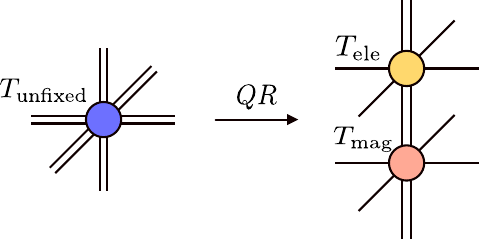}
    \caption{QR decomposition converts the fundamental tensor in Eq.~\eqref{eq:T_direct} into $T^{(n)}_{\text{ele}}$ and $T^{(n)}_{\text{mag}}$.}
    \label{fig:T_unfixed}
\end{figure}

We point out that a similar tensor-network representation can also be obtained without explicitly performing the character expansion.
As proposed in Ref.~\cite{Kuramashi:2019cgs}, the Boltzmann weights on the left-hand sides of Eqs.~\eqref{eq:char_ex} and \eqref{eq:fourier_ex} can already be regarded as local tensors.
This idea has also been applied to higher-dimensional calculations as well~\cite{Kuwahara:2022ubg,Akiyama:2023hvt,Nakayama:2024lhb}.
In our $\mathbb{Z}_{N}$ gauge theory, one can immediately define a four-leg tensor,
\begin{align}
    W^{(n, \mu, \nu)}_{a_{n,\mu}a_{n+\hat{\mu},\nu}a_{n+\hat{\nu},\mu}a_{n,\nu}}
    =
    \exp\left(
        \beta\cos \left(
            \frac{2\pi}{N}f_{n,\mu\nu}
        \right)
    \right),
\end{align}
which lives on each plaquette.
For the moment, we consider the following local tensor $T^{(n)}$ at a lattice site $n$:
\begin{align}
\label{eq:T_direct}
    &T^{(n)}_{(x_1,x_2)(y_1,y_2),(t_1,t_2)(x_1',x_2')(y_1',y_2'),(t_1',t_2')}\nonumber\\
    &=
    \sum_{a_{n,x},a_{n,y},a_{n,z}} 
    W^{(n,x,y)}_{a_{n,x} x_{2} y_{2} a_{n,y}}
    W^{(n,y,z)}_{a_{n,y} y_{1} z_{2} a_{n,z}}
    W^{(n,x,z)}_{a_{n,x} x_{1} z_{1} a_{n,z}}
    \delta_{a_{n,x},y_2',z_1'}
    \delta_{a_{n,y},z_2',x_2'}
    \delta_{a_{n,z},x_1',y_1'},
\end{align}
where $\delta_{i,j,k}$ takes the value 1 when $i=j=k$, and 0 otherwise.
We now use the QR decomposition to split $W^{(n,y,z)}$ and $W^{(n,x,z)}$ into two parts;\footnote{
Although we have described the derivation of Eqs.~\eqref{eq:another_ele} and \eqref{eq:another_mag} in terms of the QR decomposition, any decomposition that separates $a_{n,x}$ and $a_{n,y}$ from the weights $W^{(n,y,z)}$ and $W^{(n,x,z)}$ is equally valid.
}
\begin{align}
    W^{(n,y,z)}_{a_{n,y} y_{1} z_{2} a_{n,z}}
    =
    \sum_{a}Q_{a_{n,y} a}R_{a y_{1} z_{2} a_{n,z}},
    ~~~
    W^{(n,x,z)}_{a_{n,x} x_{1} z_{1} a_{n,z}}
    =
    \sum_{b}Q_{a_{n,x} b}R_{b x_{1} z_{1} a_{n,z}}.
\end{align}
We then define two tensors;
\begin{align}
\label{eq:another_ele}
    (T^{(n)}_{\text{ele}})_{x_1 y_1 z_1 z_2 x_1' y_1' a b}
    =
    \sum_{a_{n,z}}
    R_{a y_{1} z_{2} a_{n,z}}
    R_{b x_{1} z_{1} a_{n,z}}
    \delta_{a_{n,z},x_1',y_1'},
\end{align}
and
\begin{align}
\label{eq:another_mag}
    (T^{(n)}_{\text{mag}})_{x_2 y_2 ab x_2'y_2' z_1' z_2'}
    =
    \sum_{a_{n,x},a_{n,y}}
    Q_{a_{n,x} b} Q_{a_{n,y} a}
    W^{(n,x,y)}_{a_{n,x} x_{2} y_{2} a_{n,y}}
    \delta_{a_{n,x},y_2',z_1'}
    \delta_{a_{n,y},z_2',x_2'}.
\end{align}
This procedure indirectly decomposes $T^{(n)}$ in Eq.~\eqref{eq:T_direct} into $T^{(n)}_{\text{ele}}$ and $T^{(n)}_{\text{mag}}$, as illustrated in Fig.~\ref{fig:T_unfixed}.
We note that both $T^{(n)}_{\text{ele}}$ and $T^{(n)}_{\text{mag}}$ have spatial bond dimension $N$.
These two tensors give an alternative way to construct Eq.~\eqref{eq:TN_ele_mag}.
At each imaginary-time slice, the tensors $T^{(n)}_{\text{mag}}$ form a PEPO whose geometry is the same as the magnetic PEPO shown in Fig.~\ref{fig:InitialPEPOs}(a).
In contrast, the tensors $T^{(n)}_{\text{ele}}$ constitute another PEPO, whose geometry corresponds to the electric PEPO depicted in Fig.~\ref{fig:InitialPEPOs}(b).

\section{Construction of the projectors}
\label{appendix:projector}
\begin{figure}[h]
    \centering
    \includegraphics[width=0.6\linewidth]{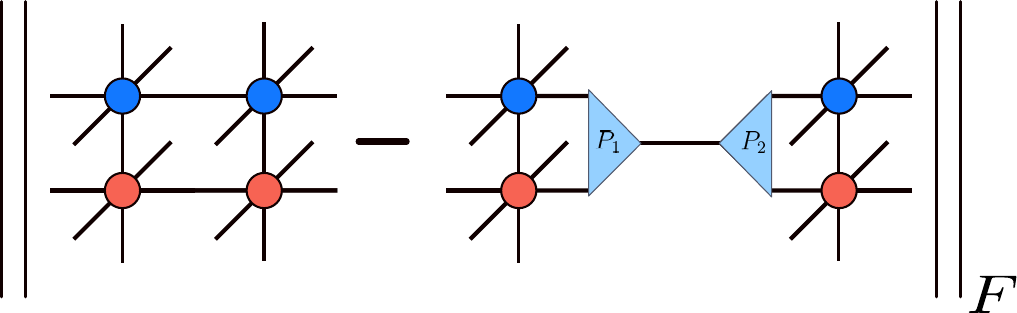}
    \caption{Cost function for the local update method.}
    \label{fig:cost_func_local}
\end{figure}

\begin{figure}[h]
    \centering
    \includegraphics[width=0.6\linewidth]{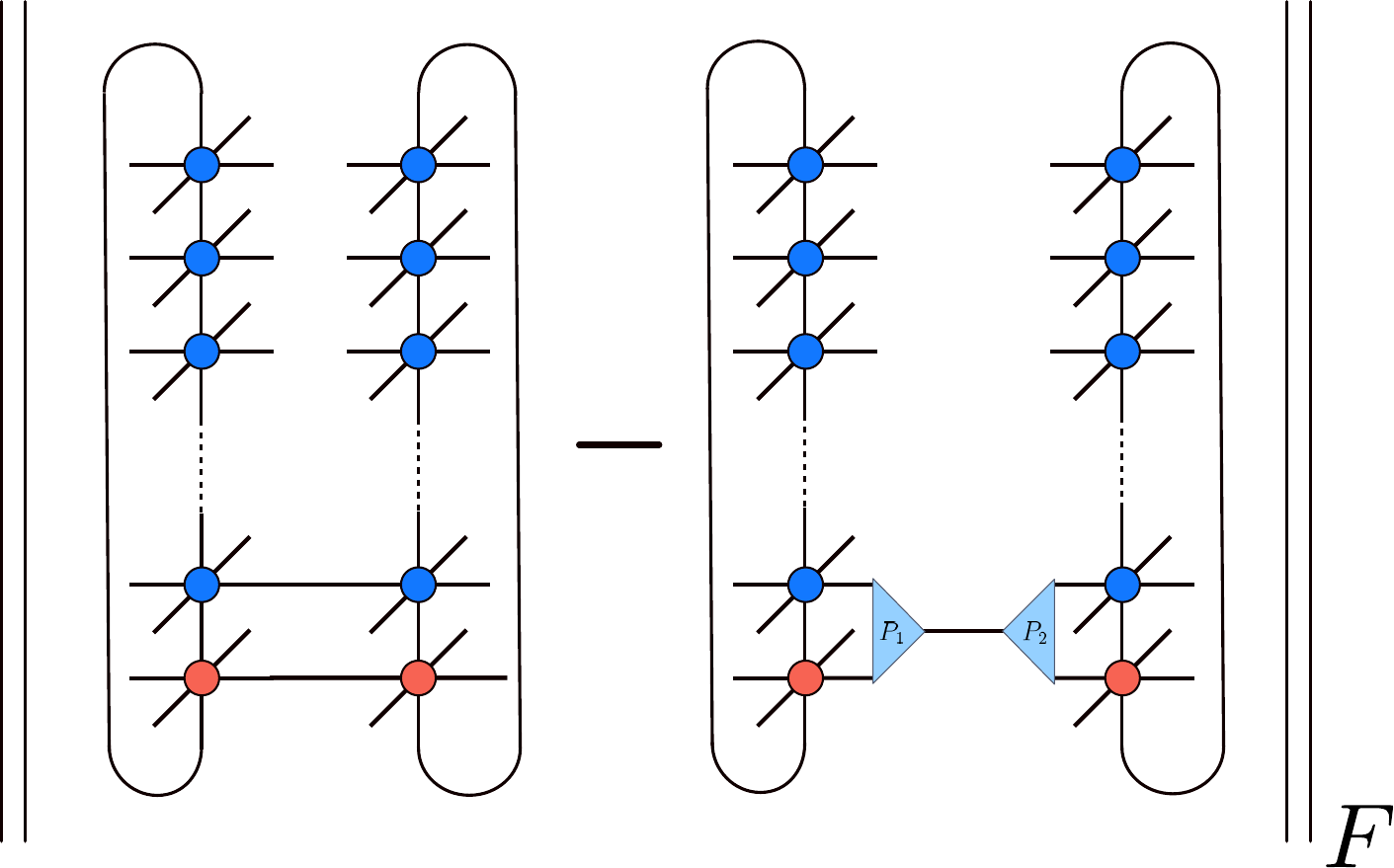}
    \caption{Cost function for the full-update method.}
    \label{fig:cost_func_full}
\end{figure}

We demonstrate our algorithm for the contraction along the temporal direction.
In the usual HOTRG method or its variants based on projectors, the cost function in Fig.~\ref{fig:cost_func_local} is minimized at every coarse-graining step, where $P_{1}$ and $P_{2}$ denote the projectors.
The red symbols in the bottom layer are the renormalized tensors at a given step of the contraction along the temporal direction, and the blue ones represent the original fundamental tensors.
We refer to this construction of projectors in the manner in Fig.~\ref{fig:cost_func_local} as the ``local update" method, since it minimizes the norm only for a local sub-network consisting of the target tensors and their adjacent tensors.
In this work, we adopt an alternative scheme, which we call the ``full-update" method, based on optimizing a larger sub-network $\Gamma$, consisting of all fundamental tensors at each spatial lattice site along the temporal direction.
We choose projectors to minimize the cost function in Fig.~\ref{fig:cost_func_full} at each approximate contraction step.


\begin{figure}[ht]
    \centering
    \includegraphics[width=0.48\linewidth]{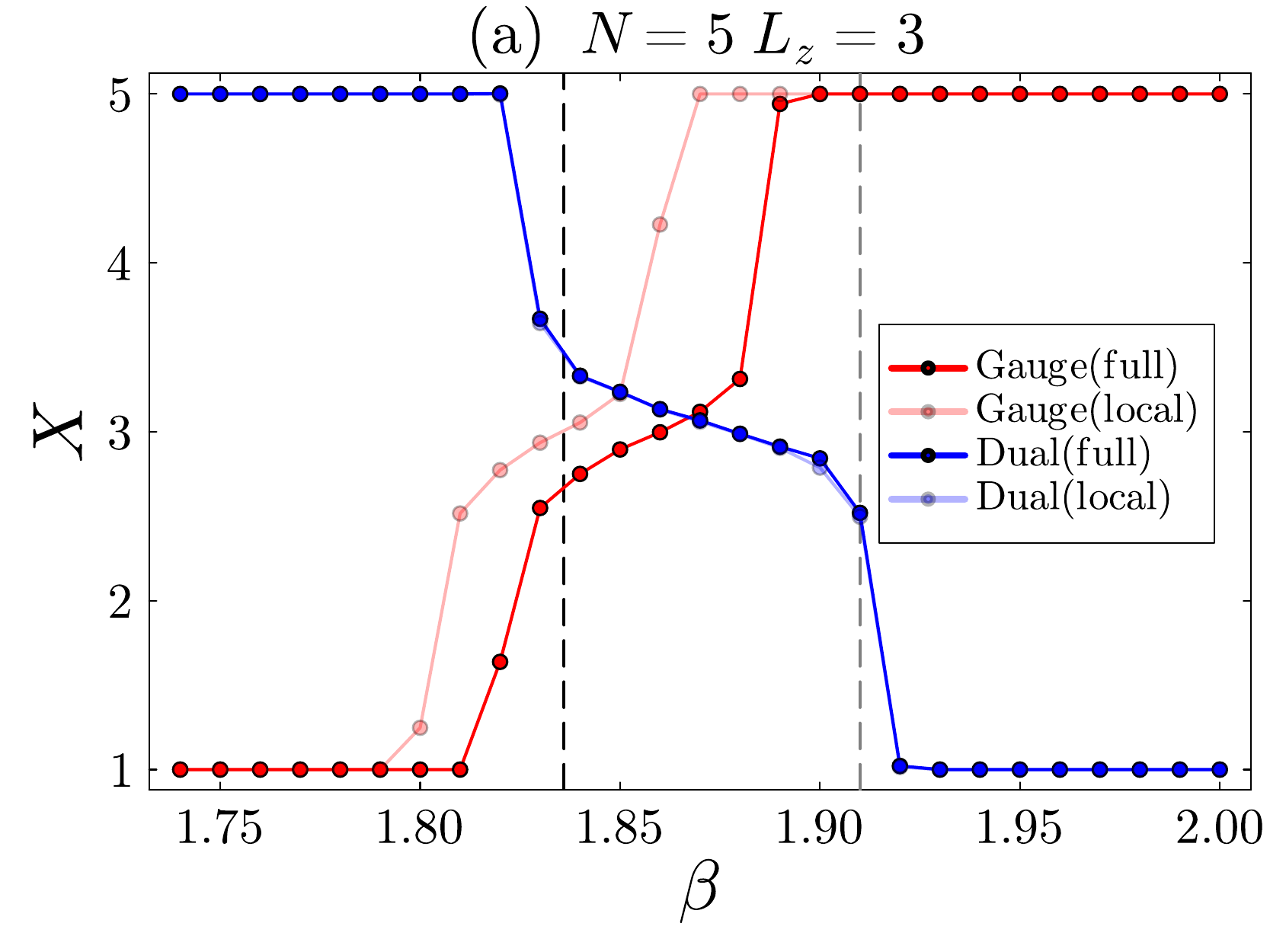}
    \includegraphics[width=0.48\linewidth]{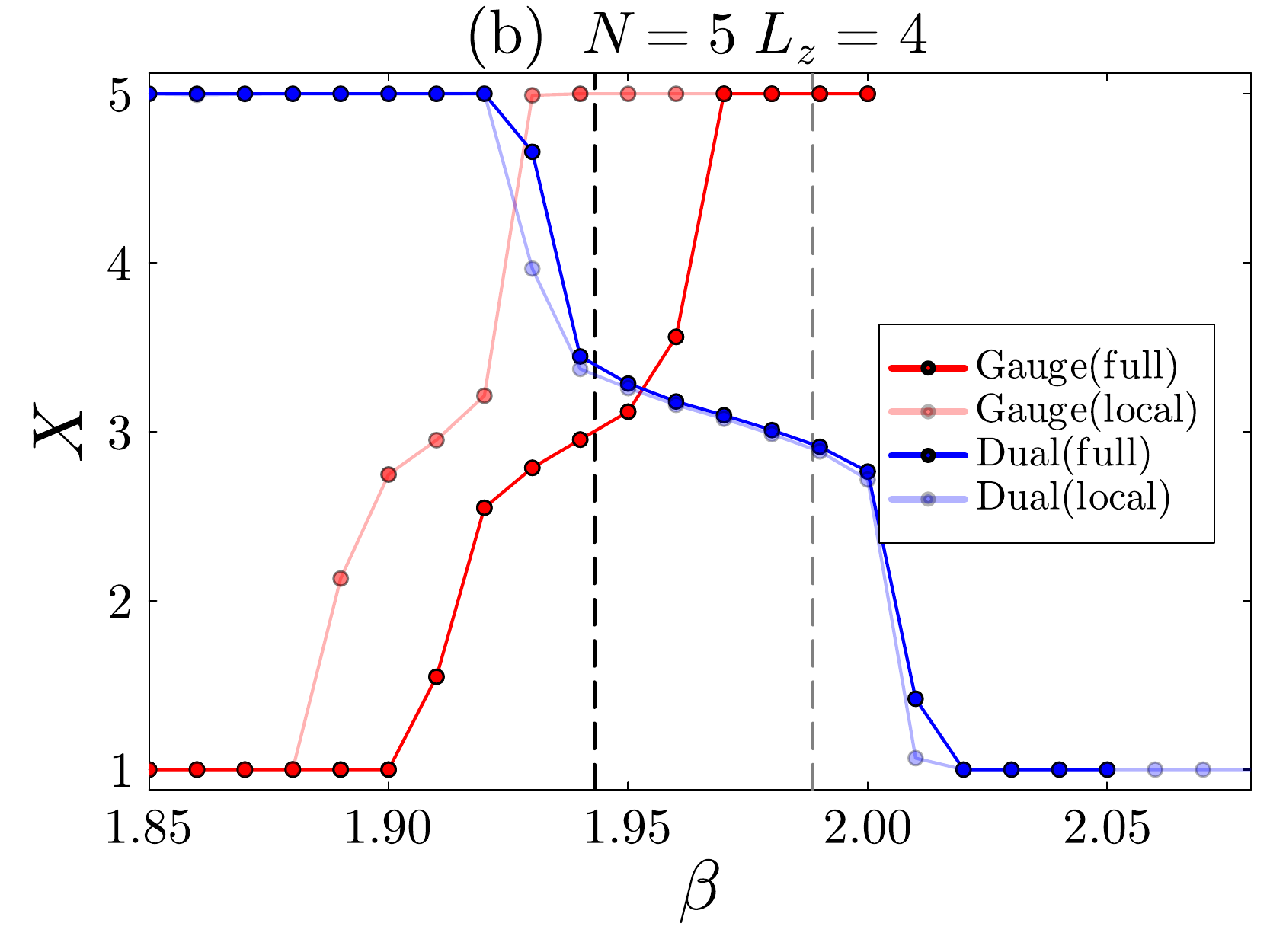}
    \caption{
        Gu--Wen ratio for $N=5$ at $L=2^{15}$ and (a) $L_z=3$, (b) $L_{z}=4$ with $(\chi_{{\rm TTNR}},\chi_{\text{BW-TRG}})=(40,81)$ for both gauge and dual theory.
    }
    \label{fig:X_Z5_full_local}
\end{figure}

There are two characteristic features of our full-update method.
First, we include the entire tensor network along the temporal direction as the environment when deriving the projectors, rather than using a local environment.
Second, we take into account the periodic boundary condition in the temporal direction, which is necessary to describe the finite-temperature effects. 
To adapt the full-update scheme, we perform an SVD of the reduced density matrix $\Gamma\Gamma^\dagger$ and then build the projectors following Refs.~\cite{boundarytrg,Pai:2024tip}. This modification can be done without changing the bottleneck cost of the HOTRG.
Fig.~\ref{fig:X_Z5_full_local} shows the Gu--Wen ratio for $N=5$, both for the original gauge theory and dual spin model at $L_{z}=3,4$.
At $L_{z}=3,4$, we have confirmed that the finite-$\chi_{{\rm TTNR}}$ and $\chi_\mathrm{BW-TRG}$ effects are well suppressed for the dual spin model in both local and full-update methods.
On the other hand, for the original gauge theory, the results obtained using the local and full-update methods differ significantly.
The transition points inferred from the local update method are not consistent with those of the dual spin model, indicating that the full-update method yields more reliable results.

\begin{figure}[t]
    \centering
    \includegraphics[width=0.48\linewidth]{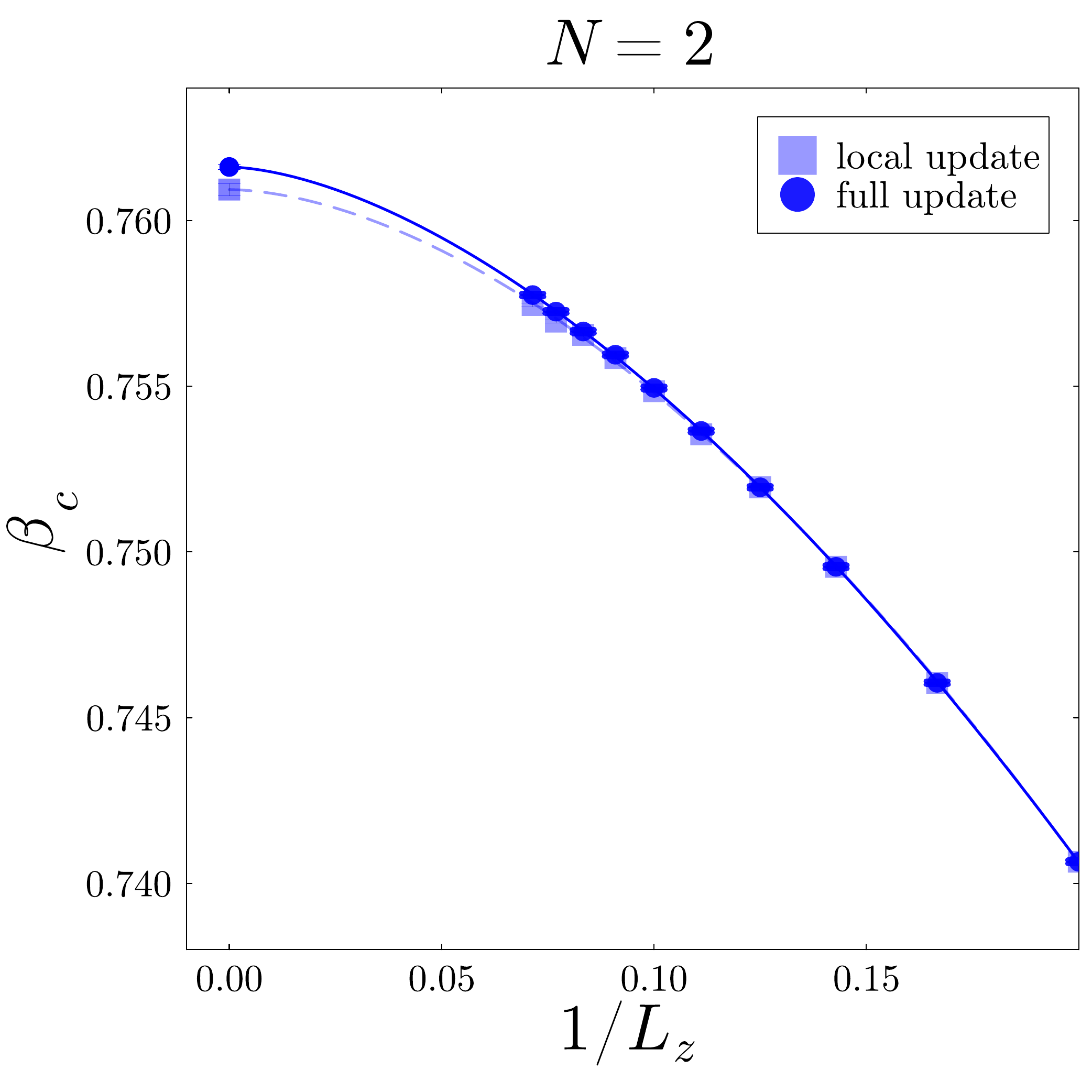}
    \caption{
         Extrapolation of the critical inverse gauge couplings at zero temperature for the $N=2$ dual spin model with $(\chi_{{\rm TTNR}},\chi_{\text{BW-TRG}})=(64,81)$.
        Gray and blue points denote the critical couplings obtained from the local update and full-update, respectively.
    }
    \label{fig:N2dual_fullvslocal}
\end{figure}

We further note that the difference between the local and full-update methods is significant even for the dual spin model in the low-temperature regime.
Fig.~\ref{fig:N2dual_fullvslocal} compares the critical couplings $\beta_c$ as a function of $1/L_{z}$ obtained by full and local update methods.
Note that the bond dimensions in Fig.~\ref{fig:N2dual_fullvslocal} are smaller than those in Table~\ref{tab:Z2}.
Employing the same fitting form as in Eq.~\eqref{eq:fit}, the full-update method yields $\beta_{c}^{}=0.76161(13)$, which is consistent with the our estimates~\eqref{eq:beta_c_gauge} and~\eqref{eq:beta_c_spin} in Sec.~\ref{subsec:zero_temp}.
This agreement confirms that our estimates of the critical inverse gauge coupling at zero temperature are robust with respect to the choice of bond dimensions.
However, the local update method gives $\beta_{c}^{}=0.76093(12)$, which is inconsistent with both the full-update results and the MC estimates.
Therefore, our full-update method is crucial in estimating the critical point in the vanishing-temperature limit.

\section{Specific heat}
\label{app:specific_heat}

\begin{figure}[h]
    \centering
    \includegraphics[width=0.48\linewidth]{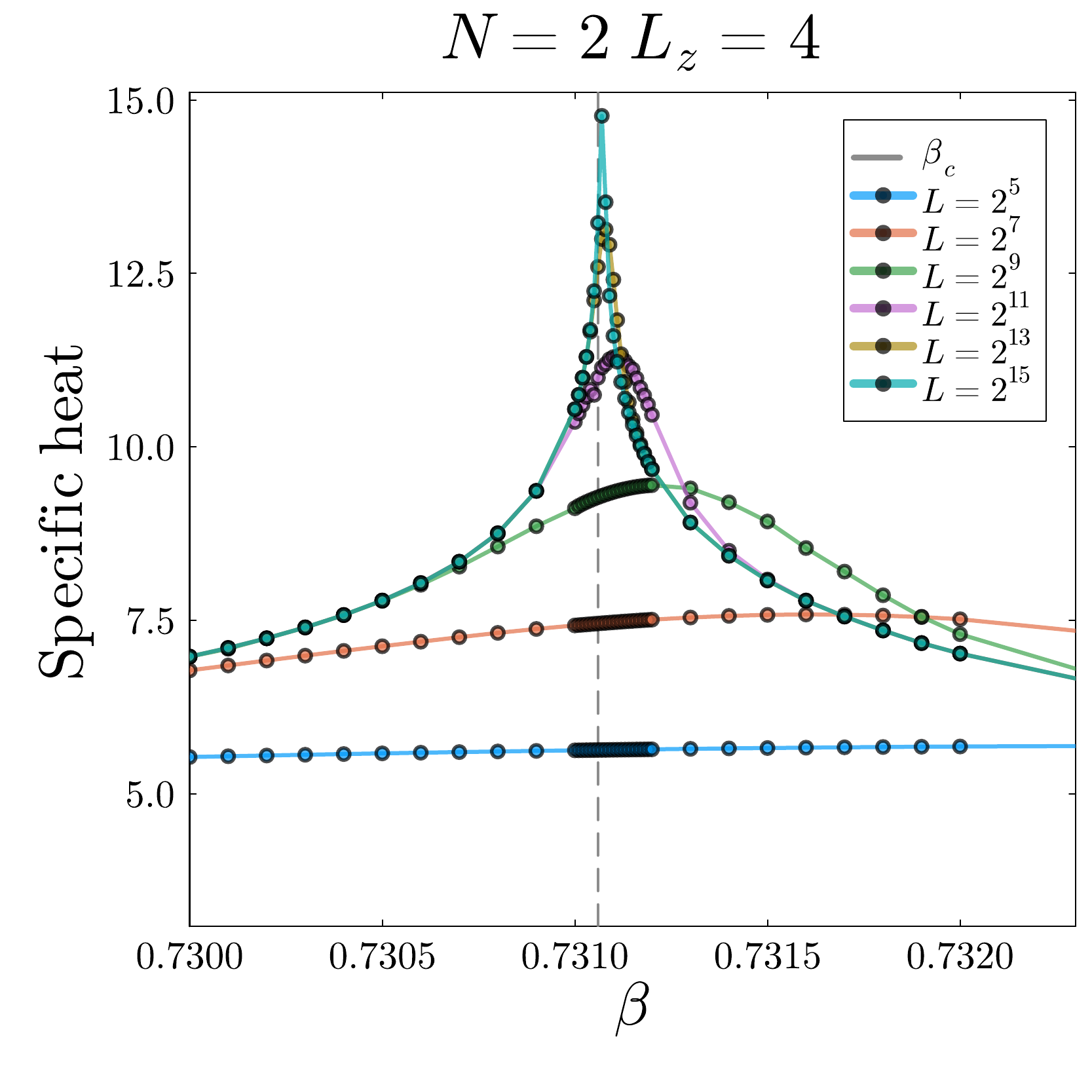}
    \caption{
        Specific heat for $N=2$, varying the spatial system size at $L_{z}=4$ with $(\chi_{{\rm TTNR}},\chi_{\text{BW-TRG}})=(64,120)$.
        Vertical dashed line denotes the critical point in Table~\ref{tab:Z2}. 
    }
    \label{fig:N2CvNt_4}
\end{figure}

\begin{figure}[h]
    \centering
    \includegraphics[width=0.48\linewidth]{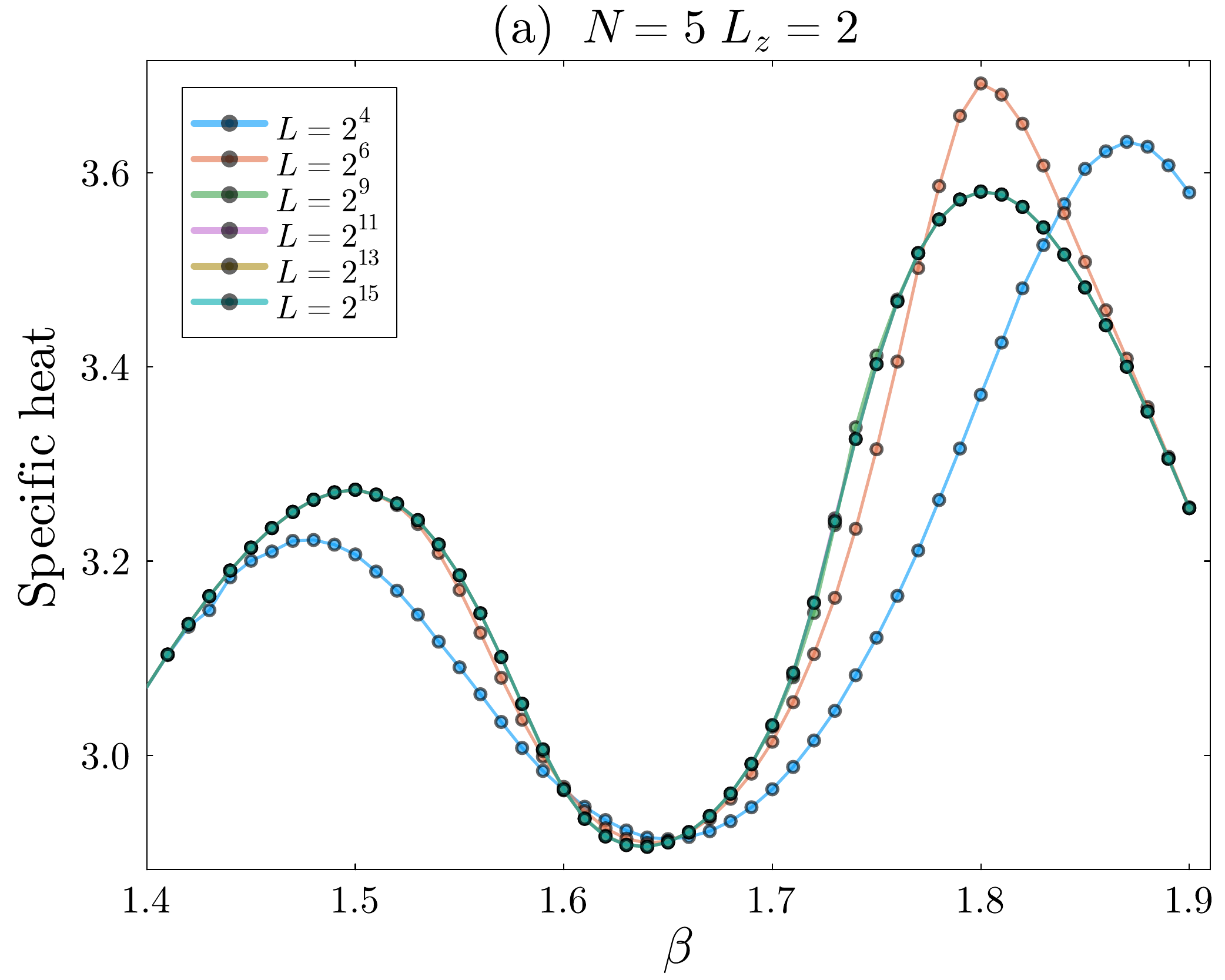}\\
    \includegraphics[width=0.48\linewidth]{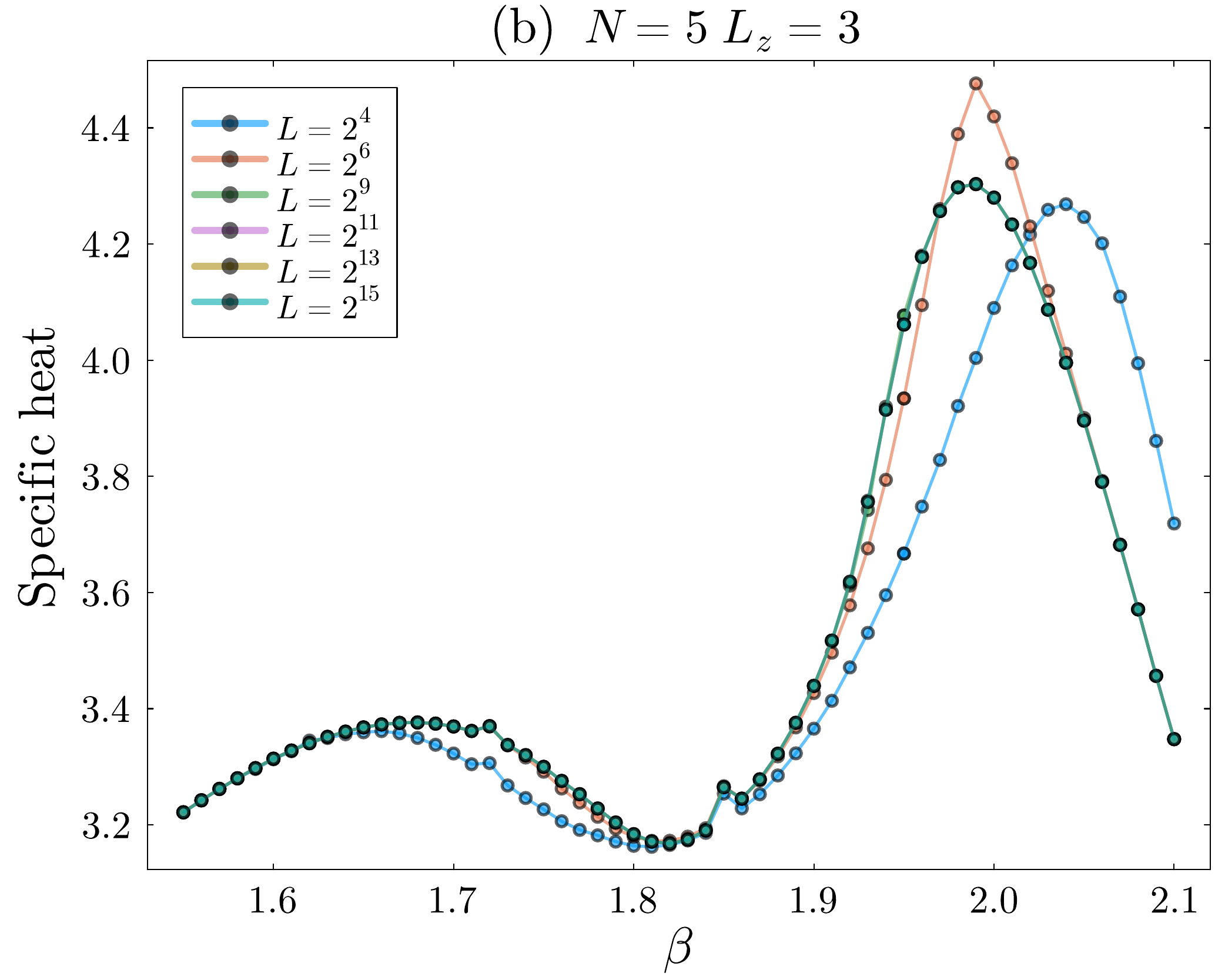}\\
    \includegraphics[width=0.48\linewidth]{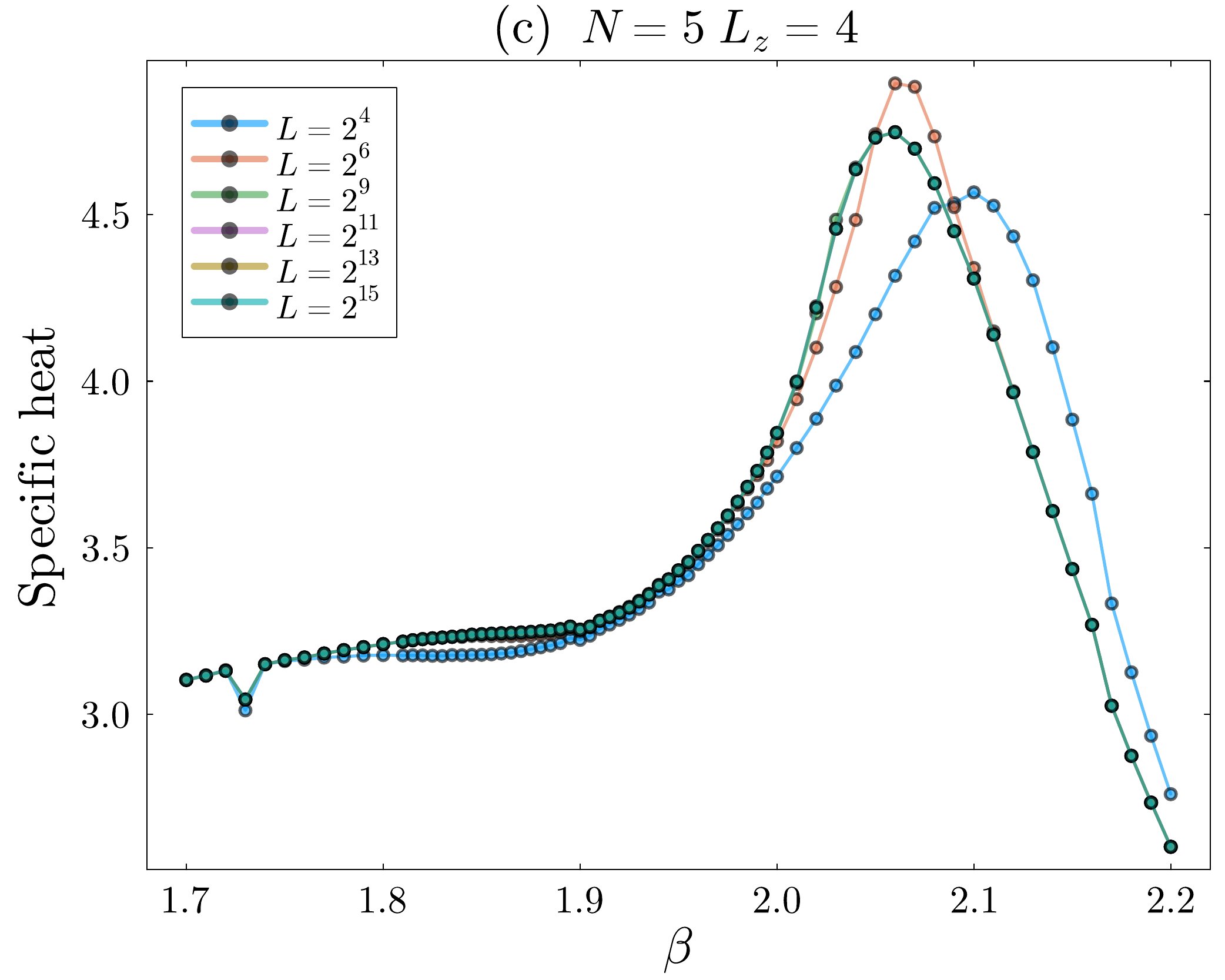}
    \caption{
        Specific heat for $N=5$, varying the spatial system size with $\chi_{\text{BW-TRG}}=81$.
        (a) $L_{z}=2$, (b) $L_{z}=3$, (c) $L_{z}=4$.
        We set $\chi_{{\rm TTNR}}=40$ for $L_{z}=2,4$ and $\chi_{{\rm TTNR}}=60$ for $L_{z}=3$.
    }
    \label{fig:N5Cv}
\end{figure}

We present supplementary results for the specific heat of the $\mathbb{Z}_N$ dual spin models.
We employ the forward-mode automatic differentiation technique~\cite{sugimoto:2026zxv}, which can be regarded as a generalization of the impurity method.
We first compute the specific heat for $N=2$. 
As shown in Fig.~\ref{fig:N2CvNt_4}, we observe that the peak becomes more pronounced with increasing the system size $L$.
The peak position is consistent with the estimate by the Gu--Wen ratio in Table~\ref{tab:Z2}.
The specific heat for $N=5$ is also computed.
Fig.~\ref{fig:N5Cv} shows the results for $L_{z}=2,3,4$ at various system sizes. 
In contrast to the case of $N=2$, the absence of divergent behavior in the specific heat is consistent with the BKT phase transitions.
The two peaks observed in Fig.~\ref{fig:N5Cv} reflect the existence of two transition points, similar to the behavior reported for the five-state clock model~\cite{Li:2019dkb}. We also observe that, as $L_z$ increases, the peak at lower $\beta$ vanishes, while the second peak remains.

\section{Bond dimension convergence}

\begin{figure}[htbp]

\includegraphics[width=0.32\linewidth]{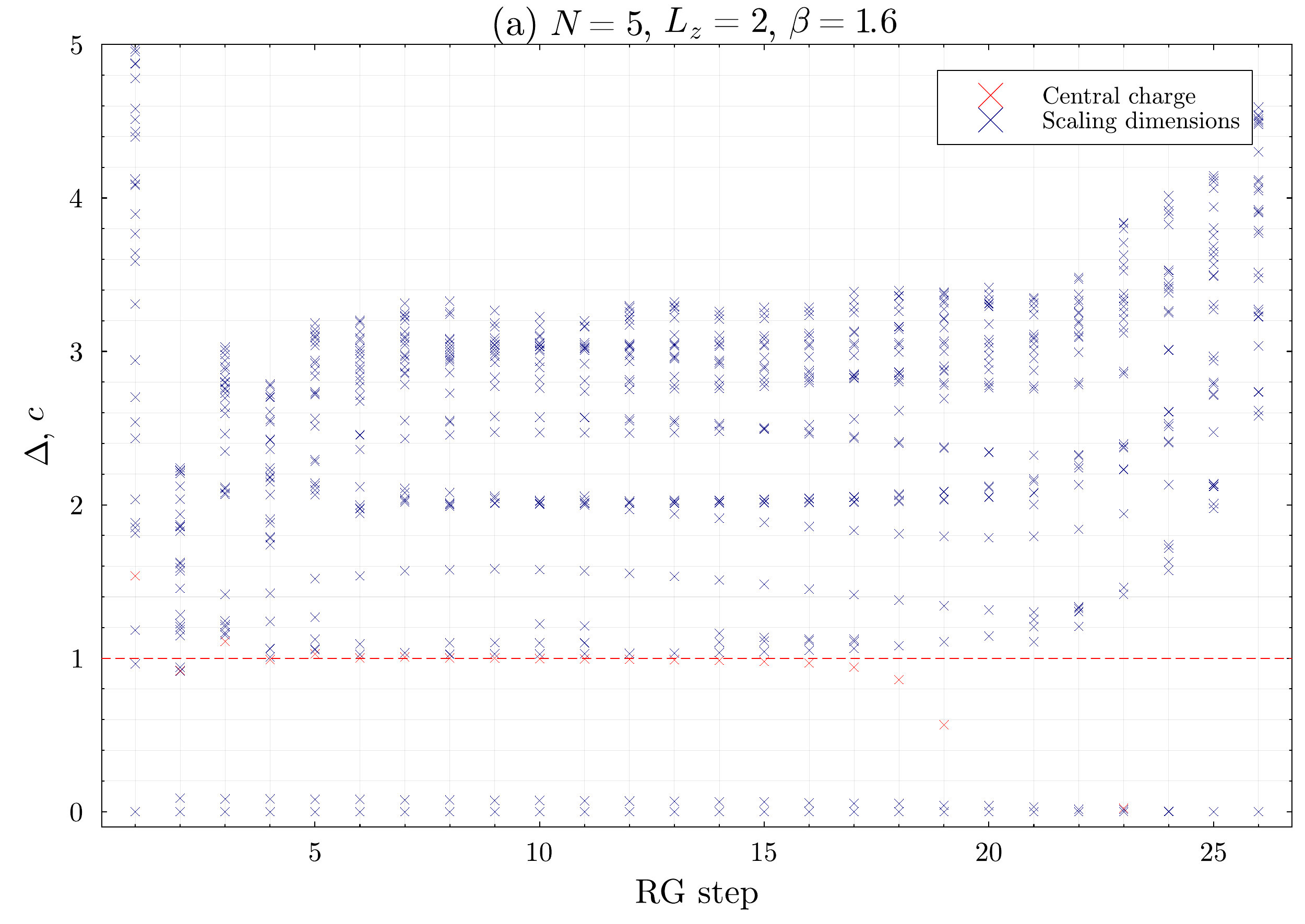}
\includegraphics[width=0.32\linewidth]{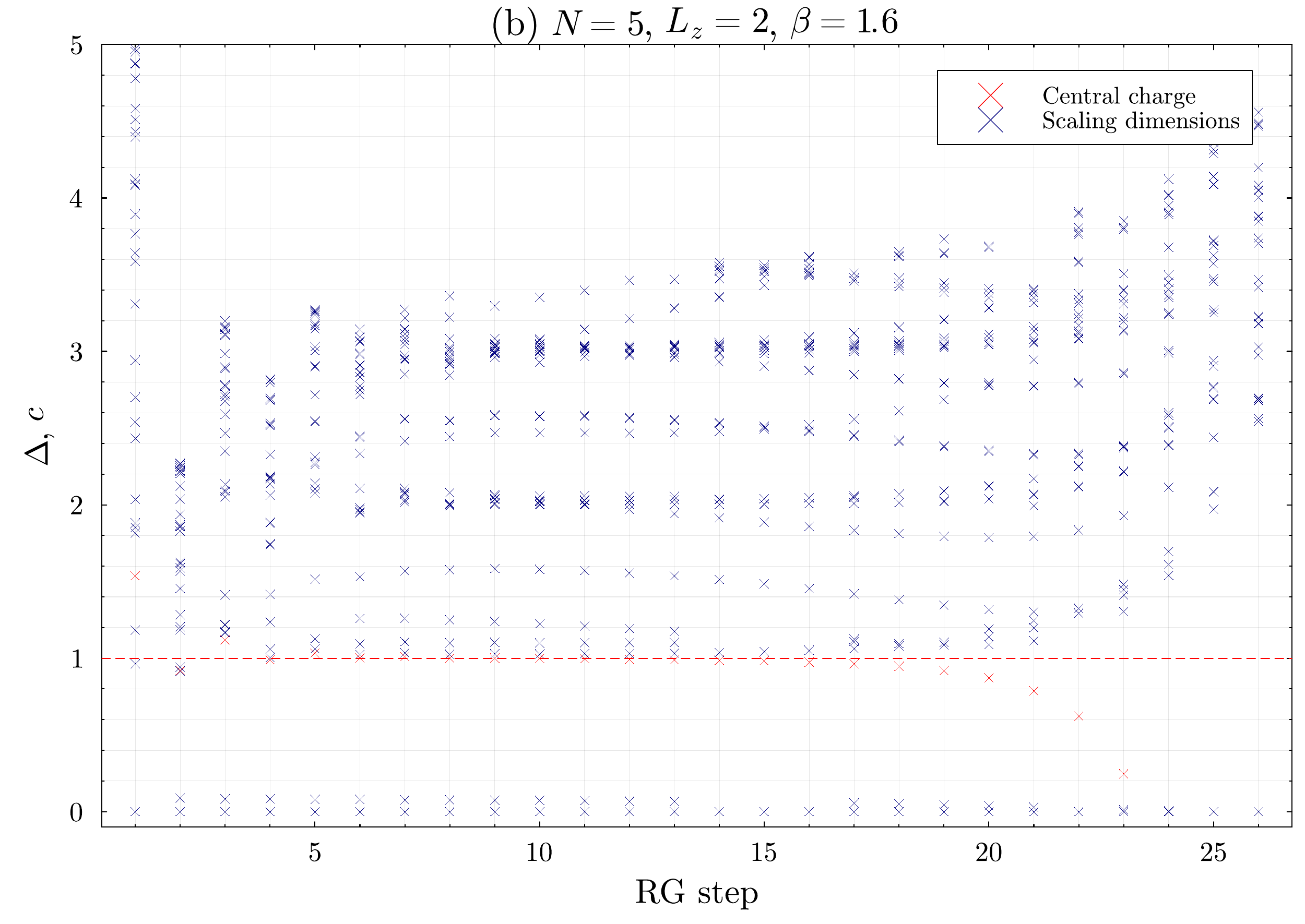}
\includegraphics[width=0.32\linewidth]{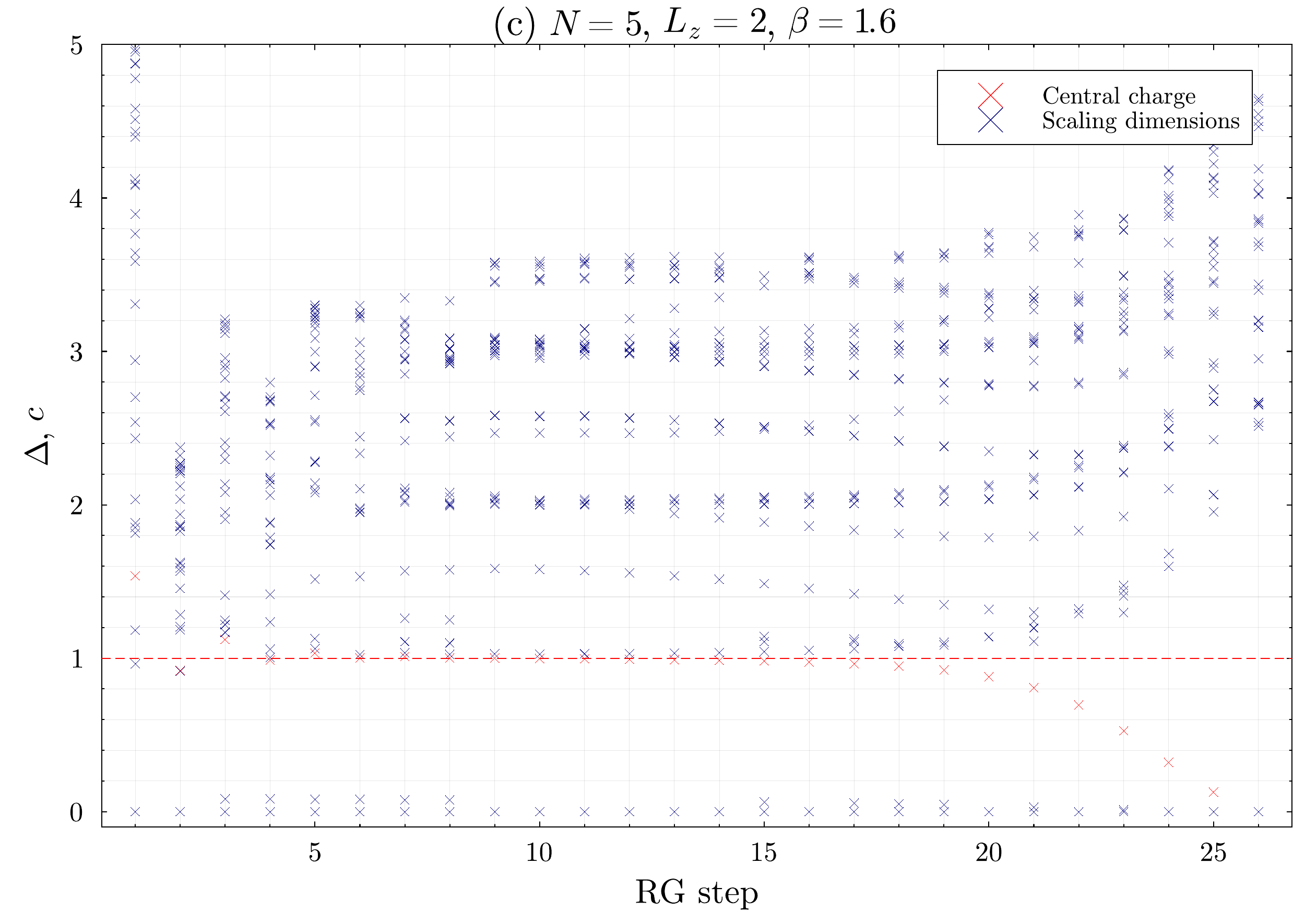}

\includegraphics[width=0.32\linewidth]{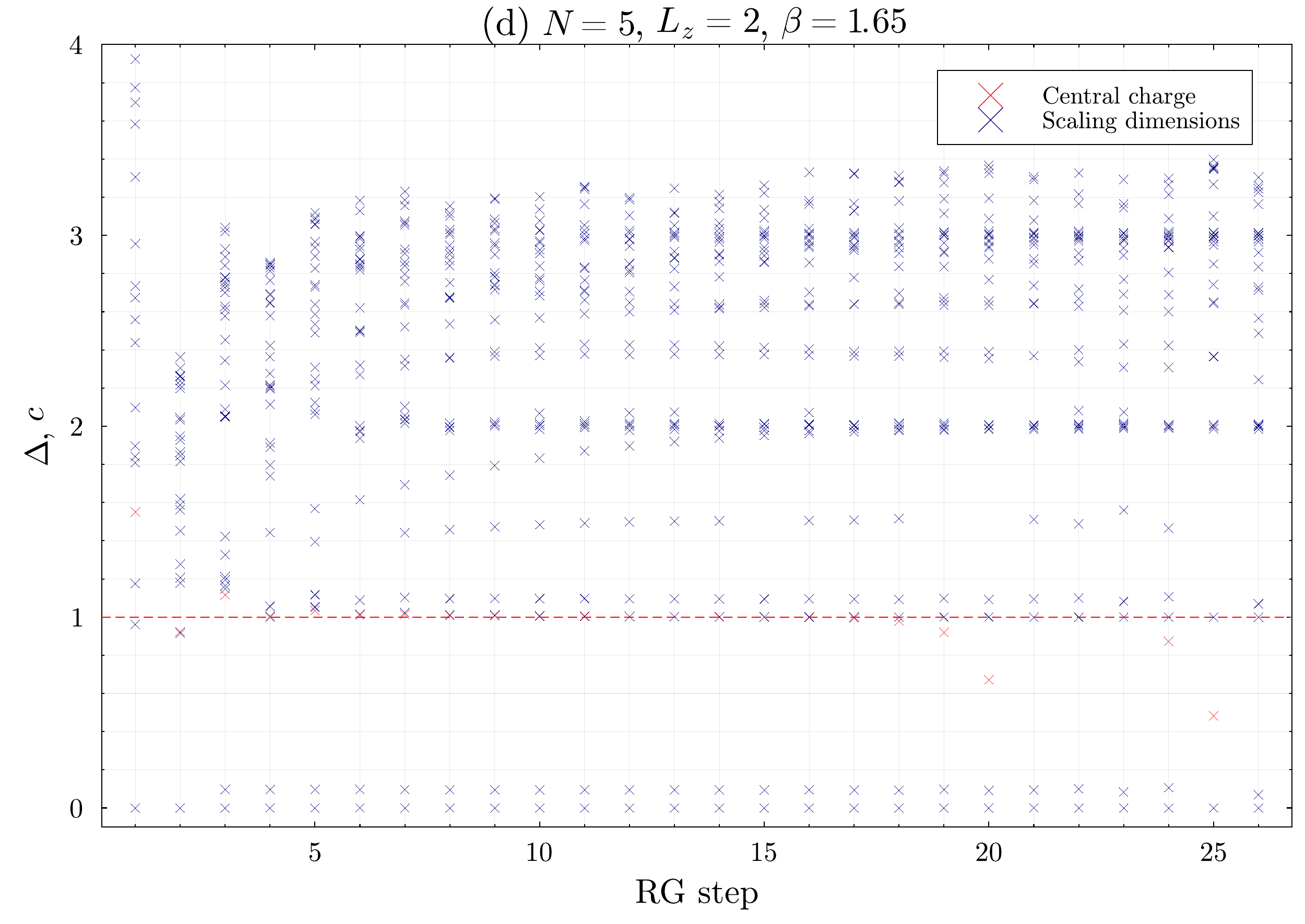}
\includegraphics[width=0.32\linewidth]{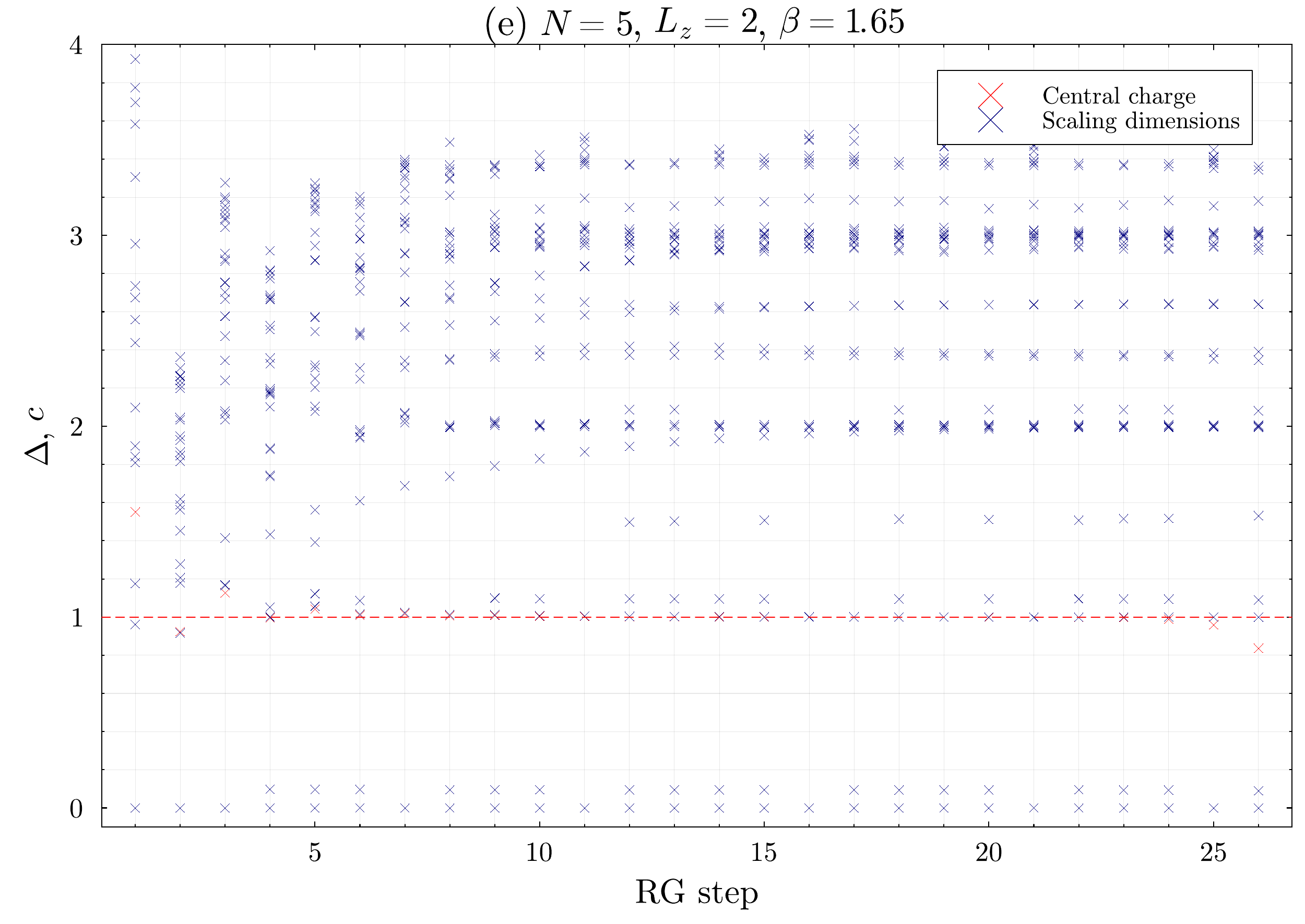}
\includegraphics[width=0.32\linewidth]{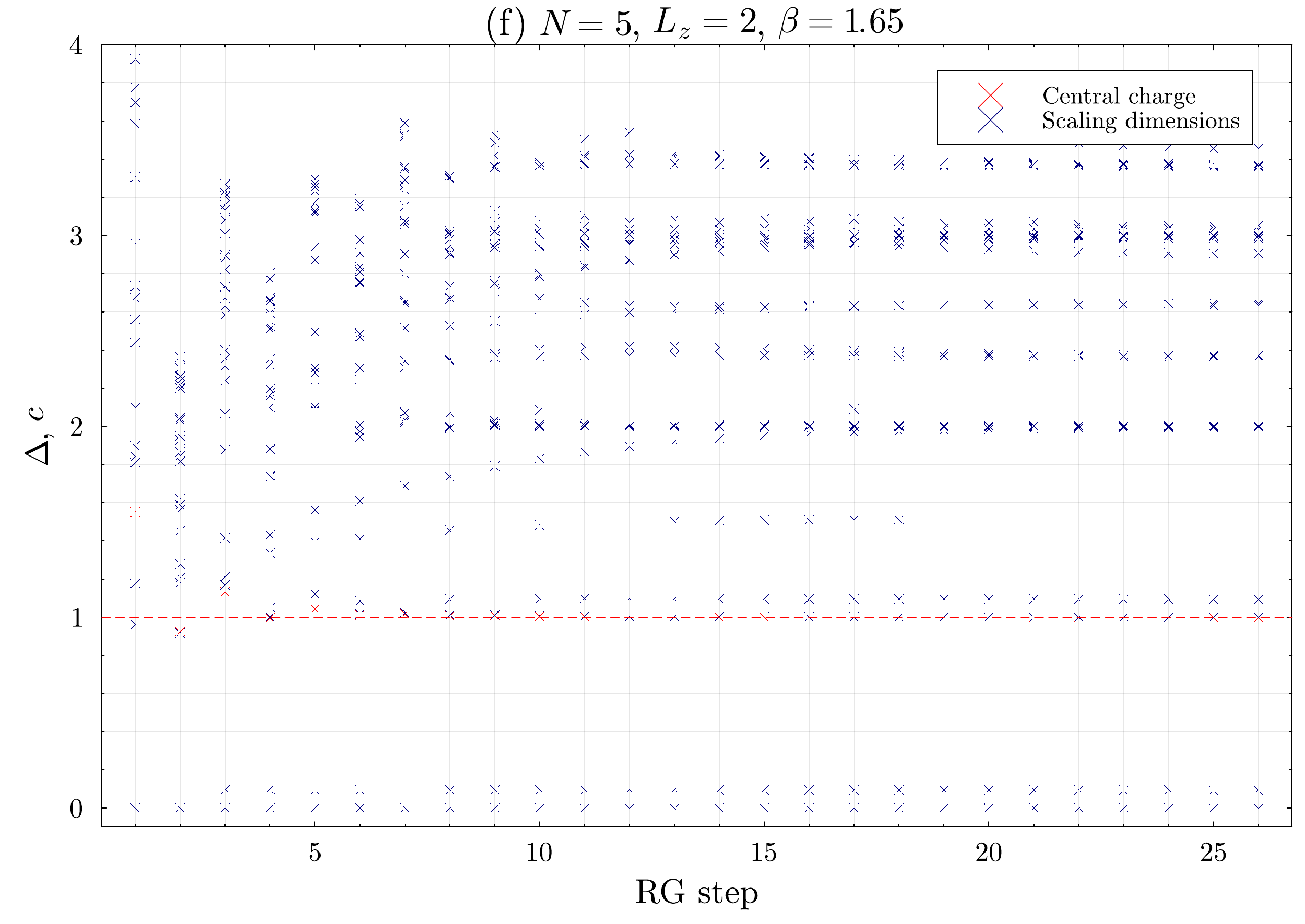}

\caption{
    Central charge $c$ (red) and scaling dimensions $\Delta$ (blue) along the RG steps for the $\mathbb{Z}_5$ lattice gauge theory at $L_z=2$, with $\beta=1.6$ (top) and $\beta=1.65$ (bottom). 
    The contraction along the temporal direction is performed exactly due to the choice of $\chi_{\rm TTNR}=25$. 
    For the spatial coarse-graining, $\chi_{\rm Loop\text{-}TNR}$ is varied as 25 (left), 40 (middle), and 50 (right).
}

\label{fig:loopbondimcftspectra}
\end{figure}

Finite-bond-dimension effects play an important role in the precise determination of the CFT data.
To address this point, we demonstrate how finite $\chi_{\rm TTNR}$ and $\chi_{\rm Loop\text{-}TNR}$ affect them, particularly for the $\mathbb{Z}_5$ theory.
Here, we consider the dual spin model with $L_{z}=2$, for which an exact contraction along the temporal direction is possible.
Figure~\ref{fig:loopbondimcftspectra} shows the critical spectra (blue) and the central charge (red) computed with $\chi_{\rm TTNR}=25$, corresponding to the exact temporal contraction, while varying $\chi_{\rm Loop\text{-}TNR}$ to examine finite-$\chi_{\rm Loop\text{-}TNR}$ effects.
The calculations are performed at $\beta=1.6$ and $\beta=1.65$, which lie slightly outside and inside the critical phase, respectively.
For the critical system, scale invariance persists as the bond dimension is increased. 
We see that the CFT data are already stable enough at $\chi_{\rm Loop\text{-}TNR}=50$.
By contrast, outside the critical phase, the system flows to $c=0$, as expected for a finitely correlated state. 

\begin{figure}[htbp]

\includegraphics[width=0.32\linewidth]{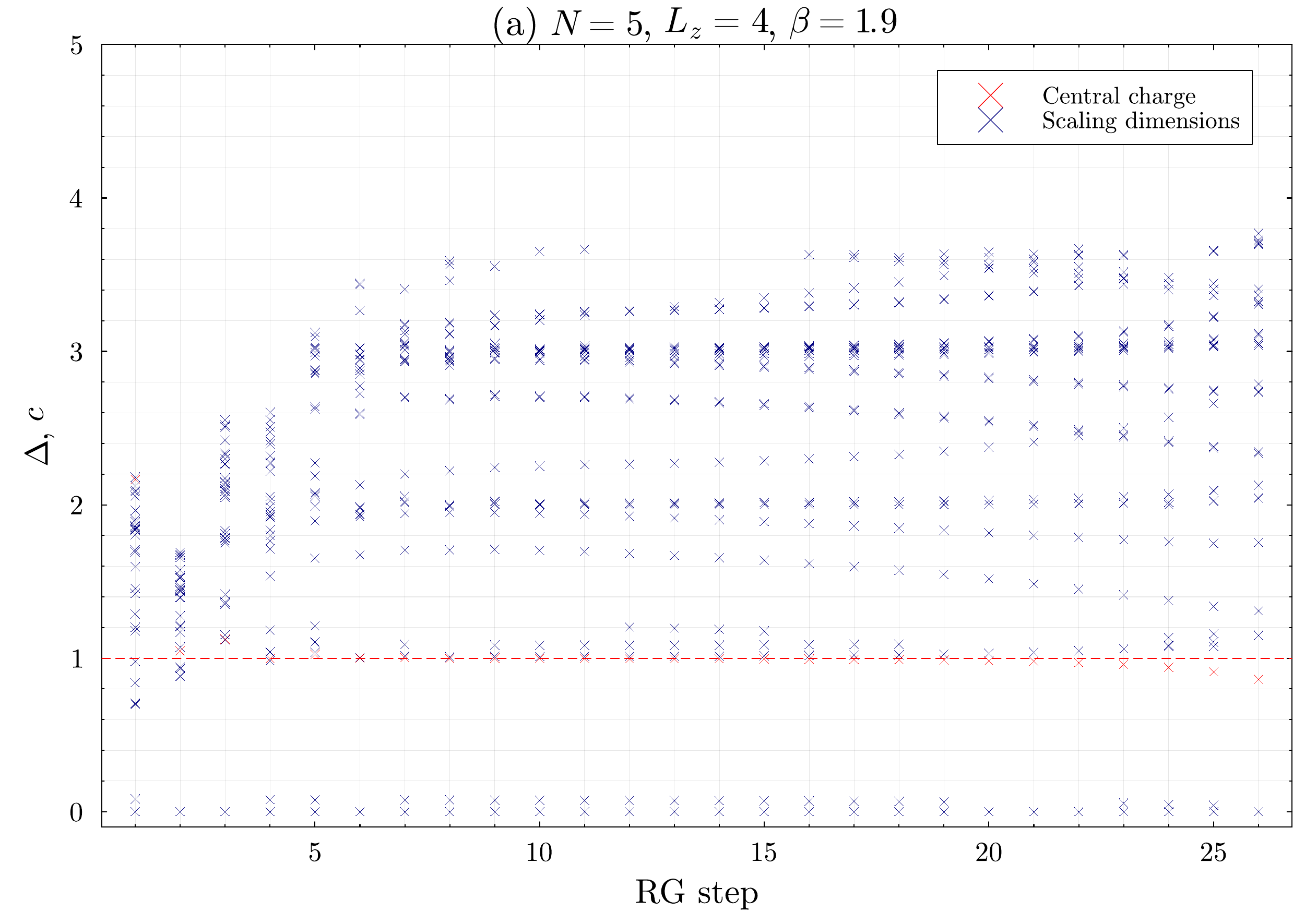}
\includegraphics[width=0.32\linewidth]{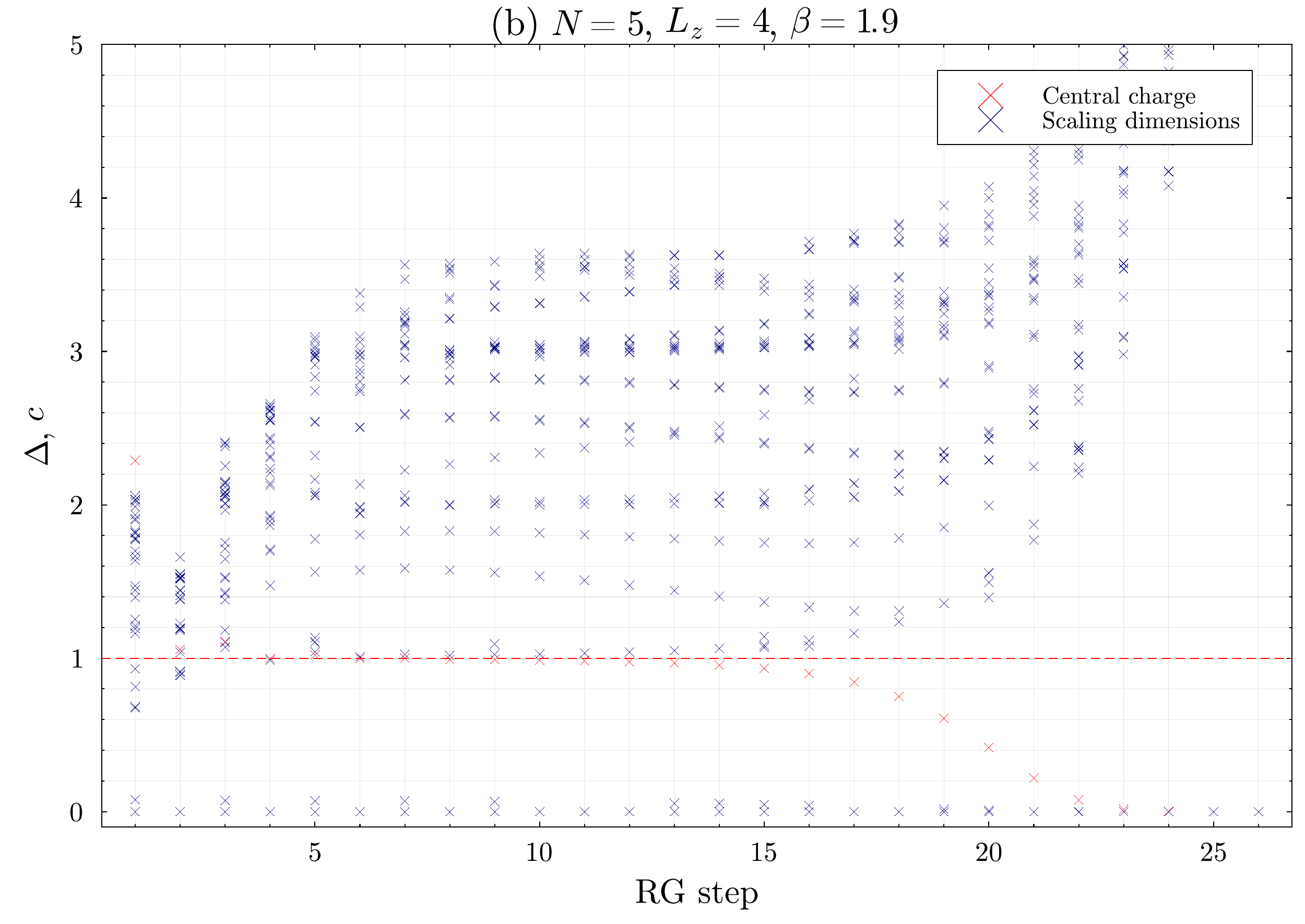}
\includegraphics[width=0.32\linewidth]{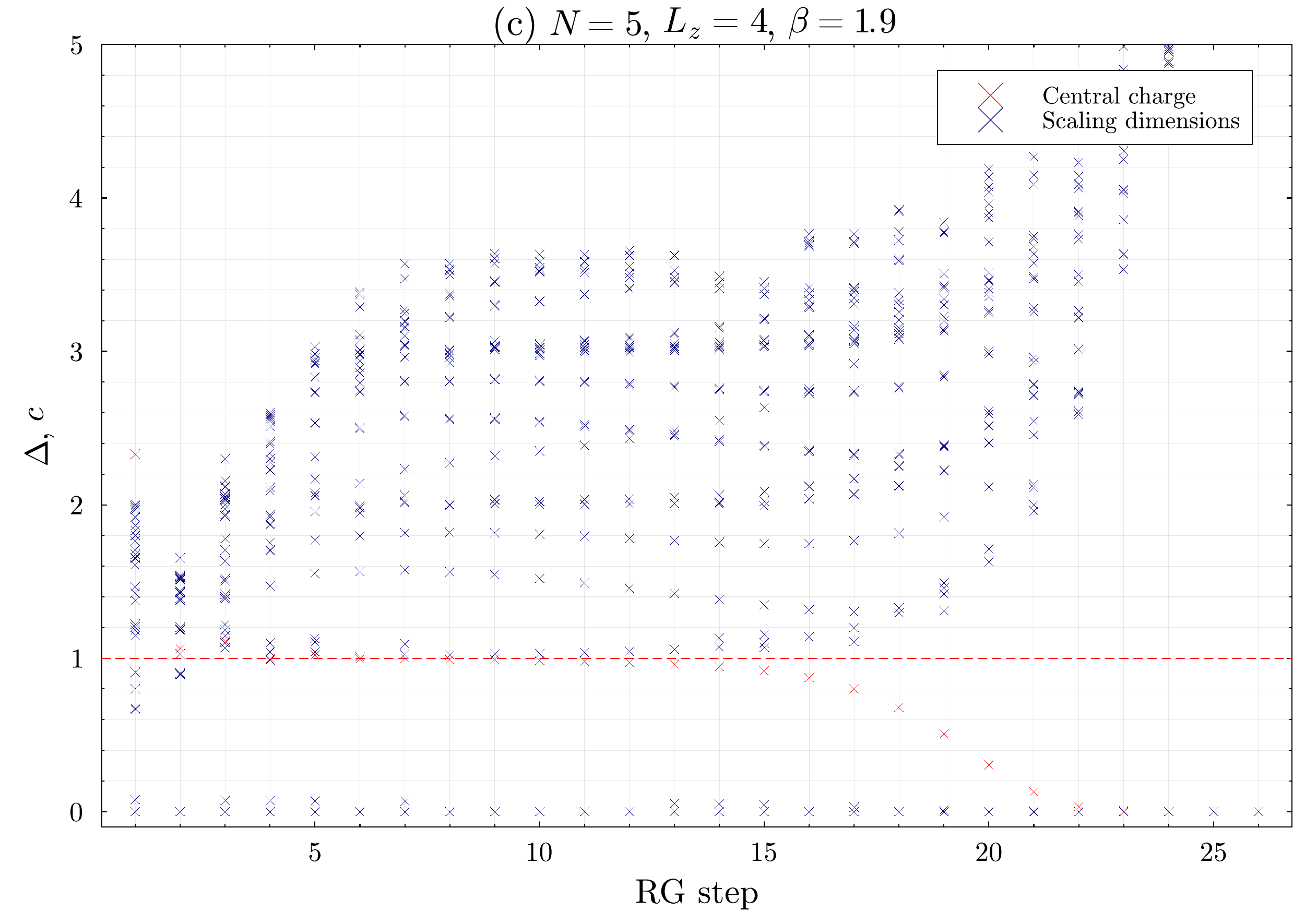}

\includegraphics[width=0.32\linewidth]{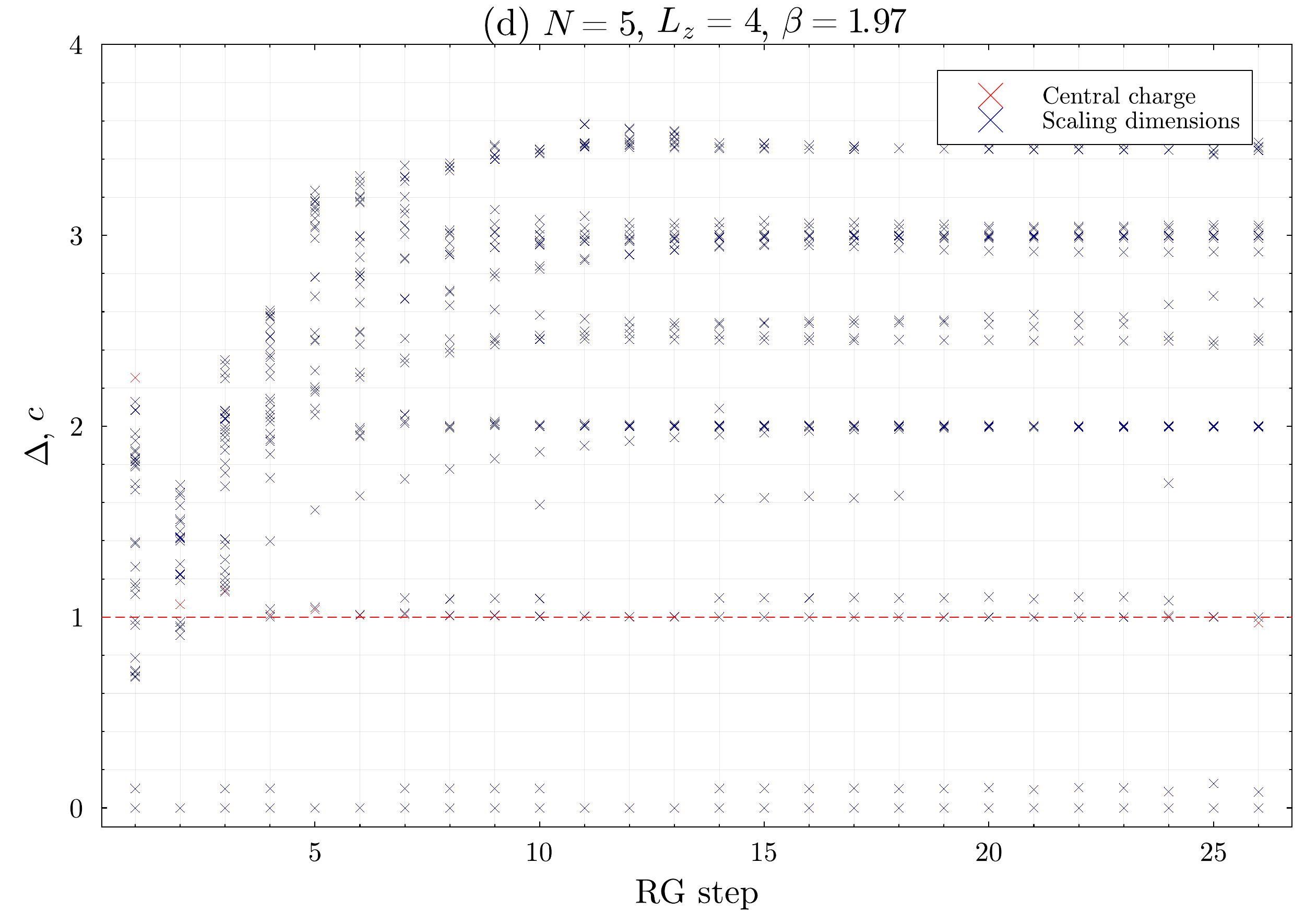}
\includegraphics[width=0.32\linewidth]{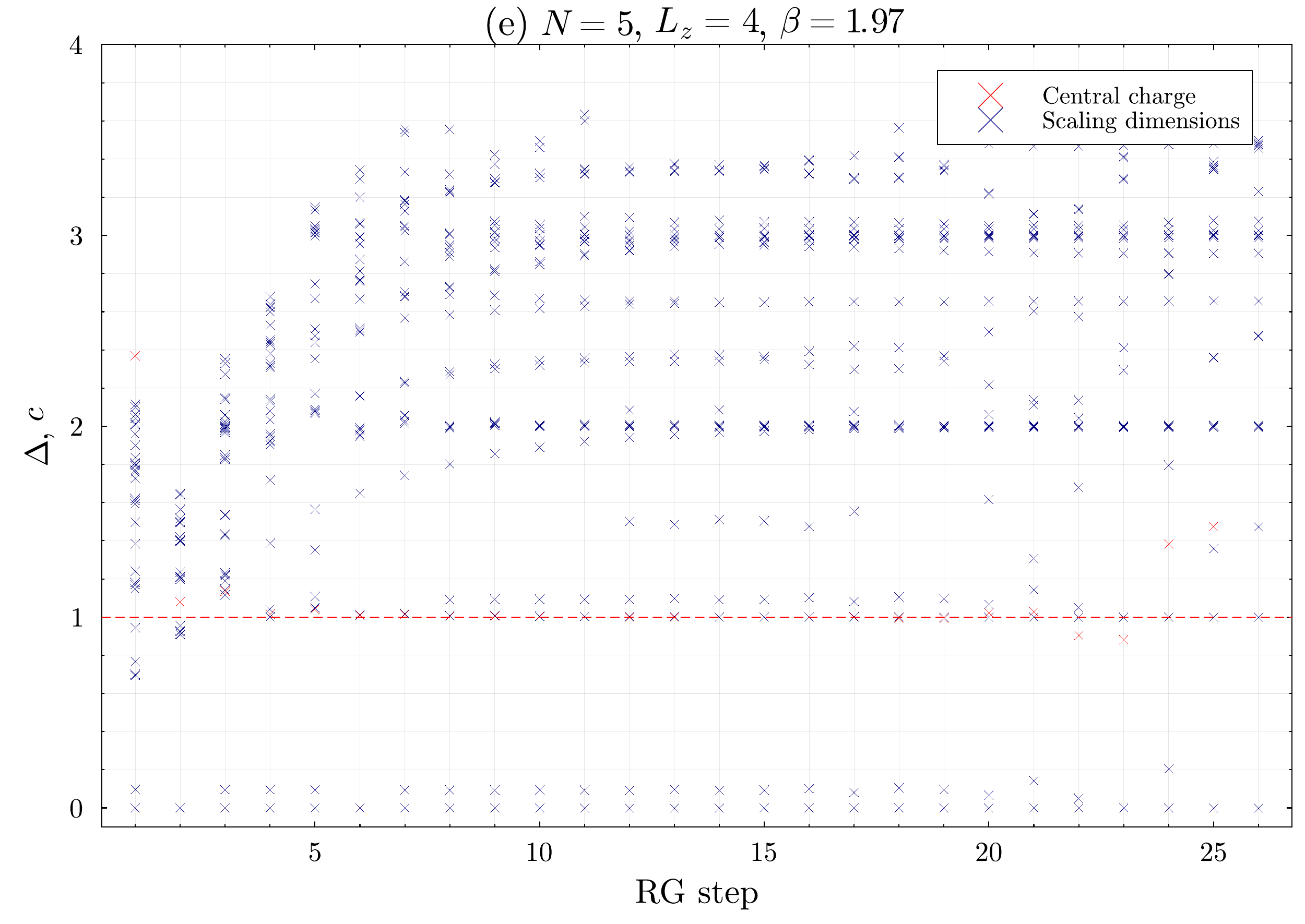}
\includegraphics[width=0.32\linewidth]{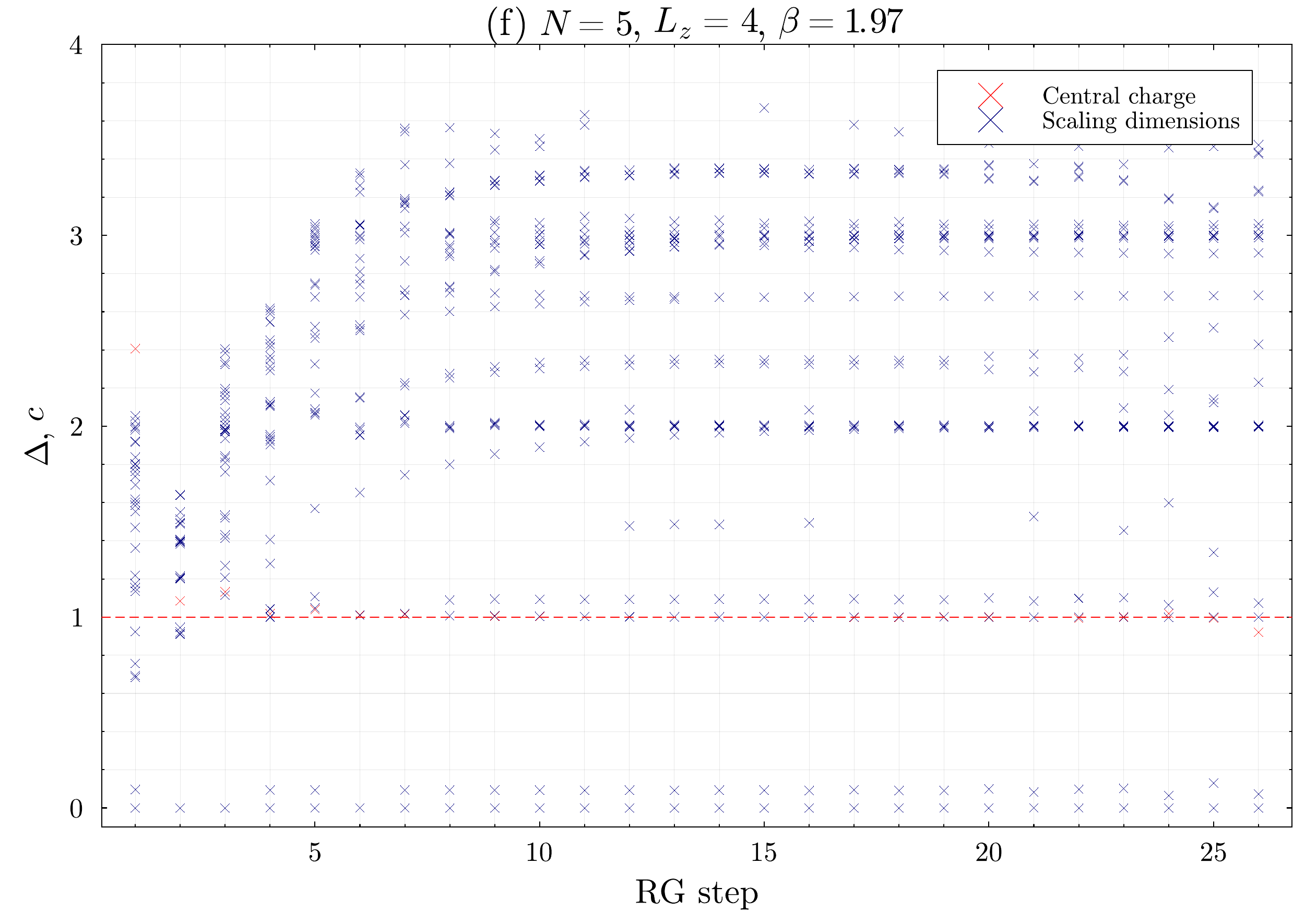}

\caption{
    Central charge $c$ (red) and scaling dimensions $\Delta$ (blue) along the RG steps for the $\mathbb{Z}_5$ lattice gauge theory at $L_z=4$, with $\beta=1.9$ (top) and $\beta=1.97$ (bottom). 
    For the spatial coarse-graining, $\chi_{\rm Loop\text{-}TNR}$ is fixed as 50. 
    For the temporal coarse-graining, $\chi_{\rm TTNR}$ is varied as 25 (left), 40 (middle), and 50 (right).
}

\label{fig:ttnrbondimcftspectra}
\end{figure}

Second, we fix $\chi_{\rm Loop\text{-}TNR}=50$ and vary $\chi_{\rm TTNR}$ in order to probe truncation effects in the temporal direction. 
We consider the same theory at $L_z=4$ with $\beta=1.9$ and $\beta=1.97$. 
As shown in Fig.~\ref{fig:ttnrbondimcftspectra}, at $\beta=1.9$ the apparent non-trivial CFT behavior observed at small $\chi_{\rm TTNR}$ disappears as $\chi_{\rm TTNR}$ is increased, indicating that it is a finite-bond-dimension artifact. 
In contrast, at $\beta=1.97$ the extracted CFT signatures are significantly more robust with respect to $\chi_{\rm TTNR}$, supporting their physical relevance.

We note that performing a controlled extrapolation to the infinite bond-dimension limit is particularly challenging in the present setup due to the interplay between the two bond dimensions, $\chi_{\rm TTNR}$ and $\chi_{\rm Loop\text{-}TNR}$, together with the extremely large correlation lengths in the critical and drifting regimes. 
In particular, we currently do not have a reliable theoretical scaling form for the finite-bond-dimension dependence of the extracted CFT data involving $\chi_{\rm TTNR}$ in this regime. 
For this reason, rather than relying on potentially uncontrolled extrapolations, we focus on systematic stability checks with respect to both bond dimensions. 
A more systematic understanding of the finite-bond-dimension scaling behavior is an important direction for future work.

\clearpage
\bibliographystyle{JHEP}
\bibliography{bib/ref}

@article{Wilson:1974sk,
    author = "Wilson, Kenneth G.",
    editor = "Taylor, J. C.",
    title = "{Confinement of Quarks}",
    reportNumber = "CLNS-262",
    doi = "10.1103/PhysRevD.10.2445",
    journal = "Phys. Rev. D",
    volume = "10",
    pages = "2445--2459",
    year = "1974"
}

@article{Elitzur:1975im,
    author = "Elitzur, S.",
    title = "{Impossibility of Spontaneously Breaking Local Symmetries}",
    doi = "10.1103/PhysRevD.12.3978",
    journal = "Phys. Rev. D",
    volume = "12",
    pages = "3978--3982",
    year = "1975"
}

@article{tHooft:1977nqb,
    author = "'t Hooft, Gerard",
    title = "{On the Phase Transition Towards Permanent Quark Confinement}",
    reportNumber = "Print-78-0099 (UTRECHT)",
    doi = "10.1016/0550-3213(78)90153-0",
    journal = "Nucl. Phys. B",
    volume = "138",
    pages = "1--25",
    year = "1978"
}

@article{Polyakov:1978vu,
    author = "Polyakov, Alexander M.",
    title = "{Thermal Properties of Gauge Fields and Quark Liberation}",
    doi = "10.1016/0370-2693(78)90737-2",
    journal = "Phys. Lett. B",
    volume = "72",
    pages = "477--480",
    year = "1978"
}

@article{Gaiotto:2014kfa,
    author = "Gaiotto, Davide and Kapustin, Anton and Seiberg, Nathan and Willett, Brian",
    title = "{Generalized Global Symmetries}",
    eprint = "1412.5148",
    archivePrefix = "arXiv",
    primaryClass = "hep-th",
    doi = "10.1007/JHEP02(2015)172",
    journal = "JHEP",
    volume = "02",
    pages = "172",
    year = "2015"
}

@article{Creutz:1980zw,
    author = "Creutz, M.",
    title = "{Monte Carlo Study of Quantized SU(2) Gauge Theory}",
    doi = "10.1103/PhysRevD.21.2308",
    journal = "Phys. Rev. D",
    volume = "21",
    pages = "2308--2315",
    year = "1980"
}

@article{Creutz:1979zg,
    author = "Creutz, Michael and Jacobs, Laurence and Rebbi, Claudio",
    editor = "Julve, J. and Ram{\'o}n-Medrano, M.",
    title = "{Monte Carlo Study of Abelian Lattice Gauge Theories}",
    reportNumber = "BNL-26307",
    doi = "10.1103/PhysRevD.20.1915",
    journal = "Phys. Rev. D",
    volume = "20",
    pages = "1915",
    year = "1979"
}

@article{Creutz:1979he,
    author = "Creutz, Michael",
    editor = "Julve, J. and Ram{\'o}n-Medrano, M.",
    title = "{Phase Diagrams for Coupled Spin Gauge Systems}",
    reportNumber = "BNL-26587",
    doi = "10.1103/PhysRevD.21.1006",
    journal = "Phys. Rev. D",
    volume = "21",
    pages = "1006",
    year = "1980"
}

@article{Bhanot:1980pc,
    author = "Bhanot, Gyan and Creutz, Michael",
    title = "{The Phase Diagram of $Z$(n) and $u$(1) Gauge Theories in Three-dimensions}",
    reportNumber = "BNL-27276",
    doi = "10.1103/PhysRevD.21.2892",
    journal = "Phys. Rev. D",
    volume = "21",
    pages = "2892",
    year = "1980"
}

@article{PhysRevB.22.3370,
  title = {Ising gauge theory at negative temperatures and spin-glasses},
  author = {Bhanot, Gyan and Creutz, Michael},
  journal = {Phys. Rev. B},
  volume = {22},
  issue = {7},
  pages = {3370--3373},
  numpages = {0},
  year = {1980},
  month = {Oct},
  publisher = {American Physical Society},
  doi = {10.1103/PhysRevB.22.3370},
  url = {https://link.aps.org/doi/10.1103/PhysRevB.22.3370}
}

@article{Balian:1974ir,
    author = "Balian, R. and Drouffe, J. M. and Itzykson, C.",
    editor = "Julve, J. and Ram{\'o}n-Medrano, M.",
    title = "{Gauge Fields on a Lattice. 2. Gauge Invariant Ising Model}",
    reportNumber = "Saclay-DPh-T/74/74",
    doi = "10.1103/PhysRevD.11.2098",
    journal = "Phys. Rev. D",
    volume = "11",
    pages = "2098",
    year = "1975"
}

@article{Ukawa:1979yv,
    author = "Ukawa, Akira and Windey, Paul and Guth, Alan H.",
    title = "{Dual Variables for Lattice Gauge Theories and the Phase Structure of Z(N) Systems}",
    reportNumber = "Print-79-0567 (PRINCETON)",
    doi = "10.1103/PhysRevD.21.1013",
    journal = "Phys. Rev. D",
    volume = "21",
    pages = "1013",
    year = "1980"
}

@article{KorthalsAltes:1978tp,
    author = "Korthals Altes, C. P.",
    title = "{Duality for $Z(N$) Gauge Theories}",
    reportNumber = "CPT-78/P-1003",
    doi = "10.1016/0550-3213(78)90207-9",
    journal = "Nucl. Phys. B",
    volume = "142",
    pages = "315--326",
    year = "1978"
}

@article{Savit:1979ny,
    author = "Savit, Robert",
    title = "{Duality in Field Theory and Statistical Systems}",
    reportNumber = "UM-HE-79-8",
    doi = "10.1103/RevModPhys.52.453",
    journal = "Rev. Mod. Phys.",
    volume = "52",
    pages = "453",
    year = "1980"
}

@article{Svetitsky:1982gs,
    author = "Svetitsky, Benjamin and Yaffe, Laurence G.",
    title = "{Critical Behavior at Finite Temperature Confinement Transitions}",
    reportNumber = "CLNS-82/530, NSF-ITP-82-53",
    doi = "10.1016/0550-3213(82)90172-9",
    journal = "Nucl. Phys. B",
    volume = "210",
    pages = "423--447",
    year = "1982"
}

@article{Byrnes:2002nv,
    author = "Byrnes, T. and Sriganesh, P. and Bursill, R. J. and Hamer, C. J.",
    title = "{Density matrix renormalization group approach to the massive Schwinger model}",
    eprint = "hep-lat/0202014",
    archivePrefix = "arXiv",
    doi = "10.1103/PhysRevD.66.013002",
    journal = "Phys. Rev. D",
    volume = "66",
    pages = "013002",
    year = "2002"
}

@article{Banuls:2015sta,
    author = "Ba{\~n}uls, M. C. and Cichy, K. and Cirac, J. I. and Jansen, K. and Saito, H.",
    title = "{Thermal evolution of the Schwinger model with Matrix Product Operators}",
    eprint = "1505.00279",
    archivePrefix = "arXiv",
    primaryClass = "hep-lat",
    reportNumber = "DESY-15-060, SFB-CPP-14-125, DESY 15-060, SFB/CPP-14-125",
    doi = "10.1103/PhysRevD.92.034519",
    journal = "Phys. Rev. D",
    volume = "92",
    number = "3",
    pages = "034519",
    year = "2015"
}

@article{Emonts:2020drm,
    author = "Emonts, Patrick and Ba{\~n}uls, Mari Carmen and Cirac, Ignacio and Zohar, Erez",
    title = "{Variational Monte Carlo simulation with tensor networks of a pure $\mathbb{Z}_3$ gauge theory in (2+1)d}",
    eprint = "2008.00882",
    archivePrefix = "arXiv",
    primaryClass = "quant-ph",
    doi = "10.1103/PhysRevD.102.074501",
    journal = "Phys. Rev. D",
    volume = "102",
    number = "7",
    pages = "074501",
    year = "2020"
}

@article{Buyens:2015tea,
    author = "Buyens, Boye and Haegeman, Jutho and Verschelde, Henri and Verstraete, Frank and Van Acoleyen, Karel",
    title = "{Confinement and string breaking for QED$_2$ in the Hamiltonian picture}",
    eprint = "1509.00246",
    archivePrefix = "arXiv",
    primaryClass = "hep-lat",
    doi = "10.1103/PhysRevX.6.041040",
    journal = "Phys. Rev. X",
    volume = "6",
    number = "4",
    pages = "041040",
    year = "2016"
}

@article{Buyens:2017crb,
    author = "Buyens, Boye and Montangero, Simone and Haegeman, Jutho and Verstraete, Frank and Van Acoleyen, Karel",
    title = "{Finite-representation approximation of lattice gauge theories at the continuum limit with tensor networks}",
    eprint = "1702.08838",
    archivePrefix = "arXiv",
    primaryClass = "hep-lat",
    doi = "10.1103/PhysRevD.95.094509",
    journal = "Phys. Rev. D",
    volume = "95",
    number = "9",
    pages = "094509",
    year = "2017"
}

@article{Dempsey:2023gib,
    author = "Dempsey, Ross and Klebanov, Igor R. and Pufu, Silviu S. and S{\o}gaard, Benjamin T. and Zan, Bernardo",
    title = "{Phase Diagram of the Two-Flavor Schwinger Model at Zero Temperature}",
    eprint = "2305.04437",
    archivePrefix = "arXiv",
    primaryClass = "hep-th",
    reportNumber = "PUPT-2640",
    doi = "10.1103/PhysRevLett.132.031603",
    journal = "Phys. Rev. Lett.",
    volume = "132",
    number = "3",
    pages = "031603",
    year = "2024"
}

@article{Itou:2024psm,
    author = "Itou, Etsuko and Matsumoto, Akira and Tanizaki, Yuya",
    title = "{DMRG study of the theta-dependent mass spectrum in the 2-flavor Schwinger model}",
    eprint = "2407.11391",
    archivePrefix = "arXiv",
    primaryClass = "hep-lat",
    reportNumber = "YITP-24-79, RIKEN-iTHEMS-Report-24",
    doi = "10.1007/JHEP09(2024)155",
    journal = "JHEP",
    volume = "09",
    pages = "155",
    year = "2024"
}

@article{ArguelloCruz:2024xzi,
    author = "Arguello Cruz, Erick and Tarnopolsky, Grigory and Xin, Yuan",
    title = "{Precision study of the massive Schwinger model near quantum criticality}",
    eprint = "2412.01902",
    archivePrefix = "arXiv",
    primaryClass = "hep-th",
    doi = "10.1103/vf7m-kzk2",
    journal = "Phys. Rev. D",
    volume = "112",
    number = "3",
    pages = "034023",
    year = "2025"
}

@article{Fujii:2024reh,
    author = "Fujii, Hirotsugu and Fujikura, Kohei and Kikukawa, Yoshio and Okuda, Takuya and Pedersen, Juan W.",
    title = "{Critical behavior of the Schwinger model via gauge-invariant variational uniform matrix product states}",
    eprint = "2412.03569",
    archivePrefix = "arXiv",
    primaryClass = "hep-lat",
    reportNumber = "UT-Komaba/24-10",
    doi = "10.1103/PhysRevD.111.094505",
    journal = "Phys. Rev. D",
    volume = "111",
    number = "9",
    pages = "094505",
    year = "2025"
}

@article{Shimizu:2014uva,
      author         = "Shimizu, Yuya and Kuramashi, Yoshinobu",
      title          = "{Grassmann tensor renormalization group approach to
                        one-flavor lattice Schwinger model}",
      journal        = "Phys. Rev. D",
      volume         = "90",
      year           = "2014",
      number         = "1",
      pages          = "014508",
      doi            = "10.1103/PhysRevD.90.014508",
      eprint         = "1403.0642",
      archivePrefix  = "arXiv",
      primaryClass   = "hep-lat",
      SLACcitation   = "%%CITATION = ARXIV:1403.0642;%%"
}

@article{Shimizu:2014fsa,
      author         = "Shimizu, Yuya and Kuramashi, Yoshinobu",
      title          = "{Critical behavior of the lattice Schwinger model with a
                        topological term at $\theta=\pi$ using the Grassmann
                        tensor renormalization group}",
      journal        = "Phys. Rev. D",
      volume         = "90",
      year           = "2014",
      number         = "7",
      pages          = "074503",
      doi            = "10.1103/PhysRevD.90.074503",
      eprint         = "1408.0897",
      archivePrefix  = "arXiv",
      primaryClass   = "hep-lat",
      SLACcitation   = "%%CITATION = ARXIV:1408.0897;%%"
}

@article{Shimizu:2017onf,
      author         = "Shimizu, Yuya and Kuramashi, Yoshinobu",
      title          = "{Berezinskii-Kosterlitz-Thouless transition in lattice
                        Schwinger model with one flavor of Wilson fermion}",
      journal        = "Phys. Rev. D",
      volume         = "97",
      year           = "2018",
      number         = "3",
      pages          = "034502",
      doi            = "10.1103/PhysRevD.97.034502",
      eprint         = "1712.07808",
      archivePrefix  = "arXiv",
      primaryClass   = "hep-lat",
      reportNumber   = "UTHEP-711, UTCCS-P-109",
      SLACcitation   = "%%CITATION = ARXIV:1712.07808;%%"
}

@article{Butt:2019uul,
    author = "Butt, Nouman and Catterall, Simon and Meurice, Yannick and Sakai, Ryo and Unmuth-Yockey, Judah",
    title = "{Tensor network formulation of the massless Schwinger model with staggered fermions}",
    eprint = "1911.01285",
    archivePrefix = "arXiv",
    primaryClass = "hep-lat",
    doi = "10.1103/PhysRevD.101.094509",
    journal = "Phys. Rev. D",
    volume = "101",
    number = "9",
    pages = "094509",
    year = "2020"
}

@article{Kanno:2024elz,
    author = "Kanno, Hayato and Akiyama, Shinichiro and Murakami, Kotaro and Takeda, Shinji",
    title = "{Grassmann tensor renormalization group for the massive Schwinger model with a {\ensuremath{\theta}} term using staggered fermions}",
    eprint = "2412.08959",
    archivePrefix = "arXiv",
    primaryClass = "hep-lat",
    reportNumber = "UTCCS-P-156, KANAZAWA 24-07, RIKEN-iTHEMS-Report-24",
    doi = "10.1007/JHEP11(2025)036",
    journal = "JHEP",
    volume = "11",
    pages = "036",
    year = "2025"
}

@article{Ohata:2023sqc,
    author = "Ohata, Hiroki",
    title = "{Monte Carlo study of Schwinger model without the sign problem}",
    eprint = "2303.05481",
    archivePrefix = "arXiv",
    primaryClass = "hep-lat",
    reportNumber = "YITP-23-29",
    doi = "10.1007/JHEP12(2023)007",
    journal = "JHEP",
    volume = "12",
    pages = "007",
    year = "2023"
}

@article{Ohata:2023gru,
    author = "Ohata, Hiroki",
    title = "{Phase diagram near the quantum critical point in Schwinger model at~$\theta = \pi$: analogy with quantum Ising chain}",
    eprint = "2311.04738",
    archivePrefix = "arXiv",
    primaryClass = "hep-lat",
    reportNumber = "YITP-23-140",
    doi = "10.1093/ptep/ptad151",
    journal = "PTEP",
    volume = "2024",
    number = "1",
    pages = "013B02",
    year = "2024"
}

@article{Levin:2006jai,
      author         = "Levin, Michael and Nave, Cody P.",
      title          = "{Tensor renormalization group approach to two-dimensional classical
                        lattice models}",
      journal        = "Phys. Rev. Lett.",
      volume         = "99",
      year           = "2007",
      number         = "12",
      pages          = "120601",
      doi            = "10.1103/PhysRevLett.99.120601",
      eprint         = "cond-mat/0611687",
      archivePrefix  = "arXiv",
      primaryClass   = "cond-mat.stat-mech",
      SLACcitation   = "%%CITATION = COND-MAT/0611687;%%"
}

@article{Evenbly:2015ucs,
    author = "Evenbly, Glen and Vidal, Guifre",
    title = "{Tensor Network Renormalization}",
    eprint = "1412.0732",
    archivePrefix = "arXiv",
    primaryClass = "cond-mat.str-el",
    doi = "10.1103/PhysRevLett.115.180405",
    journal = "Phys. Rev. Lett.",
    volume = "115",
    number = "18",
    pages = "180405",
    year = "2015"
}

@article{Felser:2019xyv,
    author = "Felser, Timo and Silvi, Pietro and Collura, Mario and Montangero, Simone",
    title = "{Two-dimensional quantum-link lattice Quantum Electrodynamics at finite density}",
    eprint = "1911.09693",
    archivePrefix = "arXiv",
    primaryClass = "quant-ph",
    reportNumber = "Phys. Rev. X 10, 041040 (2020)",
    doi = "10.1103/PhysRevX.10.041040",
    journal = "Phys. Rev. X",
    volume = "10",
    number = "4",
    pages = "041040",
    year = "2020"
}

@article{Robaina:2020aqh,
    author = "Robaina, Daniel and Ba{\~n}uls, Mari Carmen and Cirac, J. Ignacio",
    title = "{Simulating $2+1D$ $Z_3$ Lattice Gauge Theory with an Infinite Projected Entangled-Pair State}",
    eprint = "2007.11630",
    archivePrefix = "arXiv",
    primaryClass = "hep-lat",
    doi = "10.1103/PhysRevLett.126.050401",
    journal = "Phys. Rev. Lett.",
    volume = "126",
    number = "5",
    pages = "050401",
    year = "2021"
}

@article{Magnifico:2020bqt,
    author = "Magnifico, Giuseppe and Felser, Timo and Silvi, Pietro and Montangero, Simone",
    title = "{Lattice quantum electrodynamics in (3+1)-dimensions at finite density with tensor networks}",
    eprint = "2011.10658",
    archivePrefix = "arXiv",
    primaryClass = "hep-lat",
    doi = "10.1038/s41467-021-23646-3",
    journal = "Nature Commun.",
    volume = "12",
    number = "1",
    pages = "3600",
    year = "2021"
}

@article{Bender:2023gwr,
    author = "Bender, Julian and Emonts, Patrick and Cirac, J. Ignacio",
    title = "{Variational Monte Carlo algorithm for lattice gauge theories with continuous gauge groups: A study of (2+1)-dimensional compact QED with dynamical fermions at finite density}",
    eprint = "2304.05916",
    archivePrefix = "arXiv",
    primaryClass = "hep-lat",
    doi = "10.1103/PhysRevResearch.5.043128",
    journal = "Phys. Rev. Res.",
    volume = "5",
    number = "4",
    pages = "043128",
    year = "2023"
}

@article{Kuramashi:2018mmi,
      author         = "Kuramashi, Yoshinobu and Yoshimura, Yusuke",
      title          = "{Three-dimensional finite temperature Z$_{2}$ gauge
                        theory with tensor network scheme}",
      journal        = "JHEP",
      volume         = "08",
      year           = "2019",
      pages          = "023",
      doi            = "10.1007/JHEP08(2019)023",
      eprint         = "1808.08025",
      archivePrefix  = "arXiv",
      primaryClass   = "hep-lat",
      SLACcitation   = "%%CITATION = ARXIV:1808.08025;%%"
}

@article{Bazavov:2015kka,
    author = "Bazavov, Alexei and Meurice, Yannick and Tsai, Shan-Wen and Unmuth-Yockey, Judah and Zhang, Jin",
    title = "{Gauge-invariant implementation of the Abelian Higgs model on optical lattices}",
    eprint = "1503.08354",
    archivePrefix = "arXiv",
    primaryClass = "hep-lat",
    reportNumber = "INT-PUB-15-008, INT-PUB-15-008",
    doi = "10.1103/PhysRevD.92.076003",
    journal = "Phys. Rev. D",
    volume = "92",
    number = "7",
    pages = "076003",
    year = "2015"
}

@article{Unmuth-Yockey:2018xak,
    author = "Unmuth-Yockey, Judah F.",
    title = "{Gauge-invariant rotor Hamiltonian from dual variables of 3D $U(1)$ gauge theory}",
    eprint = "1811.05884",
    archivePrefix = "arXiv",
    primaryClass = "hep-lat",
    doi = "10.1103/PhysRevD.99.074502",
    journal = "Phys. Rev. D",
    volume = "99",
    number = "7",
    pages = "074502",
    year = "2019"
}

@article{Wu:2025aly,
    author = "Wu, Yantao and Liu, Wen-Yuan",
    title = "{Accurate Gauge-Invariant Tensor-Network Simulations for Abelian Lattice Gauge Theory in (2+1)D: Ground-State and Real-Time Dynamics}",
    eprint = "2503.20566",
    archivePrefix = "arXiv",
    primaryClass = "cond-mat.str-el",
    doi = "10.1103/3m3j-ds18",
    journal = "Phys. Rev. Lett.",
    volume = "135",
    number = "13",
    pages = "130401",
    year = "2025"
}

@article{Kuramashi:2019cgs,
    author = "Kuramashi, Yoshinobu and Yoshimura, Yusuke",
    archivePrefix = "arXiv",
    doi = "10.1007/JHEP04(2020)089",
    eprint = "1911.06480",
    journal = "JHEP",
    pages = "089",
    primaryClass = "hep-lat",
    title = "{Tensor renormalization group study of two-dimensional U(1) lattice gauge theory with a $\theta$ term}",
    volume = "04",
    year = "2020"
}

@article{Kuwahara:2022ubg,
    author = "Kuwahara, Takaaki and Tsuchiya, Asato",
    title = "{Toward tensor renormalization group study of three-dimensional non-Abelian gauge theory}",
    eprint = "2205.08883",
    archivePrefix = "arXiv",
    primaryClass = "hep-lat",
    doi = "10.1093/ptep/ptac103",
    journal = "PTEP",
    volume = "2022",
    number = "9",
    pages = "093B02",
    year = "2022"
}

@article{Akiyama:2022eip,
    author = "Akiyama, Shinichiro and Kuramashi, Yoshinobu",
    title = "{Tensor renormalization group study of (3+1)-dimensional Z$_{2}$ gauge-Higgs model at finite density}",
    eprint = "2202.10051",
    archivePrefix = "arXiv",
    primaryClass = "hep-lat",
    reportNumber = "UTHEP-769, UTCCS-P-143",
    doi = "10.1007/JHEP05(2022)102",
    journal = "JHEP",
    volume = "05",
    pages = "102",
    year = "2022"
}

@article{Akiyama:2023hvt,
    author = "Akiyama, Shinichiro and Kuramashi, Yoshinobu",
    title = "{Critical endpoint of (3+1)-dimensional finite density Z$_{3}$ gauge-Higgs model with tensor renormalization group}",
    eprint = "2304.07934",
    archivePrefix = "arXiv",
    primaryClass = "hep-lat",
    reportNumber = "UTHEP-780, UTCCS-P-147",
    doi = "10.1007/JHEP10(2023)077",
    journal = "JHEP",
    volume = "10",
    pages = "077",
    year = "2023"
}

@article{Sugimoto:2025vui,
    author = "Sugimoto, Yuto and Akiyama, Shinichiro and Kuramashi, Yoshinobu",
    title = "{Phase structure of (3+1)-dimensional dense two-color QCD at $T=0$ in the strong coupling limit with tensor renormalization group}",
    eprint = "2509.23637",
    archivePrefix = "arXiv",
    primaryClass = "hep-lat",
    journal = {Phys. Rev. D},
    volume = {113},
    issue = {3},
    pages = {034503},
    numpages = {10},
    year = {2026},
    month = {Feb},
    publisher = {American Physical Society},
    doi = {10.1103/jd1r-cqrc}
}

@article{Sugimoto:2026wnw,
    author = "Sugimoto, Yuto and Akiyama, Shinichiro and Kuramashi, Yoshinobu",
    title = "{Tensor renormalization group study of cold and dense QCD in the strong coupling limit}",
    eprint = "2601.20690",
    archivePrefix = "arXiv",
    primaryClass = "hep-lat",
    month = "1",
    year = "2026"
}

@article{Evenbly2017,
    author = "Evenbly, Glen",
    title = "{Algorithms for tensor network renormalization}",
    eprint = "1509.07484",
    archivePrefix = "arXiv",
    primaryClass = "cond-mat.str-el",
    doi = "10.1103/PhysRevB.95.045117",
    journal = "Phys. Rev. B",
    volume = "95",
    number = "4",
    pages = "045117",
    year = "2017"
}

@article{Hauru2018,
      author         = "Hauru, Markus and Delcamp, Clement and Mizera, Sebastian",
      title          = "{Renormalization of tensor networks using graph
                        independent local truncations}",
      journal        = "Phys. Rev. B",
      volume         = "97",
      year           = "2018",
      number         = "4",
      pages          = "045111",
      doi            = "10.1103/PhysRevB.97.045111",
      eprint         = "1709.07460",
      archivePrefix  = "arXiv",
      primaryClass   = "cond-mat.str-el",
      SLACcitation   = "%%CITATION = ARXIV:1709.07460;%%"
}

@article{Bal2017,
    author = {Bal, Matthias and Mari{\"e}n, Micha{\"e}l and Haegeman, Jutho and Verstraete, Frank},
    title = "{Renormalization group flows of Hamiltonians using tensor networks}",
    eprint = "1703.00365",
    archivePrefix = "arXiv",
    primaryClass = "cond-mat.stat-mech",
    doi = "10.1103/PhysRevLett.118.250602",
    journal = "Phys. Rev. Lett.",
    volume = "118",
    pages = "250602",
    year = "2017"
}

@article{Homma2024,
    author = "Homma, Kenji and Okubo, Tsuyoshi and Kawashima, Naoki",
    title = "{Nuclear norm regularized loop optimization for tensor network}",
    eprint = "2306.17479",
    archivePrefix = "arXiv",
    primaryClass = "cond-mat.stat-mech",
    doi = "10.1103/PhysRevResearch.6.043102",
    journal = "Phys. Rev. Res.",
    volume = "6",
    number = "4",
    pages = "043102",
    year = "2024"
}

@article{boundarytrg,
      author         = "Iino, Shumpei and Morita, Satoshi and Kawashima, Naoki",
      title          = "{Boundary Tensor Renormalization Group}",
      journal        = "Phys. Rev. B",
      volume         = "100",
      year           = "2019",
      number         = "3",
      pages          = "035449",
      doi            = "10.1103/PhysRevB.100.035449",
      eprint         = "1905.02351",
      archivePrefix  = "arXiv",
      primaryClass   = "cond-mat.stat-mech",
      SLACcitation   = "%%CITATION = ARXIV:1905.02351;%%"
}

@article{looptnr,
    author = "Yang, Shuo and Gu, Zheng-Cheng and Wen, Xiao-Gang",
    title = "{Loop Optimization for Tensor Network Renormalization}",
    eprint = "1512.04938",
    archivePrefix = "arXiv",
    primaryClass = "cond-mat.str-el",
    doi = "10.1103/PhysRevLett.118.110504",
    journal = "Phys. Rev. Lett.",
    volume = "118",
    number = "11",
    pages = "110504",
    year = "2017"
}

@article{HOTRG,
    author = "Xie, Z. Y. and Chen, J. and Qin, M. P. and Zhu, J. W. and Yang, L. P. and Xiang, T.",
    title = "{Coarse-graining renormalization by higher-order singular value decomposition}",
    eprint = "1201.1144",
    archivePrefix = "arXiv",
    primaryClass = "cond-mat.stat-mech",
    doi = "10.1103/PhysRevB.86.045139",
    journal = "Phys. Rev. B",
    volume = "86",
    number = "4",
    pages = "045139",
    year = "2012"
}

@article{Akiyama:2021xxr,
    author = "Akiyama, Shinichiro and Kuramashi, Yoshinobu",
    title = "{Tensor renormalization group approach to (1+1)-dimensional Hubbard model}",
    eprint = "2105.00372",
    archivePrefix = "arXiv",
    primaryClass = "hep-lat",
    reportNumber = "UTHEP-756, UTCCS-P-137",
    doi = "10.1103/PhysRevD.104.014504",
    journal = "Phys. Rev. D",
    volume = "104",
    number = "1",
    pages = "014504",
    year = "2021"
}

@article{Akiyama:2021glo,
    author = "Akiyama, Shinichiro and Kuramashi, Yoshinobu and Yamashita, Takumi",
    title = "{Metal-insulator transition in (2+1)-dimensional Hubbard model with tensor renormalization group}",
    eprint = "2109.14149",
    archivePrefix = "arXiv",
    primaryClass = "cond-mat.str-el",
    reportNumber = "UTHEP-760, UTCCS-P-139",
    doi = "10.1093/ptep/ptac014",
    journal = "PTEP",
    volume = "2022",
    pages = "023",
    month = "9",
    year = "2021"
}

@article{BWTRG,
  title = {Bond-weighted tensor renormalization group},
  author = {Adachi, Daiki and Okubo, Tsuyoshi and Todo, Synge},
  journal = {Phys. Rev. B},
  volume = {105},
  issue = {6},
  pages = {L060402},
  numpages = {6},
  year = {2022},
  month = {Feb},
  publisher = {American Physical Society},
  doi = {10.1103/PhysRevB.105.L060402},
  eprint         = "2011.01679",
  archivePrefix  = "arXiv",
  primaryClass   = "cond-mat.stat-mech"
}

@article{Ueda:2025mhu,
    author = "Ueda, Atsushi and De Meyer, Sander and Naravane, Adwait and Vanthilt, Victor and Verstraete, Frank",
    title = "{Global Tensor Network Renormalization for 2D Quantum systems: A new window to probe universal data from thermal transitions}",
    eprint = "2508.05406",
    archivePrefix = "arXiv",
    primaryClass = "cond-mat.str-el",
    month = "8",
    year = "2025"
}

@article{Ueda_2023,
    author = "Ueda, Atsushi and Oshikawa, Masaki",
    title = "{Finite-size and finite bond dimension effects of tensor network renormalization}",
    eprint = "2302.06632",
    archivePrefix = "arXiv",
    primaryClass = "cond-mat.stat-mech",
    doi = "10.1103/PhysRevB.108.024413",
    journal = "Phys. Rev. B",
    volume = "108",
    number = "2",
    pages = "024413",
    year = "2023"
}

@article{lyu2025latticereflectionsymmetryTNR,
    author = "Lyu, Xinliang and Kawashima, Naoki",
    title = "{Lattice-reflection symmetry in tensor-network renormalization group with entanglement filtering in two and three dimensions}",
    eprint = "2510.19428",
    archivePrefix = "arXiv",
    primaryClass = "cond-mat.stat-mech",
    doi = "10.1103/z7sk-7sq8",
    journal = "Phys. Rev. B",
    volume = "113",
    number = "14",
    pages = "144109",
    year = "2026"
}

@article{Gu:2009dr,
    author = "Gu, Zheng-Cheng and Wen, Xiao-Gang",
    title = "{Tensor-Entanglement-Filtering Renormalization Approach and Symmetry Protected Topological Order}",
    eprint = "0903.1069",
    archivePrefix = "arXiv",
    primaryClass = "cond-mat.str-el",
    doi = "10.1103/PhysRevB.80.155131",
    journal = "Phys. Rev. B",
    volume = "80",
    pages = "155131",
    year = "2009"
}

@article{Liu:2013nsa,
      author         = "Liu, Yuzhi and Meurice, Y. and Qin, M. P. and
                        Unmuth-Yockey, J. and Xiang, T. and Xie, Z. Y. and Yu, J.
                        F. and Zou, Haiyuan",
      title          = "{Exact Blocking Formulas for Spin and Gauge Models}",
      journal        = "Phys. Rev. D",
      volume         = "88",
      year           = "2013",
      pages          = "056005",
      doi            = "10.1103/PhysRevD.88.056005",
      eprint         = "1307.6543",
      archivePrefix  = "arXiv",
      primaryClass   = "hep-lat",
      reportNumber   = "FERMILAB-PUB-13-280-T",
      SLACcitation   = "%%CITATION = ARXIV:1307.6543;%%"
}

@article{Meurice:2020gcd,
    author = "Meurice, Yannick",
    title = "{Discrete aspects of continuous symmetries in the tensorial formulation of Abelian gauge theories}",
    eprint = "2003.10986",
    archivePrefix = "arXiv",
    primaryClass = "hep-lat",
    doi = "10.1103/PhysRevD.102.014506",
    journal = "Phys. Rev. D",
    volume = "102",
    number = "1",
    pages = "014506",
    year = "2020"
}

@article{Shimizu:2024ipw,
    author = "Shimizu, Haruki and Ueda, Atsushi",
    title = "{Tensor network simulations for non-orientable surfaces}",
    eprint = "2402.15507",
    archivePrefix = "arXiv",
    primaryClass = "cond-mat.str-el",
    doi = "10.48550/arXiv.2402.15507",
    month = "2",
    year = "2024"
}

@article{Morita:2024lwg,
    author = "Morita, Satoshi and Kawashima, Naoki",
    title = "{Multi-impurity method for the bond-weighted tensor renormalization group}",
    eprint = "2411.13998",
    archivePrefix = "arXiv",
    primaryClass = "cond-mat.stat-mech",
    doi = "10.1103/PhysRevB.111.054433",
    journal = "Phys. Rev. B",
    volume = "111",
    number = "5",
    pages = "054433",
    year = "2025"
}

@article{Morita:2025hsv,
    author = "Morita, Satoshi and Kawashima, Naoki",
    title = "{Tensor Renormalization Group Calculations of Partition-Function Ratios}",
    eprint = "2512.03395",
    archivePrefix = "arXiv",
    primaryClass = "cond-mat.stat-mech",
    doi = "10.7566/JPSJ.95.044001",
    journal = "J. Phys. Soc. Jap.",
    volume = "95",
    pages = "044001",
    year = "2026"
}

@article{Maeda:2025ycr,
    author = "Maeda, Jun and Tanizaki, Yuya",
    title = "{Twisted partition functions as order parameters}",
    eprint = "2505.16546",
    archivePrefix = "arXiv",
    primaryClass = "hep-th",
    reportNumber = "YITP-25-42, KUNS-3051",
    doi = "10.1007/JHEP08(2025)128",
    journal = "JHEP",
    volume = "08",
    pages = "128",
    year = "2025"
}

@article{Akiyama:2026dzg,
    author = "Akiyama, Shinichiro and Jha, Raghav G. and Maeda, Jun and Tanizaki, Yuya and Unmuth-Yockey, Judah",
    title = "{Tensor renormalization group approach to critical phenomena via symmetry-twisted partition functions}",
    eprint = "2601.02681",
    archivePrefix = "arXiv",
    primaryClass = "hep-lat",
    reportNumber = "UTHEP-814, UTCCS-P-172, KUNS-3083, YITP-25-180",
    doi = "10.1103/ywhr-lt6w",
    journal = "Phys. Rev. D",
    volume = "113",
    number = "7",
    pages = "074502",
    year = "2026"
}

@article{Li:2020sbg,
    author = "Li, Guanrong and Pai, Kwok Ho and Gu, Zheng-Cheng",
    title = "{Tensor-network renormalization approach to the q-state clock model}",
    eprint = "2009.10695",
    archivePrefix = "arXiv",
    primaryClass = "cond-mat.stat-mech",
    doi = "10.1103/PhysRevResearch.4.023159",
    journal = "Phys. Rev. Res.",
    volume = "4",
    number = "2",
    pages = "023159",
    year = "2022"
}

@article{Pai:2024tip,
    author = "Pai, Kwok Ho and Akiyama, Shinichiro and Todo, Synge",
    title = "{Grassmann tensor renormalization group approach to (1+1)-dimensional two-color lattice QCD at finite density}",
    eprint = "2410.09485",
    archivePrefix = "arXiv",
    primaryClass = "hep-lat",
    doi = "10.1007/JHEP03(2025)027",
    journal = "JHEP",
    volume = "03",
    pages = "027",
    year = "2025"
}

@article{Caselle:1995wn,
    author = "Caselle, M. and Hasenbusch, M.",
    title = "{Deconfinement transition and dimensional crossover in the 3-D gauge Ising model}",
    eprint = "hep-lat/9511015",
    archivePrefix = "arXiv",
    reportNumber = "DFTT-68-95, DAMTP-95-60",
    doi = "10.1016/0550-3213(96)00161-7",
    journal = "Nucl. Phys. B",
    volume = "470",
    pages = "435--453",
    year = "1996"
}

@article{Borisenko:2014vva,
    author = "Borisenko, Oleg and Chelnokov, Volodymyr and Gravina, Mario and Papa, Alessandro",
    title = "{Phase structure of 3D Z(N) lattice gauge theories at finite temperature: large-N and continuum limits}",
    eprint = "1408.2780",
    archivePrefix = "arXiv",
    primaryClass = "hep-lat",
    doi = "10.1016/j.nuclphysb.2014.09.004",
    journal = "Nucl. Phys. B",
    volume = "888",
    pages = "52--64",
    year = "2014"
}

@article{Borisenko:2013xna,
    author = "Borisenko, O. and Chelnokov, V. and Cortese, G. and Gravina, M. and Papa, A. and Surzhikov, I.",
    title = "{Critical behavior of 3D Z(N) lattice gauge theories at zero temperature}",
    eprint = "1310.5997",
    archivePrefix = "arXiv",
    primaryClass = "hep-lat",
    doi = "10.1016/j.nuclphysb.2013.12.003",
    journal = "Nucl. Phys. B",
    volume = "879",
    pages = "80--97",
    year = "2014"
}

@article{Ferrenberg:2018zst,
    author = "Ferrenberg, Alan M. and Xu, Jiahao and Landau, David P.",
    title = "{Pushing the limits of Monte Carlo simulations for the three-dimensional Ising model}",
    eprint = "1806.03558",
    archivePrefix = "arXiv",
    primaryClass = "physics.comp-ph",
    doi = "10.1103/PhysRevE.97.043301",
    journal = "Phys. Rev. E",
    volume = "97",
    number = "4",
    pages = "043301",
    year = "2018"
}

@article{Wada:2025ycz,
    author = "Wada, Tatsuya and Kitazawa, Masakiyo and Kanaya, Kazuyuki",
    title = "{Lee-Yang-zero ratio method in three-dimensional Ising model}",
    eprint = "2508.20422",
    archivePrefix = "arXiv",
    primaryClass = "cond-mat.stat-mech",
    reportNumber = "YITP-25-128, J-PARC-TH-0324, UTHEP-810",
    doi = "10.7566/JPSJ.95.024002",
    journal = "J. Phys. Soc. Jap.",
    volume = "95",
    pages = "024002",
    year = "2026"
}

@article{Bazavov:2007tw,
    author = "Bazavov, Alexei and Berg, Bernd A.",
    title = "{Normalized entropy density of the 3D 3-state Potts model}",
    eprint = "hep-lat/0702018",
    archivePrefix = "arXiv",
    doi = "10.1103/PhysRevD.75.094506",
    journal = "Phys. Rev. D",
    volume = "75",
    pages = "094506",
    year = "2007"
}

@article{Kos:2016ysd,
    author = "Kos, Filip and Poland, David and Simmons-Duffin, David and Vichi, Alessandro",
    title = "{Precision Islands in the Ising and $O(N)$ Models}",
    eprint = "1603.04436",
    archivePrefix = "arXiv",
    primaryClass = "hep-th",
    reportNumber = "CERN-TH-2016-050",
    doi = "10.1007/JHEP08(2016)036",
    journal = "JHEP",
    volume = "08",
    pages = "036",
    year = "2016"
}

@article{Brower:1991ip,
    author = "Brower, R. C. and Huang, Suzhou",
    title = "{Dynamical universality for Z(2) and Z(3) lattice gauge theories at finite temperature}",
    reportNumber = "BUHEP-91-1",
    doi = "10.1103/PhysRevD.44.3911",
    journal = "Phys. Rev. D",
    volume = "44",
    pages = "3911--3917",
    year = "1991"
}

@article{PhysRevLett.43.799,
  title = {First-Order Phase Transitions and the Three-State Potts Model},
  author = {Bl\"ote, H. W. J. and Swendsen, R. H.},
  journal = {Phys. Rev. Lett.},
  volume = {43},
  issue = {11},
  pages = {799--802},
  numpages = {0},
  year = {1979},
  month = {Sep},
  publisher = {American Physical Society},
  doi = {10.1103/PhysRevLett.43.799},
  url = {https://link.aps.org/doi/10.1103/PhysRevLett.43.799}
}

@article{Yosprakob:2024sfd,
    author = "Yosprakob, Atis and Okunishi, Kouichi",
    title = "{Tensor Renormalization Group Study of the 3D SU(2) and SU(3) Gauge Theories with the Reduced Tensor Network Formulation}",
    eprint = "2406.16763",
    archivePrefix = "arXiv",
    primaryClass = "hep-lat",
    doi = "10.1093/ptep/ptaf028",
    journal = "PTEP",
    volume = "2025",
    number = "3",
    pages = "033B06",
    year = "2025"
}

@article{Nakayama:2024lhb,
    author = "Nakayama, Katsumasa and Schneider, Manuel",
    title = "{Initial tensor construction and dependence of the tensor renormalization group on initial tensors}",
    eprint = "2407.14226",
    archivePrefix = "arXiv",
    primaryClass = "hep-lat",
    doi = "10.1103/PhysRevD.110.094501",
    journal = "Phys. Rev. D",
    volume = "110",
    pages = "094501",
    year = "2024"
}

@article{Fishman:2018lnr,
    author = "Fishman, M. T. and Vanderstraeten, L. and Zauner-Stauber, V. and Haegeman, J. and Verstraete, F.",
    title = "{Faster methods for contracting infinite two-dimensional tensor networks}",
    eprint = "1711.05881",
    archivePrefix = "arXiv",
    primaryClass = "cond-mat.str-el",
    doi = "10.1103/PhysRevB.98.235148",
    journal = "Phys. Rev. B",
    volume = "98",
    number = "23",
    pages = "235148",
    year = "2018"
}

@article{Luttinger:1963zz,
    author = "Luttinger, J. M.",
    title = "{An Exactly Soluble Model of a Many-Fermion System}",
    doi = "10.1063/1.1704046",
    journal = "J. Math. Phys.",
    volume = "4",
    pages = "1154--1162",
    year = "1963"
}

@article{Tomonaga:1950zz,
    author = "Tomonaga, S.",
    title = "{Remarks on Bloch's Method of Sound Waves applied to Many-Fermion Problems}",
    doi = "10.1143/PTP.5.544",
    journal = "Prog. Theor. Phys.",
    volume = "5",
    pages = "544--569",
    year = "1950"
}

@Misc{qsw,
  title        = "",
  howpublished = "\url{https://qsw.phys.s.u-tokyo.ac.jp/}"
}

@article{Sugimoto:2026zxv,
    author = "Sugimoto, Yuto",
    title = "{Forward-mode automatic differentiation for the tensor renormalization group and its relation to the impurity method}",
    eprint = "2602.08987",
    archivePrefix = "arXiv",
    primaryClass = "hep-lat",
    doi = "10.1103/9wb1-rrcn",
    journal = "Phys. Rev. D",
    volume = "113",
    number = "9",
    pages = "094502",
    year = "2026"
}

@article{Li:2019dkb,
    author = "Li, Zi-Qian and Yang, Li-Ping and Xie, Z. Y. and Tu, Hong-Hao and Liao, Hai-Jun and Xiang, T.",
    title = "{Critical properties of the two-dimensional $q$-state clock model}",
    eprint = "1912.11416",
    archivePrefix = "arXiv",
    primaryClass = "cond-mat.stat-mech",
    doi = "10.1103/PhysRevE.101.060105",
    journal = "Phys. Rev. E",
    volume = "101",
    number = "6",
    pages = "060105",
    year = "2020"
}

@misc{devos2025tensorkitjljuliapackagelargescale,
      title={TensorKit.jl: A Julia package for large-scale tensor computations, with a hint of category theory}, 
      author={Lukas Devos and Jutho Haegeman},
      year={2025},
      eprint={2508.10076},
      archivePrefix={arXiv},
      primaryClass={cs.MS},
      url={https://arxiv.org/abs/2508.10076}, 
}

@misc{vanthilt2026practicalintroductiontensornetwork,
      title={A Practical Introduction to Tensor Network Renormalization with TNRKit.jl}, 
      author={Victor Vanthilt and Adwait Naravane and Chenqi Meng and Atsushi Ueda},
      year={2026},
      eprint={2604.06922},
      archivePrefix={arXiv},
      primaryClass={cond-mat.str-el},
      url={https://arxiv.org/abs/2604.06922}, 
}

@article{Ba_uls_2020,
   title={Simulating lattice gauge theories within quantum technologies},
   volume={74},
   ISSN={1434-6079},
   url={http://dx.doi.org/10.1140/epjd/e2020-100571-8},
   DOI={10.1140/epjd/e2020-100571-8},
   number={8},
   journal={The European Physical Journal D},
   publisher={Springer Science and Business Media LLC},
   author={Bañuls, Mari Carmen and Blatt, Rainer and Catani, Jacopo and Celi, Alessio and Cirac, Juan Ignacio and Dalmonte, Marcello and Fallani, Leonardo and Jansen, Karl and Lewenstein, Maciej and Montangero, Simone and Muschik, Christine A. and Reznik, Benni and Rico, Enrique and Tagliacozzo, Luca and Van Acoleyen, Karel and Verstraete, Frank and Wiese, Uwe-Jens and Wingate, Matthew and Zakrzewski, Jakub and Zoller, Peter},
   year={2020},
   month=aug }

@misc{canals2024tensornetworkformulationlattice,
      title={A tensor network formulation of Lattice Gauge Theories based only on symmetric tensors}, 
      author={Manu Canals and Natalia Chepiga and Luca Tagliacozzo},
      year={2024},
      eprint={2412.16961},
      archivePrefix={arXiv},
      primaryClass={hep-lat},
      url={https://arxiv.org/abs/2412.16961}, 
}

@article{TagliacozzoVidalGauge,
  title = {Entanglement renormalization and gauge symmetry},
  author = {Tagliacozzo, L. and Vidal, G.},
  journal = {Phys. Rev. B},
  volume = {83},
  issue = {11},
  pages = {115127},
  numpages = {31},
  year = {2011},
  month = {Mar},
  publisher = {American Physical Society},
  doi = {10.1103/PhysRevB.83.115127},
  url = {https://link.aps.org/doi/10.1103/PhysRevB.83.115127}
}

\end{document}